\documentclass[prd, aps, twocolumn, showpacs, nofootinbib,superscriptaddress]{revtex4-1}

\usepackage{amsbsy,amsmath,amssymb,amsthm,wasysym,amsfonts}
\usepackage{graphicx}
\usepackage{psfrag}
\usepackage{epstopdf}
\usepackage{color}
\usepackage{mathrsfs}
\usepackage{comment}

\DeclareSymbolFontAlphabet{\mathrsfs}{rsfs}
\DeclareMathAlphabet\mathbfcal{OMS}{cmsy}{b}{n}

\newcommand{\be}{\begin{equation}}  
\newcommand{\ee}{\end{equation}}
\newcommand{\bea}{\begin{eqnarray}}           
\newcommand{\eea}{\end{eqnarray}} 
\newcommand{\beqn}{\begin{eqnarray*}}
\newcommand{\eeqn}{\end{eqnarray*}}
\newcommand{\ba}{\begin{align}}
\newcommand{\ea}{\end{align}}

\def\de{\partial}

\def\Fmax{{\bar{F}_{\max}}}
\def\hOmg{{\hat{\Omega}}}
\def\p4{{\psi_4}} 
\def\ha{{\tilde{a}}}

\usepackage{float}
\definecolor{cyan}{rgb}{0,0.9,0.9}
\definecolor{orange}{rgb}{0.9,0.5,0}
\definecolor{magenta}{rgb}{1,0,1}
\definecolor{purple}{rgb}{0.8,0.4,0.8}

\begin{document}

\title{Energetics and phasing of nonprecessing spinning coalescing black hole binaries}

\author{Alessandro \surname{Nagar}}
\author{Thibault \surname{Damour}}
\affiliation{Institut des Hautes Etudes Scientifiques, 91440
  Bures-sur-Yvette, France}

\author{Christian \surname{Reisswig}}
\thanks{Einstein Fellow}
\affiliation{Theoretical Astrophysics Including Relativity, California Institute of
  Technology, Pasadena, CA 91125, USA} 

\author{Denis \surname{Pollney}}
\affiliation{Department of Mathematics, Rhodes University, 6140 Grahamstown, South Africa}

\begin{abstract}
We present an improved numerical relativity (NR) calibration of the new effective-one-body (EOB) 
model for coalescing non precessing spinning black hole binaries recently introduced by 
Damour and Nagar [Physical Review D {\bf 90}, 044018 (2014)]. 
We do so by comparing the EOB predictions to both the phasing and the energetics 
provided by two independent sets of NR data covering mass ratios $1\leq q \leq 9.989$
and dimensionless spin range $-0.95\leq \chi\leq +0.994$. One set of data is a subset 
of the Simulating eXtreme Spacetimes (SXS) catalog of public waveforms; the other 
set consists of new simulations obtained with the Llama code plus Cauchy Characteristic Evolution.
We present the first systematic computation of the gauge-invariant relation between 
the binding energy and the total angular momentum, $E_{b}(j)$, for a large sample 
of, spin-aligned, SXS and Llama data.
The dynamics of the EOB model presented here involves only two free functional parameters, 
one ($a_6^c(\nu)$) entering the non spinning sector, as a 5PN effective correction to the 
interaction potential, and one ($c_3(\ha_1,\ha_2,\nu))$ in the spinning sector,
as an effective next-to-next-to-next-to-leading order correction to the spin-orbit coupling.
These parameters are determined (together with a third functional parameter $\Delta t_{\rm NQC}(\chi)$
entering the waveform) by comparing the EOB phasing with the SXS phasing,
the consistency of the energetics being checked afterwards.
The quality of the analytical model for gravitational wave data analysis purposes is assessed 
by computing the EOB/NR faithfulness. 
Over the NR data sample and when varying the total mass between
20 and 200~$M_\odot$ the EOB/NR unfaithfulness (integrated over the NR frequency range)
is found to vary between $99.493\%$ and $99.984\%$ with a median value of $99.944\%$.
\end{abstract}
\date{\today}


\maketitle

\section{Introduction}

The purpose of this paper is to present an improved version of the  new
spinning effective-one-body (EOB) model introduced in Ref.~\cite{Damour:2014sva}, 
hereafter Paper~I.  [For earlier spinning EOB 
models see~\cite{Damour:2001tu,Buonanno:2005xu,Damour:2008qf,Barausse:2009xi,Pan:2009wj,Nagar:2011fx,
Barausse:2011ys,Pan:2011gk,Taracchini:2012ig,Taracchini:2013rva,Balmelli:2013zna,Balmelli:2015lva}.]
When the mass of the binary varies between 20 and 200~$M_\odot$,
our improved model yields, when compared to a large
sample of numerical relativity (NR) waveform data,
maximal unfaithfulnesses ($\bar{F}\equiv 1-F$, integrated over the NR frequency range) 
ranging between 0.00016 and 0.00507, with median value equal to 0.00056.
The impatient data-analyst reader will find this information 
in Figs.~\ref{Fbar} and ~\ref{histo} and the text around. 
The structural elements of our improved EOB model behind such 
good faithfulnesses are: (i) the incorporation of the full 
4PN-accurate analytical knowledge of the EOB radial interaction; and
(ii) the use of a recently proposed improved ringdown description~\cite{Damour:2014yha}.
The use of the 4PN information obliged us to update the 
calibration of both the effective 5PN-coefficient $a_6^c$
entering the nonspinning sector, and the effective next-to-next-to-next-to-leading
order spin-orbit coupling coefficient $c_3$ of Paper~I.

Here, we restrict attention to the nonprecessing case where the spins are 
either aligned or antialigned with respect to the orbital angular momentum.
The EOB/NR comparisons that are used to gain new information from NR data so 
as to complete the EOB model are of two kinds: (i) on the one hand, 
we compare the gravitational wave (GW) {\it phasing} of the EOB model 
with 40  state-of-the-art (publicly available~\cite{SXS:catalog}) 
NR waveforms produced by the Caltech-Cornell-CITA 
Simulating eXtreme Spacetimes  (SXS) collaboration with the Spectral 
Einstein Code ({\tt SpEC}) code~\cite{Chu:2009md,Lovelace:2010ne,Lovelace:2011nu,Buchman:2012dw,Mroue:2012kv,Mroue:2013xna,
Hemberger:2013hsa,Lovelace:2014twa,Scheel:2014ina,Blackman:2015pia};
eleven configurations in this sample involve nonspinning binaries, 
while in the remaining 29 at least one black hole is spinning;
(ii) on the other hand we compare the EOB and NR {\it energetics} through the 
gauge-invariant relation between energy and angular momentum. To do so, we employ, 
in addition to the SXS datasets, ten newly performed simulations obtained 
with the Llama code, as a follow up of our previous work~\cite{Damour:2011fu}.

The paper is organized as follows: in Sec.~\ref{sec:NRdata} we briefly review the
origin of the NR data used in this paper, which were obtained with very different codes.
In Secs.\ref{sec:nospin}-\ref{nospin_energetics} we focus on the improved calibration 
of the nonspinning sector, its phasing and energetics, notably showing the 
good agreement between energetics when ${\cal F}_{r_*}=0$ for the mass 
ratio range $1\leq q \leq 9.989$.
In Secs.~\ref{sec:spin}-\ref{sec:energetics} we calibrate the 
effective spin-orbit parameter $c_3$ and we assess the quality of
this new spinning EOB model by computing both the EOB/NR unfaithfulness
and the energetics. In Sec.~\ref{sec:analytics} we present a preliminary
comparison of our EOB model with that of 
Alessandra Buonanno's group~\cite{Taracchini:2013rva}. Finally, 
Sec.~\ref{end} presents concluding remarks, future prospects as well
as a histogram summarizing the unfaithfulness calculations.

We use geometrized units with $G=c=1$ and the following notation:
$M=m_1+m_2$, $\mu=m_1 m_2/M$, $\nu=\mu/M$ with the convention that 
the mass ratio $q\equiv m_1/m_2\geq 1$. 

\section{Numerical relativity data}
\label{sec:NRdata}

\subsection{Data from the SXS public catalog}
We use a sample of 40 simulations from the SXS catalog.
We consider either nonspinning or spin-aligned configurations,
with mass ratios in the range $1\leq q \leq 9.989$ and dimensionless
spin magnitude $-0.95\leq \chi \leq +0.994$.
In several cases, just one of the two black holes is spinning. 
The complete information about the data we used is listed in 
Table~\ref{tab:configs}. For each run, when it is possible, we 
add an error bar on phasing computed at the NR merger. 
Here and below the instant of ``merger'' is defined as the time 
of the location of the peak of the modulus of the $\ell=m=2$ waveform. 
This error bar was estimated by computing the value at (the highest-resolution-) 
merger of the phase difference between the  highest (H) and second-highest (SH) 
resolutions available  in the catalog, 
$\delta^{\rm NR} \phi(t)\equiv \phi^{\rm H}(t)-\phi^{\rm SH}(t)$.
This difference is monotonically 
varying~\footnote{$\delta^{\rm NR} \phi(t)$ is monotonically {\it increasing} 
with $t$ except for the datasets SXS:BBH:0178 and SXS:BBH:0065 where it is 
monotonically {\it decreasing}.} 
up to merger and, by definition, taken to be zero at the start of 
the two simulations.
When only one resolution is available we do not indicate 
any phasing uncertainty in Table~\ref{tab:configs}. 
In such cases one can check the consistency by
looking at datasets with neighbouring parameters.
In SXS simulations, the data are extracted at finite radius and then extrapolated
to future null infinity (e.g.~\cite{Taylor:2013zia, Boyle:2009vi}). 
This is done by means of a fit using a 
polynomial in $1/r$, where $r$ is the extraction radius. 
The user of the catalog is free to choose between different orders of extrapolation 
($N=2,3,4$) depending on the application, with the warning that the ``best'' 
extrapolation order to use depends on the simulation and that higher 
extrapolation orders tend to do better during the inspiral and worse during ringdown.
We found experimentally that extrapolation order $N=3$ is reliable for all datasets
except for SXS:BBH:0002, where we used $N=2$ to reduce unphysical oscillations
during the late inspiral. 
In addition to the data of the catalog, we also use a $\sim14$ orbits
$q=1$, $\chi=0$ waveform~\cite{Buchman:2012dw}, that was used 
to calibrate earlier EOB models~\cite{Pan:2011gk,Damour:2012ky}.

\subsection{Data from the Llama code}
\begin{figure}[t]
\begin{center}
\includegraphics[width=0.37\textwidth]{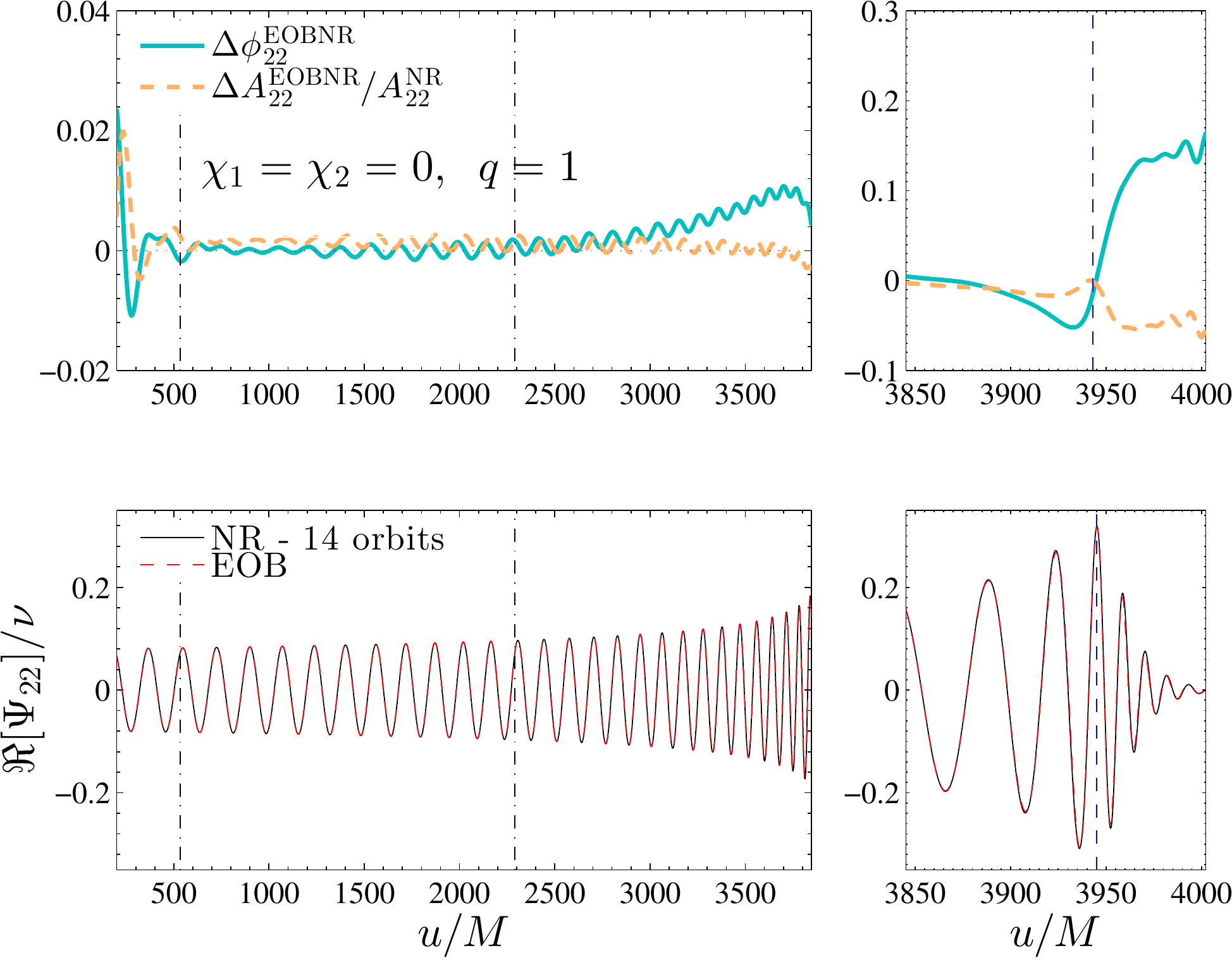}\\
\includegraphics[width=0.37\textwidth]{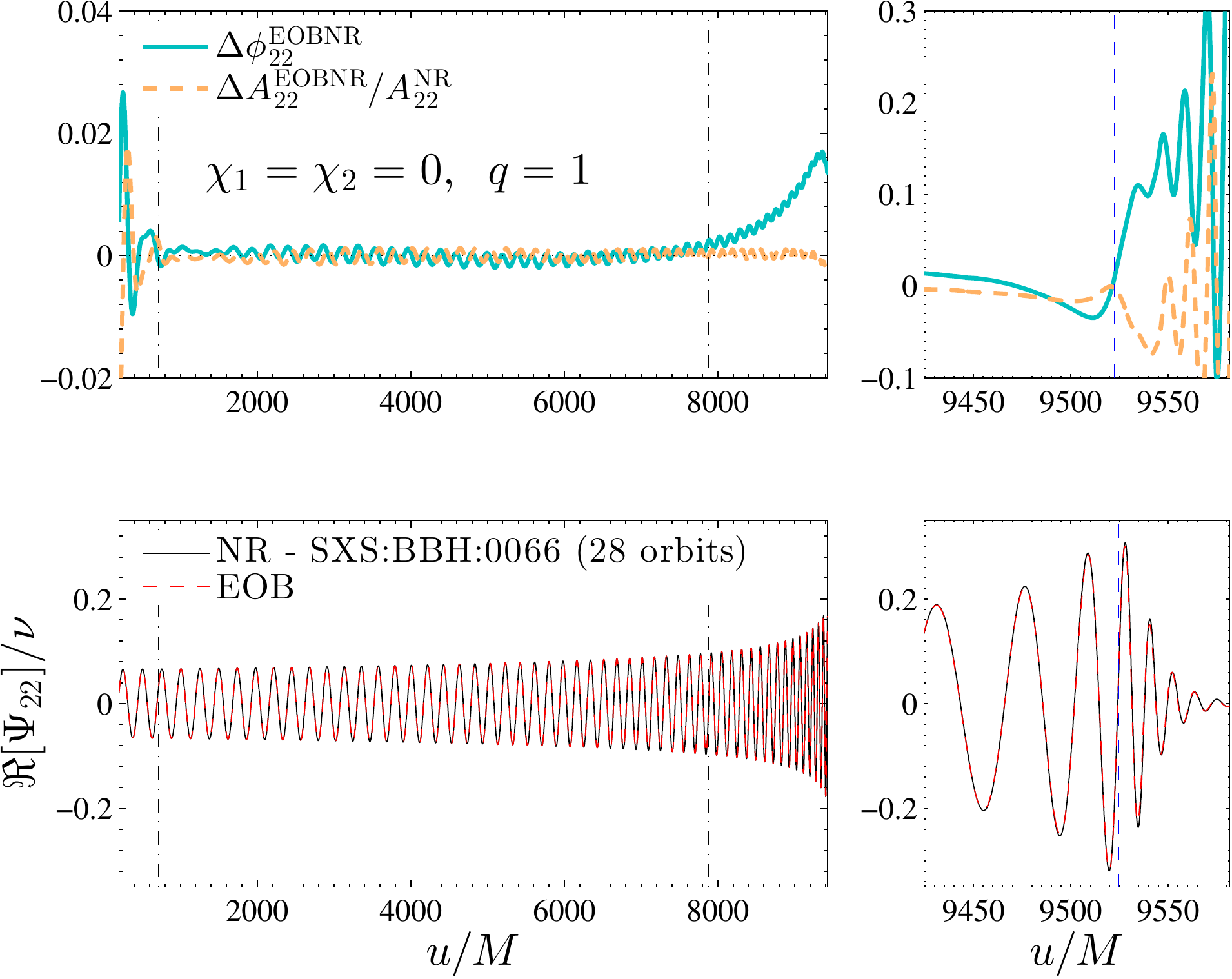}\\
\includegraphics[width=0.37\textwidth]{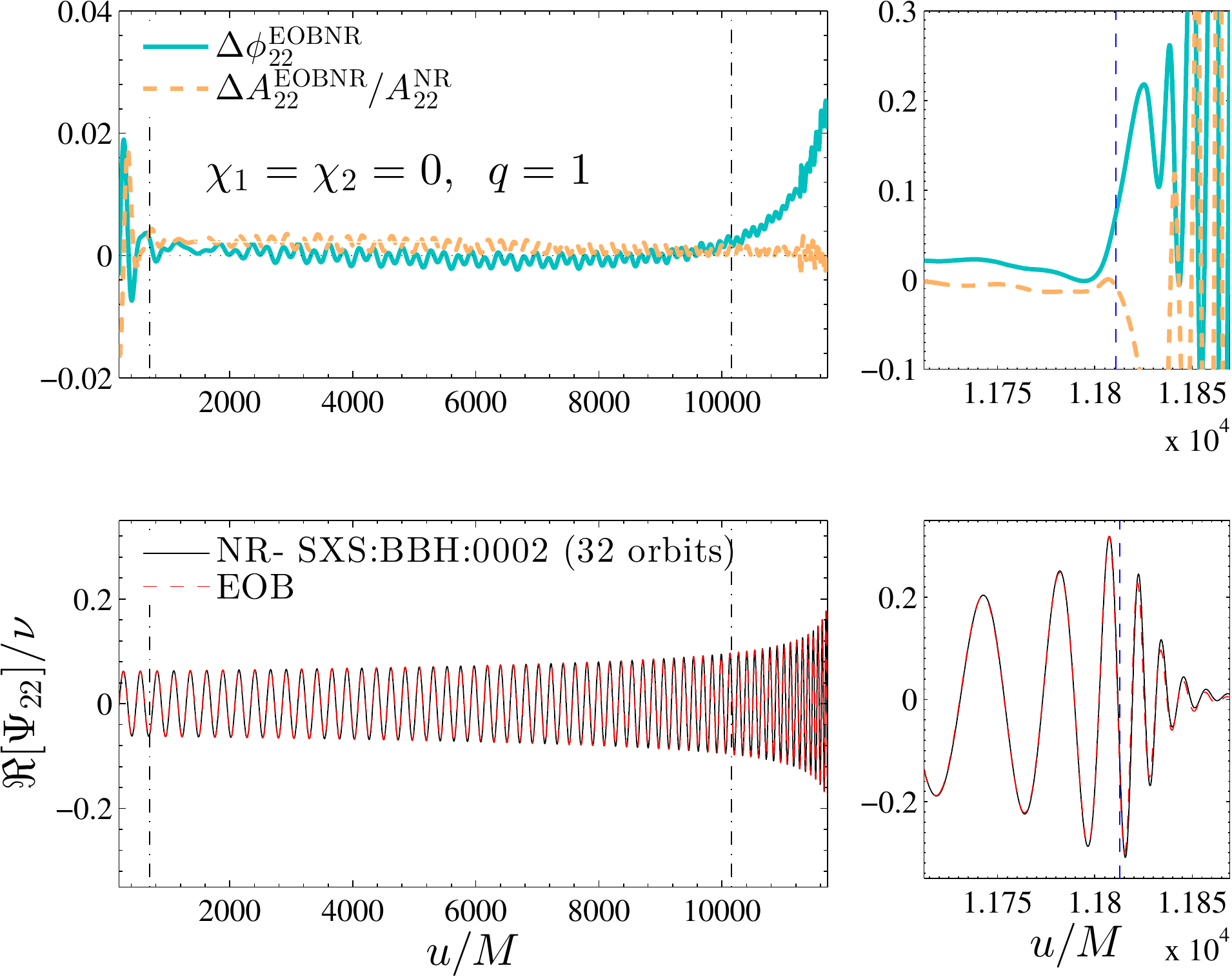}
\caption{\label{fig:phasing_q1_chi0}EOB/NR $\ell=m=2$ waveform comparison: $q=1$, $\chi_1=\chi_2=0$
and $a_6^c(0.25)\approx -57.688$ from Eq.~\eqref{eq:a6c_calibrate}. The calibration 
was done using the 14-orbit simulation, following the same procedure as in 
Ref.~\cite{Damour:2012ky} (that was also using the same dataset). The resulting EOB waveform 
is then checked for consistency against waveforms SXS:BBH:0066 ($\sim$~28 orbits, middle panel) 
and SXS:BBH:0002 ($\sim$~32 orbits, bottom panel), that is the longest NR waveform currently 
available. The EOB/NR modulus and phase agreement is excellent all through inspiral, plunge, 
merger and ringdown.}
\end{center}
\end{figure}
We use a sample of 10 configurations simulated with the Llama code,
five of them being computed with two different resolutions.
All but one of them have mass ratio $q=1$.
The dimensionless spin values that we consider are $\chi=(\pm 0.8, \pm 0.6, \pm 0.4, \pm 0.2, 0)$. 
The precise configurations (and resolutions used) 
are listed in Table~\ref{tab:llama}.
The two nonspinning configurations were first presented in 
Ref.~\cite{Damour:2011fu}; the remaining, equal-mass, spinning
configurations were first presented in Ref.~\cite{Pollney:2010hs}.
They were here simulated at higher resolutions and with improved grid setups 
to enhance accuracy. For all Llama data the waveforms at future null
infinity are estimated by using Cauchy-characteristic 
extraction (CCE)~\cite{Reisswig:2009us, Bishop:1996gt}.

The Llama simulations are relatively short (between 4 and 8 orbits). 
We use them mainly for reaching one of the main aims of this paper,
namely to check to what extent the EOB prediction for the {\it energetics} 
of  the system (as measured by  the gauge-invariant relation 
$E_b(j)$ between the dimensionless binding energy
and the angular momentum) is consistent with the corresponding NR quantity.

The longer SXS waveforms (between 12 and 32 orbits) not only were used to
check the phasing performance of the EOB model, but also its energetics
(modulo subtleties related to junk radiation discussed below, 
which are essentially absent when dealing with Llama data).
The joined use of Llama and SXS waveform data allowed us
to: (i) reach a reliable NR-calibration of the spinning 
EOB model of Paper~I; (ii) perform interesting cross checks on 
energy and angular momentum {\it at merger} between NR data obtained 
with completely different codes; 
and (iii) to compute the $E_b(j)$ energetics for a large sample of 
spin-aligned SXS data (thereby improving on Ref.~\cite{Taracchini:2013rva}
where only two $E_b(j)$ curves were computed).

\section{Improved calibration of the nonspinning EOBNR model}
\label{sec:nospin}

As mentioned in the Introduction, the spinning EOB model considered in this
paper is essentially the one introduced in Paper I (to which we refer for
our notation), in particular the waveform the waveform up to merger is defined
by Eqs.~(75)-(96) there. The only change in the theoretical frameworm concerns
the definition of the EOB radial potential $A$. In the present work, we use of 
the full 4PN-accurate analytical knowledge  of the EOB radial interaction (see below).
In addition we {\it extended} (without any additional theoretical modification) 
the application of the new ringdown description of Ref.~\cite{Damour:2014yha}
to the full SXS datasets of Table~\ref{tab:configs}.
For the equal-mass, equal-spin configurations, the ringdown description
is exactly as described in  Ref.~\cite{Damour:2014yha}, including the use
of the fitting coefficients listed in Table~II there.
On the other hand, for the remaining configurations in our Table~\ref{tab:configs}
we do not make use of the latter fit, but rather we apply the ringdown 
modelization methodology of~\cite{Damour:2014yha} separately to each dataset. 
We postpone to future work the construction of a global analytical extrapolation 
of the latter, discrete, ringdown data.

We use here the 5PNlog-accurate 
post-Newtonian expansion of the orbital $A_{\rm orb}^{\rm PN}$ 
function
\begin{align}
&A_{\rm orb}^{\rm PN}(u_c) = 1 - 2u_c + 2\nu u_c^3 +\nu a_4 u_c^4 \nonumber \\
& + \nu [a_5^c(\nu) + a_5^{\rm log} \ln u_c] u_c^5 +\nu[a_6^c(\nu) + a_6^{\log} \ln u_c]u_c^6 ,
\end{align}
where
\be
\nu\equiv \dfrac{\mu}{M}= \dfrac{m_1 m_2}{(m_1+m_2)^2}=\dfrac{q}{(1+q)^2}
\ee
is the main deformation parameter of EOB theory, which varies between $\nu=0$,
in the large-mass-ratio limit ($q=m_1/m_2\gg 1$) and $\nu=1/4$ in the equal-mass
case. The dimensionless gravitational potential $u_c$ is defined 
as $u_c=M/r_c$ in terms of the EOB centrifugal radius
\be
r_c\equiv \sqrt{r^2+a^2 + \dfrac{2 M a^2}{r}+\delta a^2(r)} \ .
\ee
Here the (next-to-leading-order) correction of the Kerr parameter, 
$\delta a^2(r)$ \cite{Balmelli:2013zna}, is defined in Eq.~(59) of Paper~I.
Contrary to Paper~I, where we had phenomenologically fixed the 4PN 
coefficient $a_5^c(\nu)$ to the ($\nu$-independent) fiducial value 
$a_5^c=23.5$, we use here the exact, $\nu$-dependent, analytical 
expression of $a_5^c(\nu)$ obtained in Ref.~\cite{Bini:2013zaa}
(see also~\cite{Damour:2015isa}).
We recall that the orbital EOB radial potential $A_{\rm orb}$ is defined
by Pad\'e resumming $A_{\rm orb}^{\rm PN}(u_c)$ as 
\be
\label{eq:Anonspinning}
A_{\rm orb}(u_c;\,\nu;\,a_6^c) =P^1_5[A^{\rm PN}_{\rm orb}(u_c)].
\ee
In view of the change in the analytical expression of $A_{\rm orb}$, our first task
will be to provide a new calibration of the single, effective 5PN functional 
parameter  $a_6^c(\nu)$ entering $A_{\rm orb}$. 
We perform this calibration by means of a sample of 
{\it nonspinning} waveforms. [This nonspinning-calibrated orbital potential will
then be used as is in our spinning EOB model]. We use as calibrating waveforms
eight SXS simulations with mass ratio $q=(1,1.5,2,3,4,5,6,8)$ 
[the $q=9.989$ one is used just as a cross check]. 
For $q=1$ we use the same $\sim 14$~orbits waveform that was used in the calibration
procedure of Ref.~\cite{Damour:2012ky}. We tune $a_6(\nu)$ so that the EOB and 
NR phasing agree (after a suitable alignment) within the NR phasing error 
at NR merger. Following the footsteps of Ref.~\cite{Damour:2012ky} (and in
particular the cross-check of the time-domain analysis with the $Q_\omega$ analysis) 
we got 
\be
\label{eq:a6c_calibrate}
a_6^c(\nu)=3097.3\,\nu^2 - 1330.6\,\nu + 81.38 .
\ee
Note that $a_6^c(\nu)$ varies between $-57.69$ for $\nu=0.25$ (equal-mass case)
and $-7.432$ for $q=9.989$ ($\nu\approx 0.0827$).
Moreover the parameter $\Delta t_{\rm NQC}$, giving the time-lag between the 
peak of the pure orbital frequency $\Omega_{\rm orb}$ [see Eq.~(100) of Paper~I] 
and the NR/EOB matching point on the time axis (which is used to determine 
next-to-quasi-circular (NQC) corrections to the waveform) is fixed
(as discussed in Sec.~V of Paper~I) as $\Delta t_{\rm NQC}(\chi=0)=1M$ 
in the nonspinning case\footnote{
By contrast, in the case of high, positive spins, $\chi\geq 0.85$, it 
is useful to allow $\Delta t_{\rm NQC}$ to depend on the spin (see below).}.
Finally, as mentioned in the introduction, we use the new modelization of
the ringdown introduced in Ref.~\cite{Damour:2014yha}. 
In this paper, we use the actual fitting
coefficients as obtained applying the procedure of Ref.~\cite{Damour:2014yha}
to each single waveform. The performance of suitable ``fits of fits'' extending 
the validity of the procedure of Ref.~\cite{Damour:2014yha} outside
the domain of known waveforms will be discussed elsewhere. 

Figure~\ref{fig:phasing_q1_chi0} 
(which plots the Regge-Wheeler-Zerilli normalized waveform $\Psi_{22}\equiv R h_{22}/\sqrt{24}$)
illustrates the EOB/NR phase agreement, for $q=1$ when using: (i) the 14-orbit (up to merger) 
calibration waveform (top-panel); 
(ii) the $\sim 28$-orbit SXS:BBH:0066 waveform of the catalog; 
and (iii) the longest NR waveform available in the catalog, SXS:BBH:0002, 
that corresponds to about 32 orbits up to merger. We stress that the 
actual calibration of $a_6^c(\nu)$ made use neither of  
SXS:BBH:0066 nor of SXS:BBH:0002, which are used here just to 
provide an independent check of the calibration procedure.
The vertical lines in the plots highlight the alignment time intervals, 
corresponding to dimensionless GW frequencies  $(M\omega_L,M\omega_R)$.
To check consistency, we fixed $M\omega_R=0.045$ for each data set.
Then, we use $M\omega_L=0.035$ for the 14-orbit run, 
$M\omega_L=0.023$ for SXS:BBH:0066 and $M\omega_L=0.025$ for SXS:BBH:0002.
In the latter case, the phase difference between EOB and NR, 
accumulated up to NR merger,
 $\Delta\phi^{\rm EOBNR}_{\rm mrg}\equiv \phi^{\rm EOB}_{\rm mrg}-\phi^{\rm NR}_{\rm mrg}=0.073$~rad
is comparable to the NR uncertainty at merger, 
$\approx 0.066$~rad~(see Table~\ref{tab:configs}). Similarly, when using 
$a_6^c(\nu)$ as given by Eq.~\eqref{eq:a6c_calibrate}, one finds that 
the EOB-NR phase difference is comparable to 
(though in some cases slightly larger than) the 
NR uncertainty at merger for all 9 nonspinning data
(compare columns seven and eight in Table~\ref{tab:configs}). 
\begin{figure}[t]
\begin{center}
\includegraphics[width=0.45\textwidth]{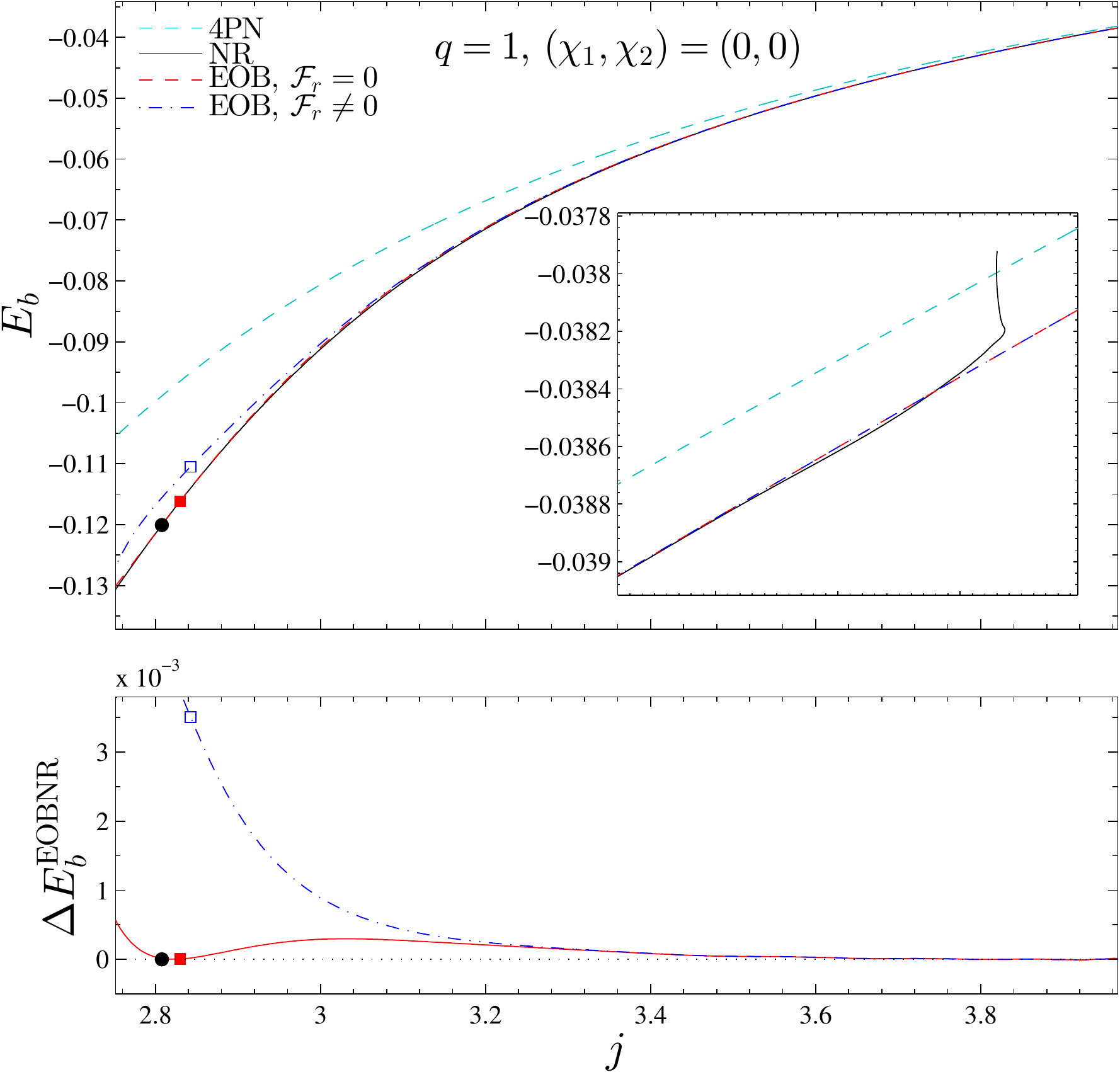}
\caption{\label{fig:ej0_no_Fr} Energetics for $q=1$, nonspinning binary: comparison 
between NR Llama data, the 4PN, Taylor-expanded, curve and two EOB curves, 
one with and the other without a radial part of the radiation reaction ${\cal F}_{r_*}$. 
The (effective) choice ${\cal F}_{r_*}=0$ displays the smallest ($\sim 10^{-4}$) 
discrepancy with NR data up to merger (indicated by colored markers).
The inset focuses on the initial transient driven by junk radiation.}
\end{center}
\end{figure}

\section{Energetics of nonspinning coalescences: the choice ${\cal F}_{r_*}=0$}
\label{nospin_energetics}

\subsection{Energetics with (nonspinning) Llama data}
Having used SXS phasing data to determine the relation 
Eq.~\eqref{eq:a6c_calibrate}, let us now turn to discussing 
the energetics of the model, motivating, in particular, 
the choice of a vanishing radial component of radiation 
reaction that we made here.
The choice  ${\cal F}_{r_*}=0$, was introduced
in Ref.~\cite{Buonanno:2000ef} and used in several
previous EOB works~\cite{Damour:2009kr,Damour:2011fu,Bernuzzi:2014owa})
as well as in Paper~I.
As in previous works~\cite{Damour:2011fu,Bernuzzi:2012ci,Bernuzzi:2014owa}, 
the analysis of the energetics is done via the gauge-invariant 
relation between the dimensionless binding energy $E_b$ and 
the dimensionless total angular momentum, $j$, $E_b(j)$. 
These quantities are computed as
\begin{align}
E_b      & \equiv \dfrac{M_{\rm ADM}^0-\Delta {\cal E}_{\rm rad} -M}{\mu}, \\
j        & \equiv  \dfrac{{\cal J}^0_{\rm ADM}-\Delta{\cal J}_{\rm rad}}{M\mu}.
\end{align}
Here  $(M_{\rm ADM}^0,{\cal J}^0_{\rm ADM})$ denote the total, 
initial Arnowitt-Deser-Misner (ADM) mass-energy and 
angular momentum of the system (including the contribution of the individual spins), $(\Delta {\cal E}_{\rm rad},\Delta{\cal J}_{\rm rad})$ denote the
energy and angular momentum radiated in GWs, while $M=m_1+m_2$
and $\mu=m_1m_2/M$, where $m_1$ and $m_2$ are the NR measured 
initial Christodoulou masses. 
We recall that, while $\Delta {\cal E}_{\rm rad}$ is obtained by time-integrating
the square of the news ($\dot{h}$), $\Delta{\cal J}_{\rm rad}$ is obtained 
by time-integrating a bilinear expression in the news and the strain $h$. 
To obtain the strain we start from the Fourier transform of $\psi_4$ and
use the method of~\cite{Reisswig:2010di}.
\begin{figure}[t]
\begin{center}
\includegraphics[width=0.45\textwidth]{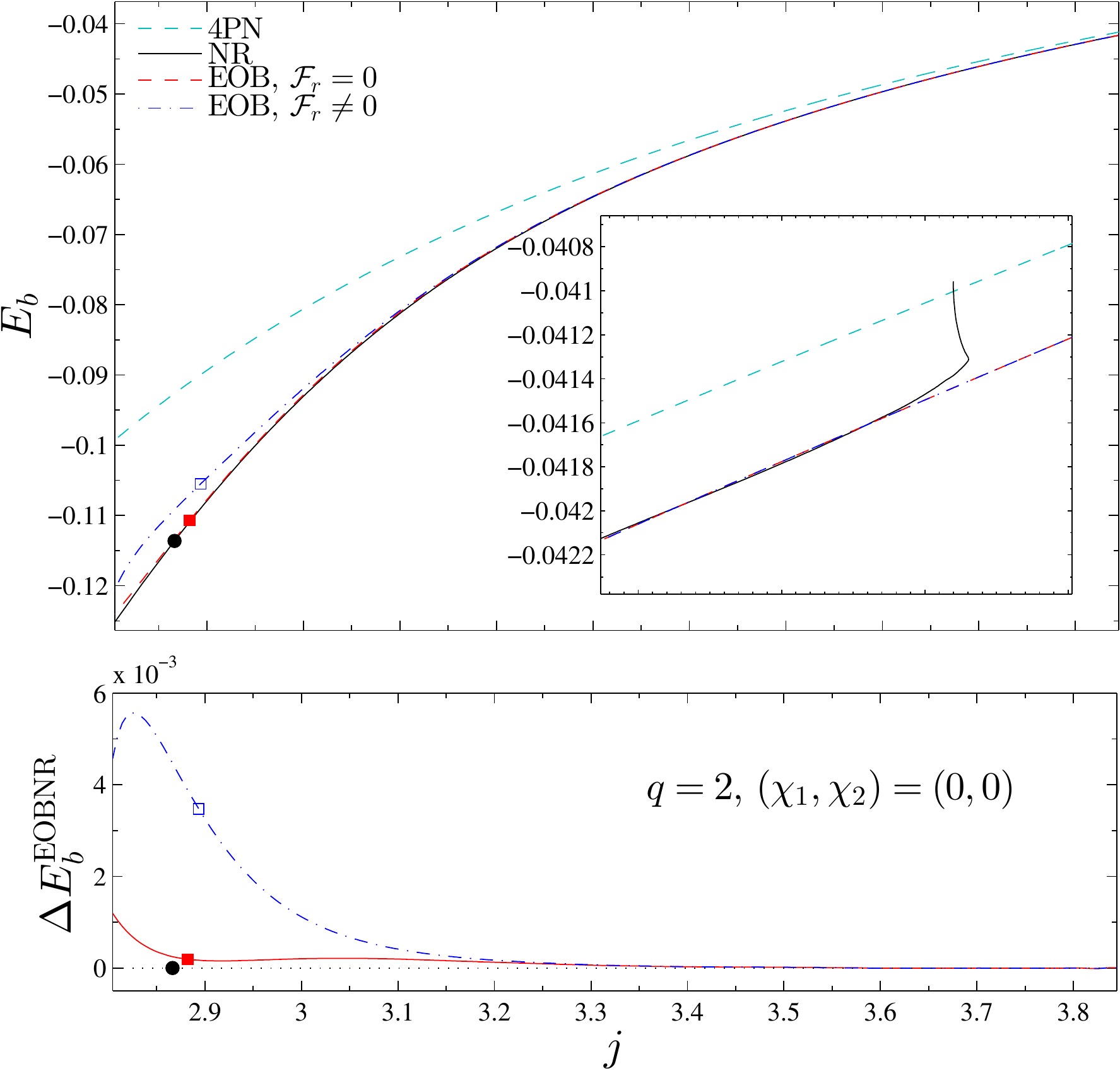}
\caption{\label{fig:ej0_q2} Energetics comparison, as in Fig.~\ref{fig:ej0_no_Fr}, but  for $q=2$,
nonspinning, Llama NR data. The ${\cal F}_{r_*}=0$ curve displays the smallest discrepancy with NR 
data up to merger (indicated by colored markers).}
\end{center}
\end{figure}
The $E_b(j)$ relation can be straightforwardly obtained from the results of the
BBH simulations obtained with the Llama code~\cite{Pollney:2009yz, Reisswig:2012nc}, 
as we did in a previous study~\cite{Damour:2011fu} limited to nonspinning BBHs.
On the other hand, when using SXS data, several subtleties have to be
properly addressed in order to accurately estimate the $E_b(j)$ 
relation (see Appendix for details).

Figure~\ref{fig:ej0_no_Fr} displays the $E_b(j)$ relation in the $q=1$, nonspinning case. 
It shows the triple comparison between the NR curve and: (i) the PN-expanded
$E_b(j)$ relation; (ii) the EOB model with $a_6^c$ given by Eq.~\eqref{eq:a6c_calibrate}
and with the choice ${\cal F}_{r_*}=0$; (iii)exactly  the same EOB model (in particular, 
with the value of $a_6^c$ given by Eq.~\eqref{eq:a6c_calibrate}) but with 
the (nonresummed), 2.5PN accurate, expression of ${\cal F}_{r_*}$ of Ref.~\cite{Bini:2012ji}.
Note that the latter expression was already used in previous 
EOB work~\cite{Damour:2012ky}, though its effect on $E_b(j)$ 
is investigated here for the first time.

The general structure of the PN-expanded $E_b(j;\chi_1,\chi_2)$ relation is 
(in the spinning case)
\be
\label{eq:Epn}
E^{\rm PN}_b = \sum_{n=2}^{10} \dfrac{c_n(\nu;\,\chi)}{\ell^n} \ ,
\ee
where $\ell=j-\left(\dfrac{m_1}{m_2}\chi_1 + \dfrac{m_2}{m_1}\chi_2\right)$ denotes
the dimensionless {\it orbital} angular momentum, $\ell={L}/{(\mu M)}$ and the
coefficients $c_n(\nu;\,\chi)$ depend at most quadratically on the dimensionless
spins $\chi_1\equiv S_1/m_1^2$, $\chi_2\equiv S_2/m_2^2$. The accuracy we use in 
$E^{\rm PN}_b$ is: 4PN in the nonspinning sector~\cite{Damour:2014jta}; NNLO for
the spin-orbit and NLO for the spin-spin part (see e.g.~\cite{Levi:2014sba}).

As shown in Fig.~\ref{fig:ej0_no_Fr} (see especially the bottom panel, which 
displays $\Delta E_b^{\rm EOBNR}(j)\equiv E_b^{\rm EOB}(j)-E_b^{\rm NR}(j)$) the use 
of a vanishing radial radiation reaction leads, from an effective point of view,
to a better agreement between the EOB  energetics and the NR one up to 
merger\footnote{We found that a different calibration of $a_6^c$ 
is unable to displace the ${\cal F}_{r_*}\neq 0$ curve on top of the NR one}.
The NR and EOB merger points are indicated, respectively, by a filled circle 
(NR), by a filled square (EOB, ${\cal F}_{r_*}=0$) and by an empty square  
(EOB, ${\cal F}_{r_*}\neq 0$). By contrast, the PN curve shows the largest 
deviation from NR results, especially at low $j$'s.
Finally, Fig.~\ref{fig:ej0_q2} shows that the same conclusions hold true
also for the $q=2$ Llama data.

As discussed in Refs.~\cite{Iyer:1995rn,Buonanno:2000ef}, there exists a coordinate
gauge where ${\cal F}_{r_*}=0$. Further work~\cite{Bini:2012ji} has shown that the 
requirement of the vanishing of the Schott contribution $J^{\rm Schott}$ to the 
angular momentum required a specific, nonzero, value of ${\cal F}_{r_*}$
(which was then used in Ref.~\cite{Damour:2012ky} because the vanishing
of $J^{\rm Schott}$ seemed apriori required by the definition of the azimuthal
component ${\cal F}_\varphi$ of the radiation reaction within the EOB formalism).
However, because of the various approximations made in defining ${\cal F}_\varphi$ 
in the EOB formalism, it is not actually required to choose the 
${\cal F}_{r_*}$ that is implied by a vanishing $J^{\rm Schott}$. Our present
work experimentally shows that the condition ${\cal F}_{r_*}=0$ is a 
simple and effective way to accomplish a (rather surprisingly) 
good agreement between the numerical and EOB $E_b(j)$ curves. 
We leave to future work a theoretical study of why such a condition,
in conjunction with the current approximate value of  ${\cal F}_\varphi^{\rm EOB}$, 
happens to lead to rather small values of both $J^{\rm Schott}$ and $E_b^{\rm Schott}$
(as shown in Figs.~\ref{fig:ej0_no_Fr}-\ref{fig:ej0_q2} above).

 \begin{table*}[t]
   \caption{EOB/NR phasing comparison. The columns report: the number of the dataset; 
   the name of the configuration in the SXS catalog; the mass ratio $q=m_1/m_2$; the symmetric mass 
   ratio $\nu$; the dimensionless spins $\chi_1$ and $\chi_2$; the phase difference 
   $\Delta\phi^{\rm EOBNR}\equiv \phi^{\rm EOB}-\phi^{\rm NR}$ computed at NR merger; the NR phase 
   uncertainty at NR merger $\delta\phi^{\rm NR}_{\rm mrg}$ (when available) measured taking the 
   difference between the two highest resolution levels (see text); the maximum value of the 
   unfaithfulness $\bar{F}\equiv 1-F$ as per Eq.~\eqref{eq:barF}. 
   The $\Delta\phi^{\rm EOBNR}$'s in brackets for $\chi_1=\chi_2>+0.85$ 
   were obtained using Eq.~\eqref{eq:DTnqcFit} for $\Delta t^{\rm NQC}(\chi)$.} 
  \centering  
\begin{ruledtabular}
  \begin{tabular}{c|cccccclccc}        
$\#$ &  Name & N orbits & $q$ & $\nu$ & $\chi_1$  & $\chi_2$ & $\Delta\phi^{\rm EOBNR}_{\rm mrg}$  [rad] & $\delta\phi^{\rm NR}_{\rm mrg}$ [rad] & $\max(\bar{F})$ \\
\hline
1 & SXS:BBH:none   & 14    &  1   &  0.25 &   0.0       &   0.0     & $-0.016$  & $\dots$ & 0.00087 \\
2 & SXS:BBH:0066  & 28     &  1   &  0.25 &   0.0       &   0.0     & $+0.010$   & $\dots$ & 0.00068 \\
3 & SXS:BBH:0002  & 32.42  &  1   &  0.25 &   0.0       &   0.0     & +0.073     & 0.066   & 0.00101  \\
4 & SXS:BBH:0007  & 29.09  &  1.5 &  0.24 &   0         &   0       & +0.05     & 0.018   & 0.00201  \\
5 & SXS:BBH:0169  & 15.68  &  2   &  $0.\bar{2}$   &  0      &   0     & $-0.15$ & 0.02   & 0.00045  \\
6 & SXS:BBH:0030  & 18.22  &  3   & 0.1875         &  0       &   0    & $-0.074$ & 0.087  & 0.00035  \\
7 & SXS:BBH:0167  & 15.59  &  4   & 0.16           &    0     &  0     & $-0.059$ & 0.52   & 0.00035 \\
8 & SXS:BBH:0056  & 28.81  &  5   & $0.13\bar{8}$  &    0     &  0     & $-0.089$ & 0.44   & 0.00038 \\
9 & SXS:BBH:0166  & 21.56  &  6   & 0.1224         &   $ 0$   &  0     & $-0.198$ & $\dots$& 0.00037 \\
10 & SXS:BBH:0063  & 25.83 &  8   & 0.0987         &     0    &  0     & $-0.453$ & 1.01   & 0.00292 \\         
11 & SXS:BBH:0185  & 24.91 & 9.98911 & 0.0827       &     0    &  0     & $-0.0051$ & 0.376 & 0.00066\\      
\hline
12 & SXS:BBH:0004  & 30.19 &  1   &  0.25 &   $-0.50$   &   0.0       & $-0.017$    & 0.068   & 0.00403\\
13 & SXS:BBH:0005  & 30.19 &  1   &  0.25 &   $+0.50$   &   0.0       & +0.08       & 0.28    & 0.00052\\
14 &SXS:BBH:0156  & 12.42 &  1   &  0.25 &   $-0.95$   &  $-0.95$     &  +0.32      & 2.17    & 0.00058\\
15 &SXS:BBH:0159  & 12.67 &  1   &  0.25 &   $-0.90$   &  $-0.90$     & +0.06       & 0.38    & 0.00047\\
16 &SXS:BBH:0154  & 13.24 &  1   &  0.25 &   $-0.80$   &  $-0.80$     & +0.11       & $\dots$ & 0.00044\\
17 &SXS:BBH:0151  & 14.48 &  1   &  0.25 &   $-0.60$   &  $-0.60$     & $-0.049$    & 0.14     & 0.00042\\
18 &SXS:BBH:0148  & 15.49 &  1   &  0.25 &   $-0.44$   &  $-0.44$     & +0.14       & 0.72           & 0.00043\\
19 &SXS:BBH:0149  & 17.12 &  1   &  0.25 &   $-0.20$   &  $-0.20$     & +0.45       & 0.90           & 0.00085\\
20 &SXS:BBH:0150  & 19.82 &  1   &  0.25 &   $+0.20$   &  $+0.20$     & +0.94             & 0.99    & 0.00275\\
21 &SXS:BBH:0152  & 22.64 &  1   &  0.25 &   $+0.60$   &  $+0.60$     & +0.01             & 0.36    & 0.00068\\
22 &SXS:BBH:0155  & 24.09 &  1   &  0.25 &   $+0.80$   &  $+0.80$     & $-0.39$           & 0.26    & 0.00110  \\
23 &SXS:BBH:0153  & 24.49 &  1   &  0.25 &   $+0.85$   &  $+0.85$     & +0.06             & $\dots$ & 0.00059 \\
24 &SXS:BBH:0160  & 24.83 &  1   &  0.25 &   $+0.90$   &  $+0.90$     & +0.41 (+0.41)     & 0.80    & 0.00117\\
25 &SXS:BBH:0157  & 25.15 &  1   &  0.25 &   $+0.95$   &  $+0.95$     & +0.37 (+0.83)     & 1.18    & 0.00295 \\
26 &SXS:BBH:0158  & 25.27 &  1   &  0.25 &   $+0.97$   &  $+0.97$     & +0.37 (+0.49)     & 1.26    & 0.00325\\
27 &SXS:BBH:0172  & 25.35 &  1   &  0.25 &   $+0.98$   &  $+0.98$     & +0.99 (+0.46)     & $2.02$  & 0.00422 \\
28 &SXS:BBH:0177  & 25.40 &  1   & 0.25  &   $+0.99$   &  $+0.99$     & +0.22 (+0.48)     & $0.40$  & 0.00507  \\
29 &SXS:BBH:0178  & 25.43 &  1   & 0.25  &   $+0.994$  &  $+0.994$    & +0.24 (+0.23)     & $-0.53$ & 0.00506 \\
30 &SXS:BBH:0013  & 23.75 &  1.5 &  0.24          &  $+0.5$  &   0    & $+0.31$           & $\dots$ & 0.00058\\
31 &SXS:BBH:0014  & 22.63 &  1.5 &  0.24          &  $-0.5$  &   0    & $-0.15$           & $0.15$  & 0.00046\\
32 &SXS:BBH:0162  & 18.61 &  2   &  $0.\bar{2}$   & $+0.6$   &   0    & $-0.20$           & 0.71    & 0.00027\\
33 &SXS:BBH:0036  & 31.72 &  3   &  0.1875        & $-0.5$   &   0    & +0.08             & 0.065   & 0.00040\\
34 &SXS:BBH:0031  & 21.89 &  3   &  0.1875        & $+0.5$   &   0    & +0.12             & 0.034   & 0.00023 \\
35 &SXS:BBH:0047  & 22.72 &  3   &  0.1875        & $+0.5$   & $+0.5$ & $-0.034$          & $\dots$ & 0.00030\\
36 &SXS:BBH:0046  & 14.39 &  3   &  0.1875        & $-0.5$   & $-0.5$ & +0.36             & $\dots$ & 0.00054 \\
37 &SXS:BBH:0110  & 24.24 &  5   & $0.13\bar{8}$  &   $+0.5$ &  0     & $+0.24$           & $\dots$ & 0.00016 \\
38 &SXS:BBH:0060  & 23.17 &  5   & $0.13\bar{8}$  &   $-0.5$ &  0     & $+0.21$           & $0.8$   & 0.00034 \\
39 &SXS:BBH:0064  & 19.16 &  8   & 0.0987         &   $-0.5$ &  0     & $+0.026$          & 0.8     & 0.00042\\         
40 &SXS:BBH:0065  & 33.97 &  8   & 0.0987         &   $+0.5$ &  0     & $+1.33$           &  $-3.0$    & 0.00040\\    
  \end{tabular}
\end{ruledtabular}
  \label{tab:configs}
\end{table*}

\begin{figure}[t]
\begin{center}
\includegraphics[width=0.47\textwidth]{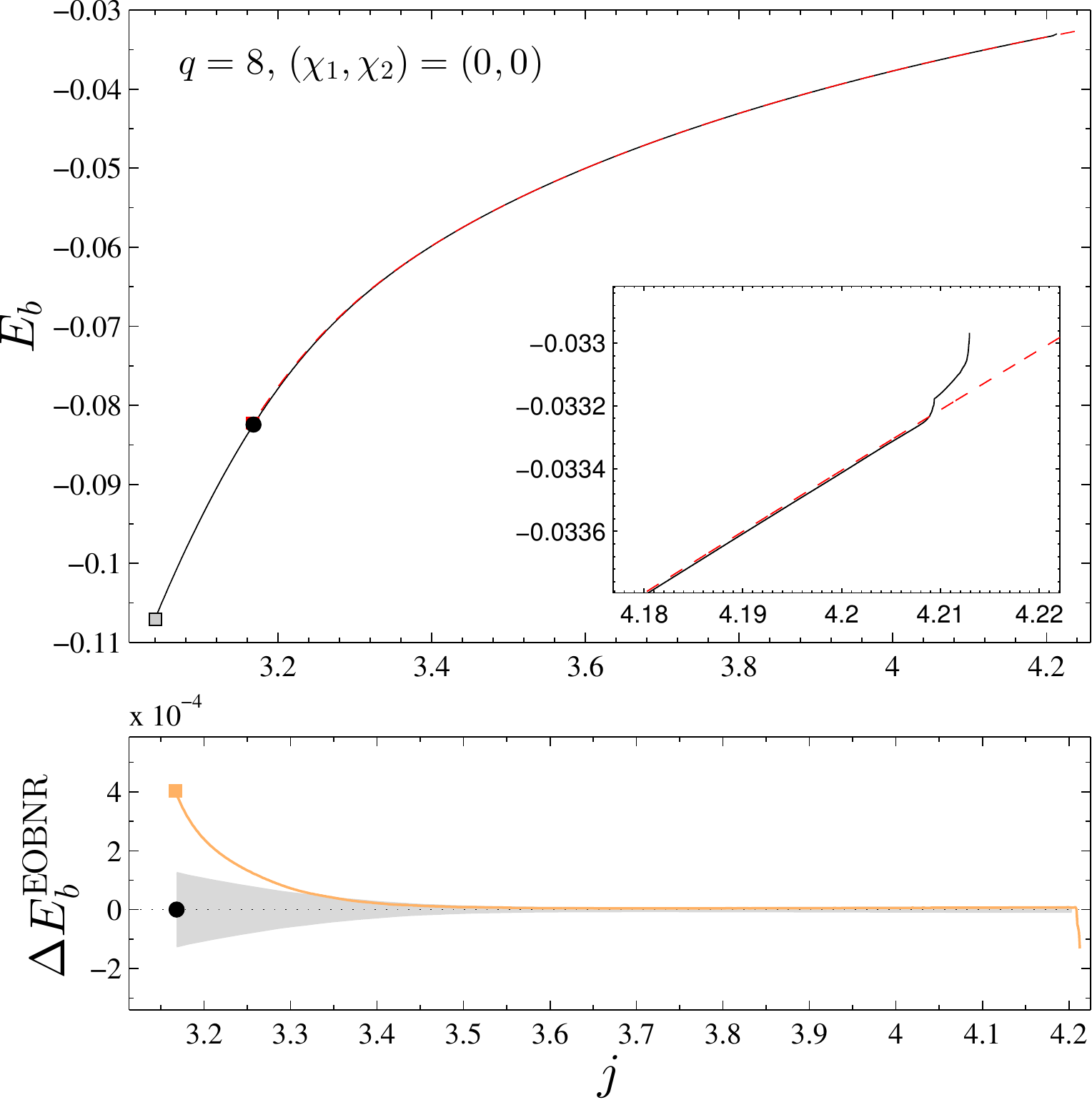}
\caption{\label{fig:ejq8} Comparison between EOB and SXS energetics for the $q=8$ case. 
 The shaded area indicates numerical uncertainty measured taking the difference between 
 the highest and second highest resolution data. Note that the EOB merger state (red, filled, square marker) 
 and the NR merger state (black, filled, circle) are barely distinguishable in the upper panel.}
\end{center}
\end{figure}

\begin{figure}[t]
\begin{center}
\includegraphics[width=0.47\textwidth]{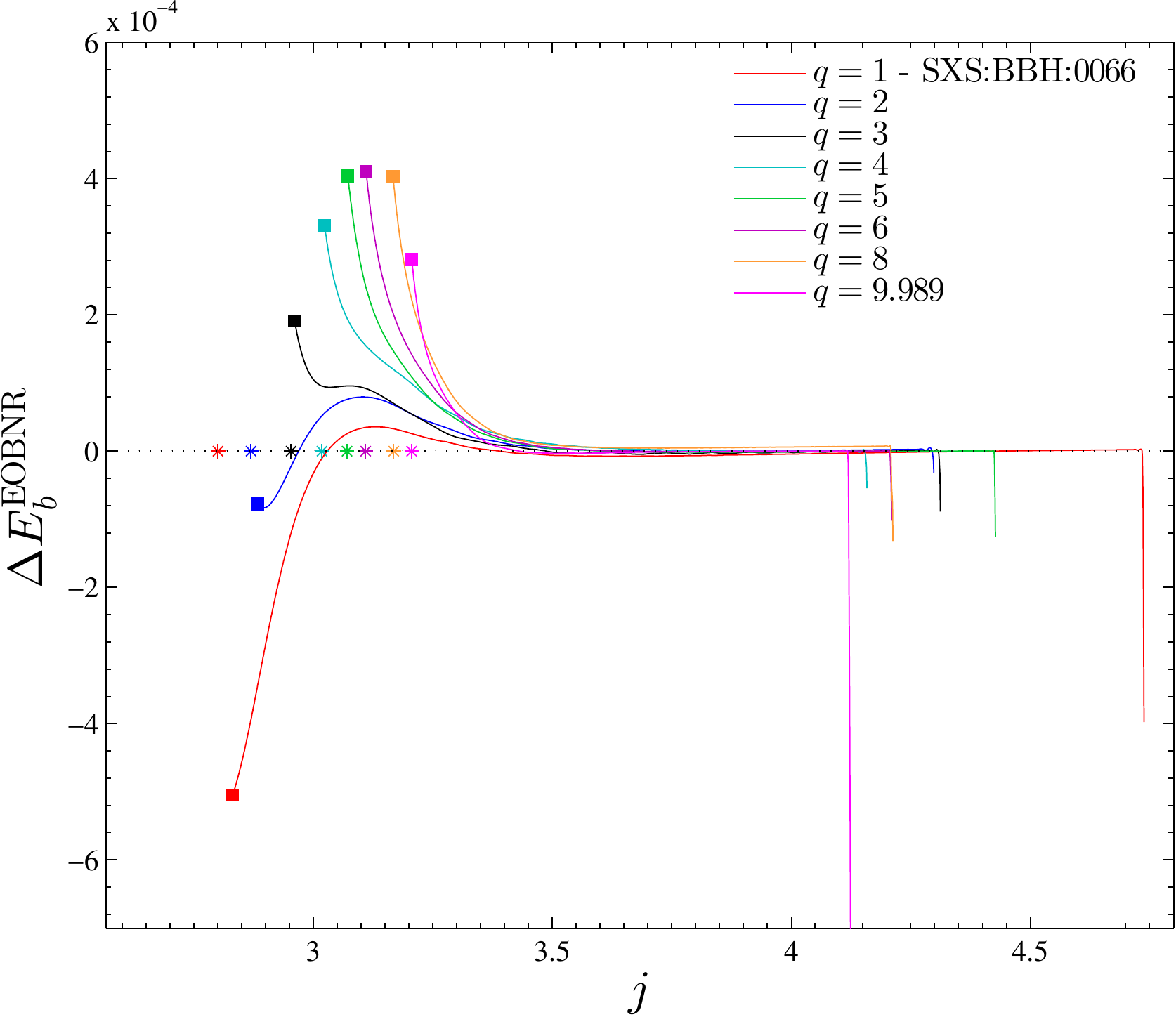}
\caption{\label{fig:diff} EOB-NR differences between $E_b(j)$ curves for all but one $(q=1.5$) 
nonspinning SXS data of Table~\ref{tab:configs}. The difference is compatible with 
NR uncertainties up to merger ($\sim 10^{-4}$, see Fig.~\ref{fig:ejq8} for the illustrative $q=8$ case).}
\end{center}
\end{figure}

\begin{figure}[t]
\begin{center}
\includegraphics[width=0.45\textwidth]{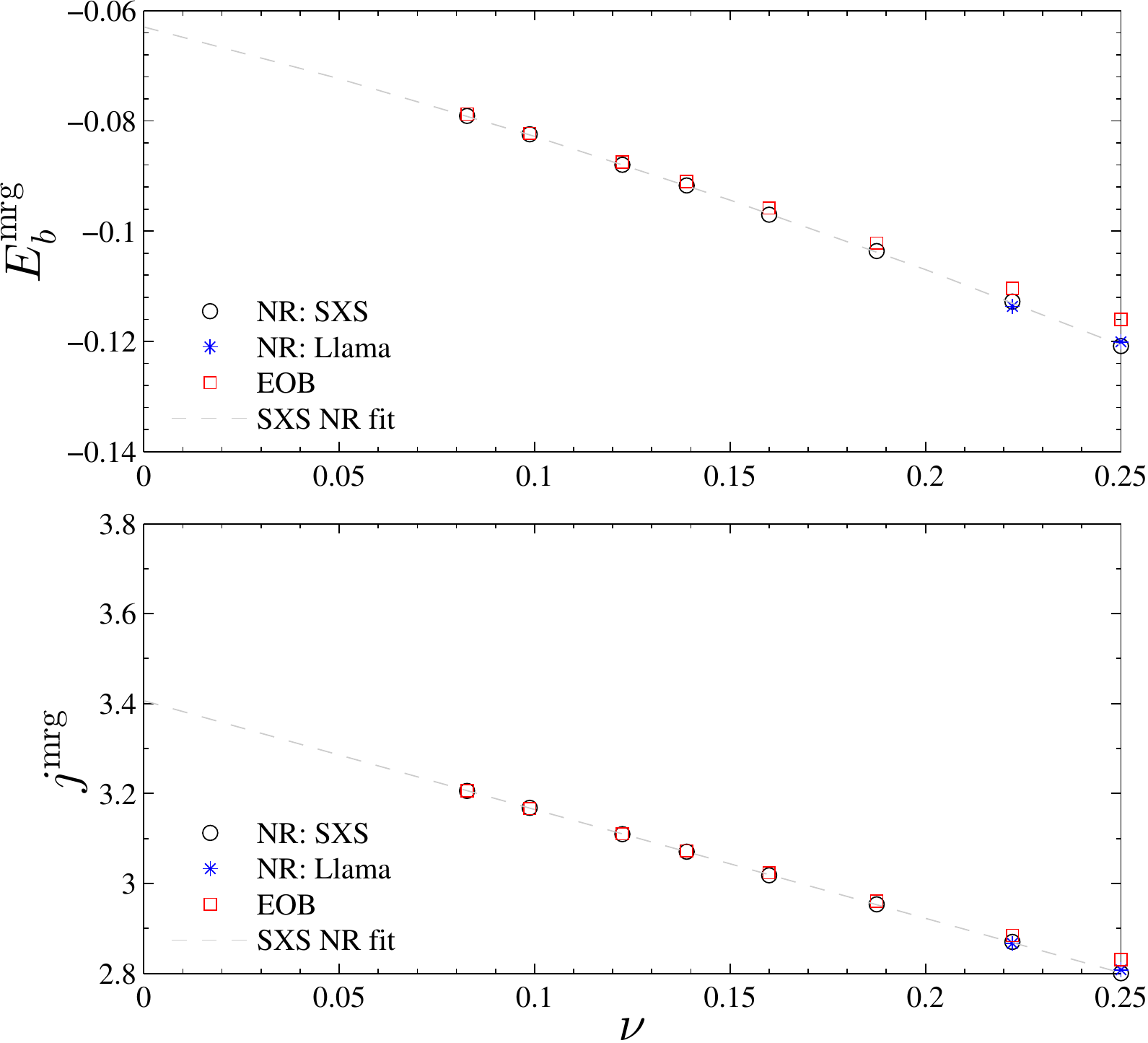}
\caption{\label{fig:mrg}Binding energy and angular momentum {\it at merger} (i.e. at the maximum of $|h_{22}|$) 
for nonspinning binaries versus the symmetric mass ratio $\nu$. The plot shows: (i) the highly accurate numerical agreement 
between SXS and Llama final configurations for $q=1$ and $q=2$; (ii) the good numerical consistency 
between EOB predictions and actual NR states.}
\end{center}
\end{figure}

 \begin{table*}[t]
   \caption{BBH configurations, evolved using the Llama code, used to compute the gauge-invariant relation between
    energy and angular momentum. From left to right: name of the run, number of orbits up to merger, mass ratio $q\equiv m_1/m_2$, 
    symmetric mass ratio $\nu$, dimensionless spins of the two black holes, $(\chi_1,\chi_2)$; the initial ADM mass $M^0_{\rm ADM}$; 
    initial total angular momentum ${\cal J}_{\rm ADM}^0$.}
  \centering  
\begin{ruledtabular}
  \begin{tabular}{cccccccc}        
    Name & N orbits & $q$ & $\nu$ & $\chi_1$  & $\chi_2$ & $M_{\rm ADM}^0$ & ${\cal J}_{\rm ADM}^0$ \\    
\hline
q1\_s0            &    8.09   &  1  & 0.25         & $0.0$     & $0.0$   & 0.99051968 & 0.99325600 \\
q2\_s0            &    6.70   &  2  & $0.\bar{2}$  & $0.0$     & $0.0$   & 0.990898   &  0.85599600\\
q1\_s-8D10\_h96   &    3.83   &  1  & 0.25         & $-0.8$    & $-0.8$  & 0.989412   &  0.61736   \\
q1\_s-8D10\_h1152 &    3.84   &  1  & 0.25         & $-0.8$    & $-0.8$  & 0.989412   &  0.61736   \\
q1\_s-6D12\_h64   &    8.25   &  1  & 0.25         & $-0.6$    & $-0.6$  & 0.986161   &  0.7552392 \\
q1\_s-6D12\_h512  &    8.23   &  1  & 0.25         & $-0.6$    & $-0.6$  & 0.986161   &  0.7552392  \\
q1\_s-4D10\_h64   &    5.67   &  1  & 0.25         & $-0.4$    & $-0.4$  & 0.990138   &  0.791588   \\
q1\_s-2D10\_h64   &    5.88   &  1  & 0.25         & $-0.2$    & $-0.2$  & 0.989941   &  0.877499   \\
q1\_s-2D10\_h512  &    5.88   &  1  & 0.25         & $-0.2$    & $-0.2$  & 0.989941   &  0.877499   \\
q1\_s2D8\_h64     &    4.93   &  1  & 0.25         & $+0.2$    & $+0.2$  & 0.987658   &  0.984992   \\
q1\_s2D8\_h512    &    4.93   &  1  & 0.25         & $+0.2$    & $+0.2$  & 0.987658   &  0.984992   \\
q1\_s4D8\_h64     &    5.67   &  1  & 0.25         & $+0.4$    & $+0.4$  & 0.987619   &  1.072176   \\
q1\_s6D8\_h64     &    6.38   &  1  & 0.25         & $+0.6$    & $+0.6$  & 0.987649   &  1.160296   \\
q1\_s6D8\_h512    &    6.48   &  1  & 0.25         & $+0.6$    & $+0.6$  & 0.987649   &  1.160296   \\
q1\_s8D8\_h64     &    7.24   &  1  & 0.25         & $+0.8$    & $+0.8$  & 0.987744   &  1.249296   \\
  \end{tabular}
\end{ruledtabular}
  \label{tab:llama}
\end{table*}

\subsection{Energetics with (nonspinning) SXS data}
\label{Sec:ejSXS}

We found that a similar excellent agreement among energetics 
holds, up to merger for all nonspinning SXS configurations 
at our disposal, which cover the mass-ratio range $1\leq q\leq 9.989$. 
[For a previous comparison of nonspinning merger characteristics
see Ref.~\cite{Damour:2013tla}].

So far, the only published computation of the energetics from SXS data was presented
in Ref.~\cite{Taracchini:2013rva}. It was limited to two, quasi extremal, 
spinning configurations with spin either aligned or antialigned with the orbital 
angular momentum, with $\chi_1=\chi_2=-0.95$ and $\chi_1=\chi_2=+0.98$.

We present here the first systematic computation of $E_b(j)$ curves from SXS 
data covering a large sample of nonprecessing, spinning (and nonspinning) 
configurations. We computed $E_b(j)$ curves from the nonspinning binaries 
in Table~\ref{tab:configs} applying the same standard procedure 
we used for Llama data. Note, however, that, while the time-integration of 
energy and angular momentum losses from Llama data could be meaningfully
performed starting at the beginning of the simulation, in the case of SXS
data, we found necessary to correct the time-integration of the early
``junk-radiation'' losses by adding a vectorial shift $(\Delta j^0,\Delta E_b^0)$ 
in the $(j,E_b)$ plane. This shift was essentially determined so as to 
minimize the EOB/NR difference $\Delta E_b^{\rm EOBNR}(j)$ during the early inspiral.
All technical details of our computation are explained in 
the Appendix [see in particular Table~\ref{tab:shifts}, that lists the values 
of the vectorial shifts $(\Delta j^0,\Delta E_b^0)$ we used].

The main results of our SXS-data analysis are: (i) for $q=1$ and $q=2$, the 
energetics computed from SXS data confirms the results based on Llama data given above 
in Figs.~\ref{fig:ej0_no_Fr}-\ref{fig:ej0_q2}; (ii) for the other values of $q$,
$3\leq q \leq 9.989$, not covered by Llama data, we again find an excellent consistency
between the energetics of our EOB model and the SXS NR one; (iii) it is in particular
remarkable that the binding energy and angular momentum {\it at merger} 
$(E^{\rm mrg}_b,j^{\rm mrg}_b)$ provided by our EOB model are in accurate 
agreement with the NR one~\footnote{As mentioned above, we recall that the instant of 
the merger is defined, both in EOB and NR, as the time 
when the of the modulus of the $\ell=m=2$ waveform reaches its maximum.}.
These results are exemplified in Figs.~\ref{fig:ejq8}, \ref{fig:diff} and Fig.~\ref{fig:mrg}.

For illustrative purposes, Fig.~\ref{fig:ejq8} explicitly shows the  $q=8$ case,
dataset SXS:BBH:0063. The bottom panel shows the EOB-NR difference. 
The vectorial shift in the  $(j,E_b)$ 
plane used here is $\Delta j^0=-4\times 10^{-3}$, 
$\Delta E_b^0 = -7.7709\times 10^{-4}$ (so that $E_b(j)\equiv E_b^{\rm raw}(j-\Delta j^0)+\Delta E_b^0$.
As discussed in detail in the Appendix these (quite small) shifts are 
determined through the following two-step procedure. Step 1:  $\Delta j^0$ is determined 
first, by requiring that the difference between the EOB and NR dimensionless orbital frequencies 
(defined as $\hat{\Omega}=\de E_b/\de j$)  $\Delta \hat{\Omega}^{\rm EOBNR}(j)=\hat{\Omega}^{\rm EOB}(j)-\hat{\Omega}^{\rm NR}(j)$ 
oscillates around zero on the largest possible $j$-interval. 
With a so chosen $\Delta j_0$, one finds that $\Delta\hat{\Omega}^{\rm EOBNR}$ 
oscillates between $\pm 5\times 10^{-5}$ over the interval $3.6\lesssim j \lesssim 4.2$.
Step 2: the constant $\Delta E_b^0$ is then determined by minimizing the difference 
$\Delta E_b^{\rm EOBNR}(j)\equiv E_b^{\rm EOB}(j)-E_b^{\rm raw}(j-\Delta j^0)$, which is 
essentially constant on a large $j$ interval, during the early inspiral.
After adding these small shifts to the raw data, one finds that 
$\Delta E_b^{\rm EOBNR}(j)$ is of the order of $\pm 10^{-5}$ for $3.5\lesssim j \lesssim 4.2$.

The shaded area in the bottom panel of Fig.~\ref{fig:ejq8} indicates the numerical uncertainty,
as estimated by taking the difference between the (raw) $E_b(j)$ curves obtained from the 
highest (Lev5) and second-highest (Lev4) resolution data present in the SXS catalog. 
In order to get a conservative uncertainty estimate, we do not apply any vectorial shift to the raw data.
Let us emphasize the remarkable agreement between EOB and NR 
energetics {\it at merger}. The difference at the NR merger point is 
of order  $4\times 10^{-4}$, which is barely visible in the top panel of Fig.~\ref{fig:ejq8}
(and it is approximately twice the estimated uncertainty there). 

For the other nonspinning configurations in Table~\ref{tab:configs}, one obtains 
curves quite analogous to the $q=8$ case. The relevant information is displayed
in Fig.~\ref{fig:diff} that collects the 
$\Delta E_b^{\rm EOBNR}(j)\equiv E_b^{\rm EOB}(j)-E_b^{\rm NR}(j)$ 
differences for all nonspinning data but one (the $q=1.5$ one). 
Note that the end points indicated by square markers 
in Fig.~\ref{fig:diff}, record the values of the function 
$\Delta E_b^{\rm EOBNR}(j)$ at the corresponding EOB merger
values of $j$, $j_{\rm mrg}^{\rm EOB}$. On the other hand, the
stars on the $x$-axis of the figure mark the corresponding
NR merger values of $j$, $j^{\rm NR}_{\rm mrg}$.

For $q=1$, we use here the SXS:BBH:0066 dataset with $\sim 28$ orbits.
The vectorial shifts $(\Delta j_b^0,\Delta E_b^0)$ we use are listed
in Table~\ref{tab:shifts} in the Appendix. For all configurations, one finds 
that the EOB/NR agreement is of the order $10^{-5}$ at the beginning of the inspiral 
and grows only up to a few parts in $10^{-4}$ around merger.
These differences are comparable to (though slightly larger than)
the corresponding error bars computed, as above, taking the difference 
between the two highest resolutions available.

In  Fig.~\ref{fig:mrg} we complement the above results by focusing
on the EOB/NR comparison for 
merger characteristics, $(j_{\rm mrg},E_{\rm mrg})$ as a function of $\nu$.
The figure shows that the EOB/NR disagreement increases with $\nu$:
from a difference of $3.07\times 10^{-4}$ 
for $E_b^{\rm mrg}$ and $1.7\times 10^{-4}$ for $j^{\rm mrg}$ for $q=9.989$ ($\nu=0.0827$)
the EOB model ends up slightly overestimating the NR values in the equal-mass
case, $\nu\to 0.25$.
In particular, for $\nu=0.25$ we see that the EOB prediction for $E_b^{\rm mrg}$
is larger than the NR one by $4.9\times 10^{-3}$ and that the EOB prediction 
for $j^{\rm mrg}$ is larger than the NR one by  $3.1\times 10^{-2}$.
[These differences are larger than those apparent in the corresponding 
endpoints of Fig.~\ref{fig:diff} because we are now computing
$E^{\rm EOB}_b(j_{\rm mrg}^{\rm EOB})-E_b^{\rm NR}(j_{\rm mrg}^{\rm NR})$ while
the vertical displacement of the endpoints of Fig.~\ref{fig:diff} corresponded
to $E^{\rm EOB}_b(j_{\rm mrg}^{\rm EOB})-E_b^{\rm NR}(j_{\rm mrg}^{\rm EOB})$].
In the same figure we also display Llama merger data, denoted by asterisks. 
We find that Llama and SXS data are fully consistent within the $10^{-4}$ level. 
Finally, the $\nu$-dependence of the SXS NR $E_b^{\rm mrg}$ can 
be fitted by means of a second order polynomial, while
$j^{\rm mrg}(\nu)$ is well fitted by a straight line, namely
\begin{align}
E_{b,{\rm SXS}}^{\rm mrg} & = -0.062951 -0.176949\nu -0.216264\nu^2,\\
j^{\rm mrg}_{\rm SXS}   & =  3.406583-2.421052\nu.
\end{align}

\section{NR calibration of the EOB spin-orbit coupling}
\label{sec:spin}

Let us now turn to discussing the numerical completion of the EOB spinning model. 
As mentioned above, the spinning sector of the model we use here is exactly the 
same as described in Paper~I. In the latter work it was shown that it was possible
to obtain a good EOBNR phasing agreement by tuning a {\it single} effective functional 
parameter $c_3(\chi)$ that was entering the spin-orbit coupling at 
next-to-next-to-next leading order\footnote{The analysis of Paper~I was limited to the
equal-mass, equal-spin case.}. The best values of the parameter $c_3$, for various spin values, 
are listed in Table~I of Paper~I. Since the orbital $A_{\rm orb}$ function has 
changed, in the present work, because  of (i) the use of the 
complete analytical  4PN information and (ii) the consequently modified 
functional dependence of $a_c^6(\nu)$ given by 
Eq.~\eqref{eq:a6c_calibrate}, we now need to look for a new determination 
of $c_3(\chi)$. In doing so, we also consider here more NR simulations than in Paper~I, 
notably taking into account all available nonprecessing data in the SXS catalog. 
The spinning SXS configurations we use are listed in Table~\ref{tab:configs}: 
the mass ratio varies in the range $1\leq q \leq 8$ and there
are several configurations where only one of the two black-holes is spinning.
A priori, one expects the NNNLO effective parameter $c_3$ to be a function of both 
the mass ratio and the spins of the binary. The determination of $c_3$ is done
in two steps. In a first step, we separately considered each binary configuration
and determine a preliminary best value of $c_3$ by minimizing the EOB-NR phase
difference, after alignment in the inspiral phase, so to be compatible with 
(and typically smaller than) the NR uncertainty. This procedure is rather 
straightforward, as it is just a one parameter search.
In a second step, we look for a global, analytical representation that approximately
reproduces the latter preliminary best values of $c_3$ as a function of symmetric
mass ratio and spins.
We found that one can represent, with sufficient accuracy, the values 
(determined by minimizing the EOB/NR phase difference as explained above) 
of $c_3$ for the entire sample of configurations listed in 
Table~\ref{tab:configs}, by means of the following simple functional relation
\begin{align}
\label{eq:c3fun}
&c_3(\ha_1,\ha_2,\nu)  = p_0 \dfrac{1 + n_1(\ha_1+\ha_2) + n_2(\ha_1+\ha_2)^2}{1 + d_1(\ha_2+\ha_2)}\nonumber\\
               & + (p_1\nu  + p_2\nu^2 + p_2\nu^3) (\ha_1 + \ha_2)\sqrt{1-4\nu} \nonumber\\
              & + p_4 (\ha_1-\ha_2)\nu^2,
\end{align}
where
\begin{align}
p_0 & = 44.786477,\\
n_1 & = -1.879350,\\
n_2 & =  0.894242,\\
d_1 & =  -0.797702,\\
p_1 & =  1222.36, \\
p_2 & =  -12764.4, \\
p_3 & =  36689.6,\\
p_4 & = -358.086.
\end{align}
\begin{figure}[t]
\begin{center}
\includegraphics[width=0.47\textwidth]{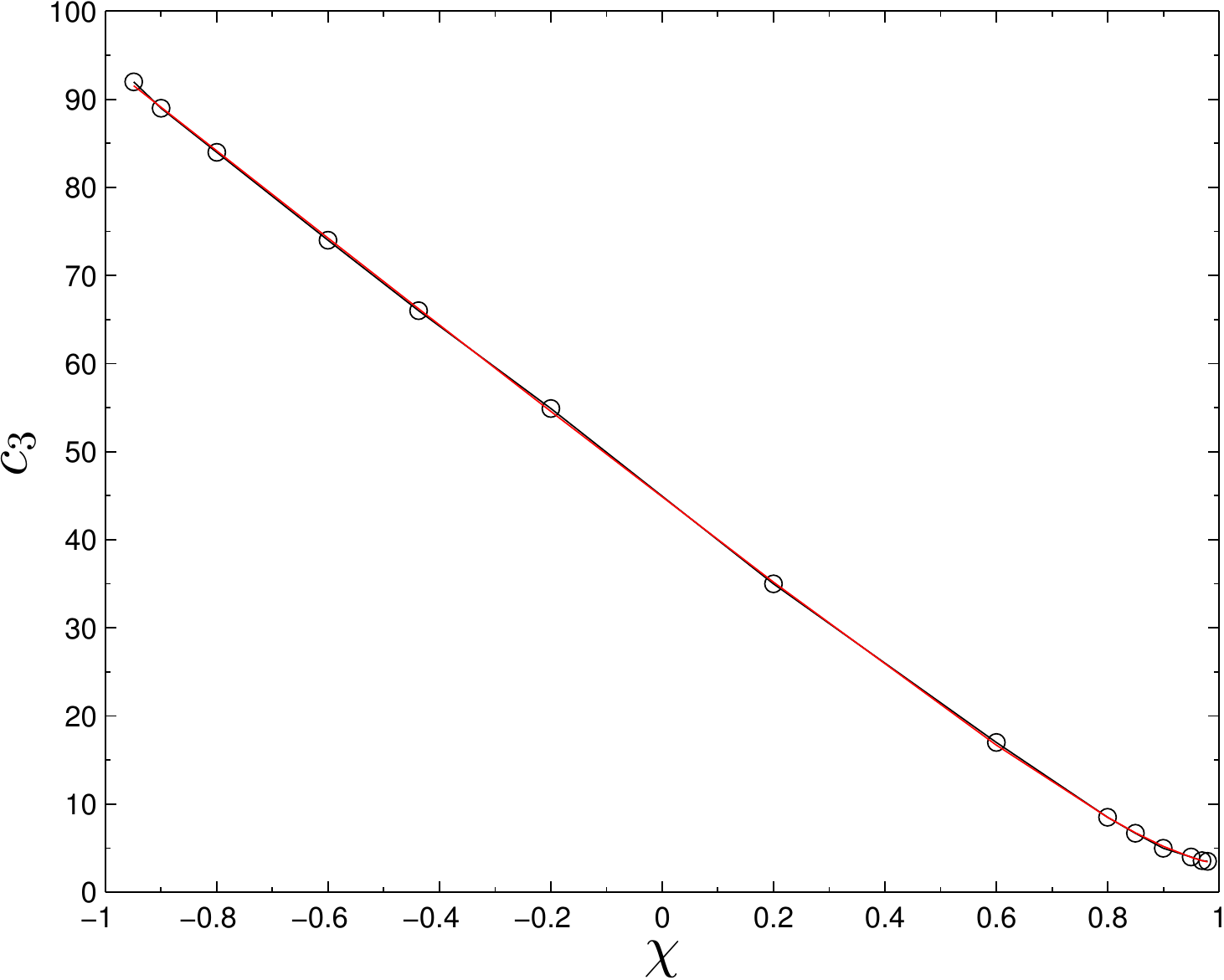}
\caption{\label{fig:c3_vs_chi} Quasi-linear behavior of the NNNLO
effective spin-orbit coefficient $c_3(\hat{a}_1,\hat{a}_2,\nu)$, 
Eq.~\eqref{eq:c3fun}, in the equal-mass ($\nu=0.25$), equal-spin
$\hat{a}_1=\hat{a}_2=\chi$ case. The tuning of this single dynamical 
parameter allows to get an excellent EOB/NR phasing agreement 
throughout inspiral, plunge, merger and ringdown.}
\end{center}
\end{figure}
and where we found convenient to introduce the spin quantities $\ha_{1,2}\equiv X_{1,2}\,\chi_{1,2}$, with
$X_{1,2}\equiv m_{1,2}/M$, and $M=m_1+m_2$. With our convention that
$m_1>m_2$, in terms of the symmetric mass ratio $\nu$ we have
\be
X_1 = \dfrac{1}{2}\left(1+\sqrt{1-4\nu}\right),
\ee
and $X_2 = 1-X_1$. The terms in Eq.~\eqref{eq:c3fun} that vanish in
the equal-mass, equal-spin case were chosen, for simplicity, to be 
linear in the spins. Similarly, the polynomial dependence in $\nu$ was 
found necessary to properly fit the values of $c_3$ yielding a 
good NR/EOB phasing agreement for $q=8$, 
$(\chi_1,\chi_2)=(+0.5,0)$, SXS:BBH:0065 configuration.
In the equal-mass, equal-spin (aligned, or anti-aligned with the orbital 
angular momentum) case, one finds that the dependence of $c_3$ on the spin 
is nearly linear (see Fig.~\ref{fig:c3_vs_chi}).

Finally, let us mention that, as already discussed 
in Paper~I (see Table~I there) we found it necessary 
to flex the simple choice $\Delta t_{\rm NQC}=1M$, uniformly 
used in the nonspinning case, so as to allow it to depend on spin
for large, positive, spins. Actually, the only available simulations
where we found the need of flexing $\Delta t_{\rm NQC}$ are the six
equal-mass, equal-spin configurations with $\chi=\chi_1=\chi_2> \chi_0=+0.85$. 
In practice, for spins $\chi=\{+0.90,+0.95,+0.97,+0.98,+0.99,+0.994\}$ we
found as good choices 
$\Delta t_{\rm NQC} = \{0.2,-1.2,-1.7,-2.0,-3.0,-3.2\}$, respectively.
In the EOB numerical evolution we use a time-resolution $\Delta t^{\rm EOB}=0.1M$,
and $\Delta t_{\rm NQC}$ is chosen as an integer multiple of $\Delta t^{\rm EOB}$.
The values of $\Delta t_{\rm NQC}$ listed above are accurately fitted by 
using a $(1,1)$ Pad\'e approximant:
\be
\label{eq:DTnqcFit}
\Delta t_{\rm NQC}(\chi) = \dfrac{1 + n_1(\chi-\chi_0)}{1+d_1(\chi-\chi_0)}
\ee
with $n_1=-16.06288$ and $d_1=-4.04266$ and $\chi_0=0.85$.

The quality of the fit yielded by Eq.~\eqref{eq:c3fun} (together with
the discrete values of $\Delta t_{\rm NQC}$ listed above)
is quantitatively assessed by measuring the EOB-NR  phase 
difference at NR merger after having aligned (in time and phase)  the EOB waveforms 
to the NR waveform during the early inspiral. Such differences are listed 
as $\Delta\phi_{\rm mrg}^{\rm EOBNR}$ in Table~\ref{tab:configs}. 
The same table also clearly illustrates the compatibility of the 
EOBNR model with the numerical phase uncertainties 
$\delta\phi^{\rm NR}_{\rm mrg}$ at merger all over the waveform sample considered. 
The use, in addition to Eq.~\eqref{eq:c3fun}, of the fit \eqref{eq:DTnqcFit}
slightly worsens $\Delta\phi_{\rm mrg}^{\rm EOBNR}$ as indicated by the values
in parantheses in Table~\ref{tab:configs}.
Anyway, both values are approximately within (half of) the numerical error bar. 

\begin{figure}[t]
\begin{center}
\includegraphics[width=0.47\textwidth]{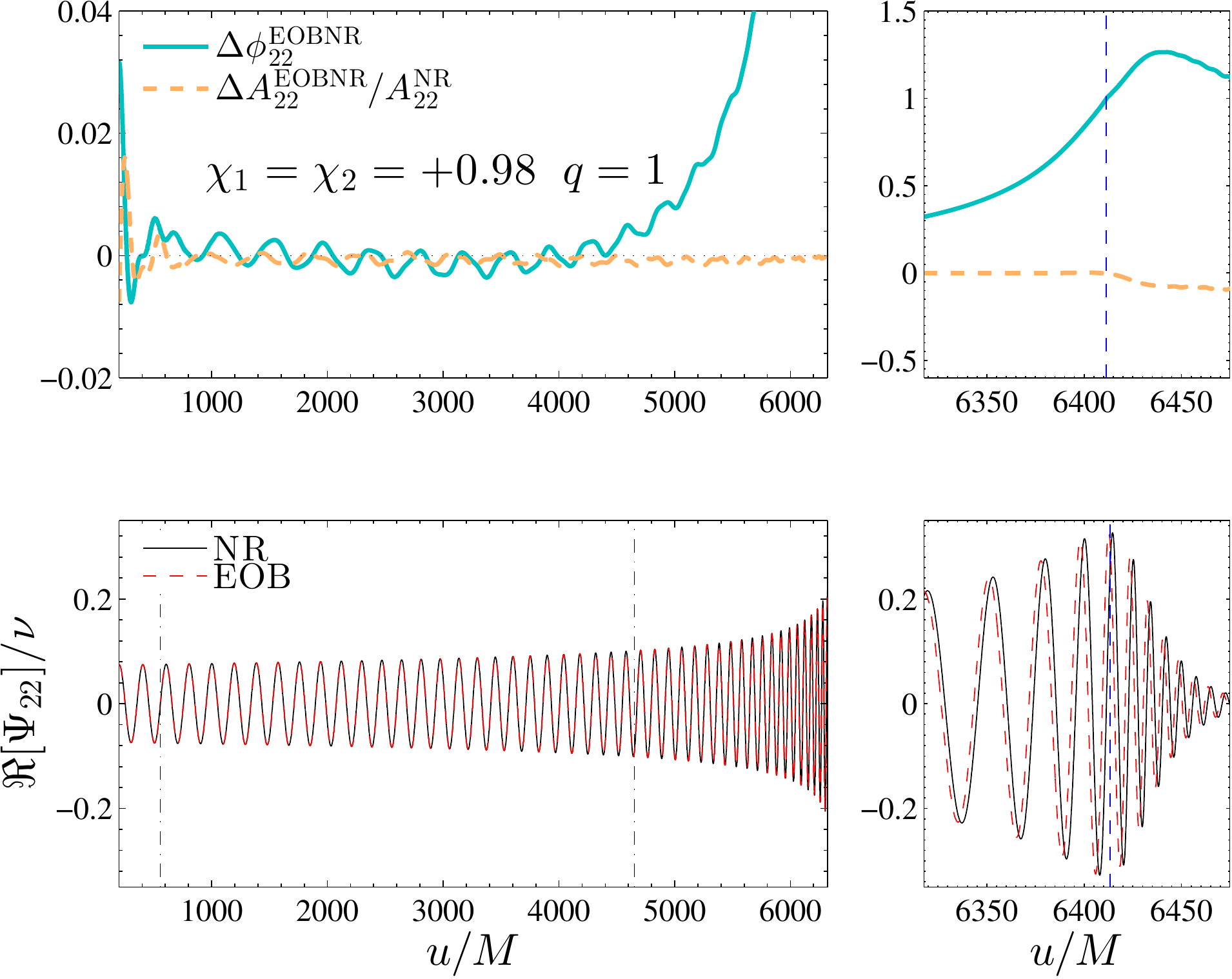}\\
\includegraphics[width=0.47\textwidth]{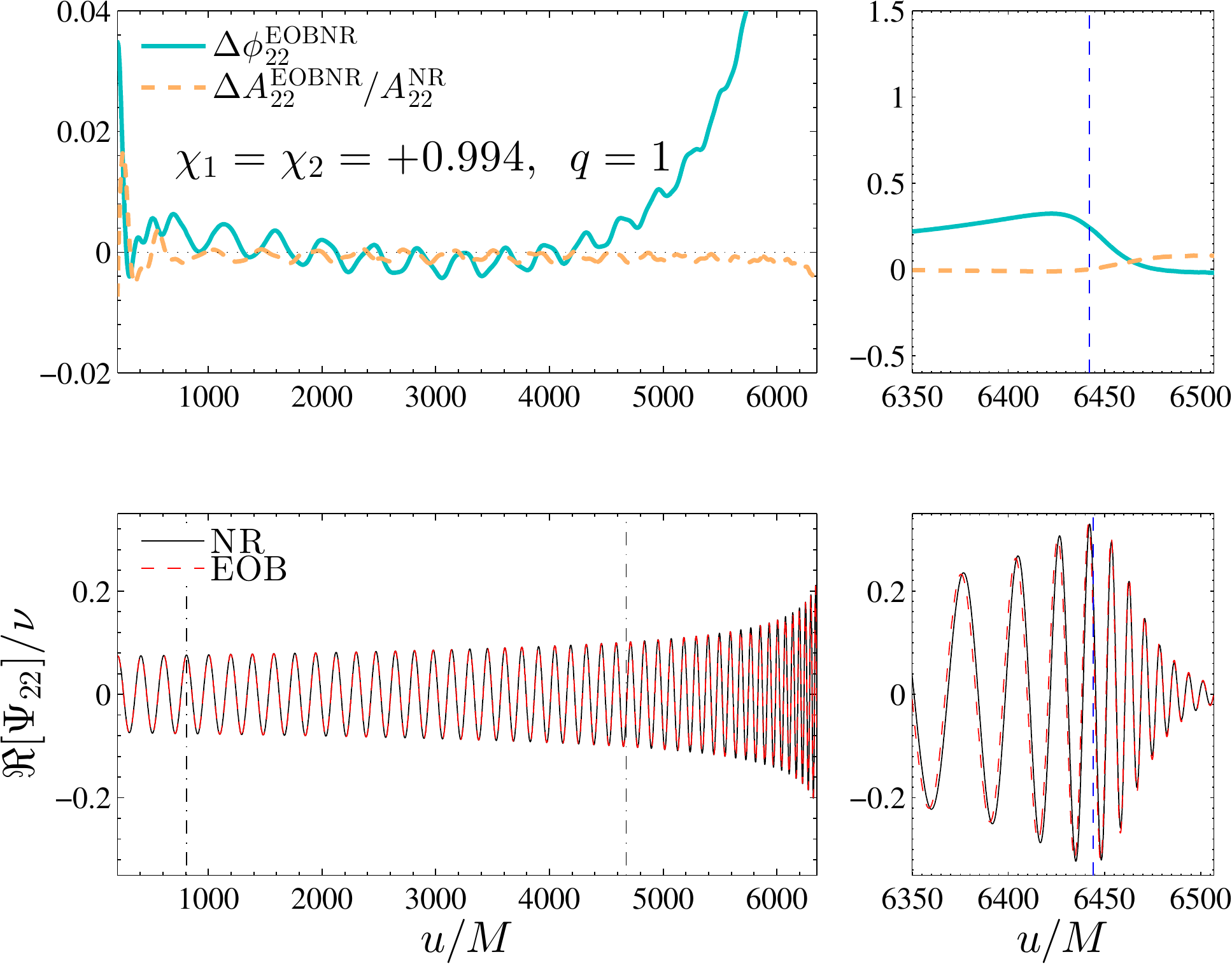}
\caption{\label{phasing_q1098} Illustrative EOBNR time-domain phasing comparison for
quasi-extremally spinning binaries: \hbox{$(q,\chi_1,\chi_2)=(1,+0.98,+0.98)$}, 
dataset SXS:BBH:0172 (top panel), and \hbox{$(q,\chi_1,\chi_2)=(1,+0.994,+0.994)$}, 
dataset SXS:BBH:0178 (bottom panel). In both cases, the EOBNR difference at merger 
(dashed vertical line) is within the corresponding NR  uncertainty, that is 
$\sim 2$~rad for $\chi=+0.98$ and $\sim 0.53$~rad for$\chi=+0.994$ (see Table~\ref{tab:configs}).}
\end{center}
\end{figure}

Let us emphasize that the EOB/NR agreement remains excellent for the 
near-extremal spinning cases $\chi_1=\chi_2=+0.98$ and $\chi_1=\chi_2=+0.994$.
Figure~\ref{phasing_q1098} shows the EOB/NR comparison 
obtained by performing the usual time-domain comparison when aligning the waveforms 
on the frequency interval $(M\omega_{L},M\omega_{R})$ corresponding 
to the time-interval indicated by the two vertical dashed line in the plot.
Analogous (or better) plots are found for all other configurations.
As an indicator of this good EOB/NR agreement we just give in Table~\ref{tab:configs}.
the values of the phase difference at merger.

\begin{figure}[t]
\begin{center}
\includegraphics[width=0.48\textwidth]{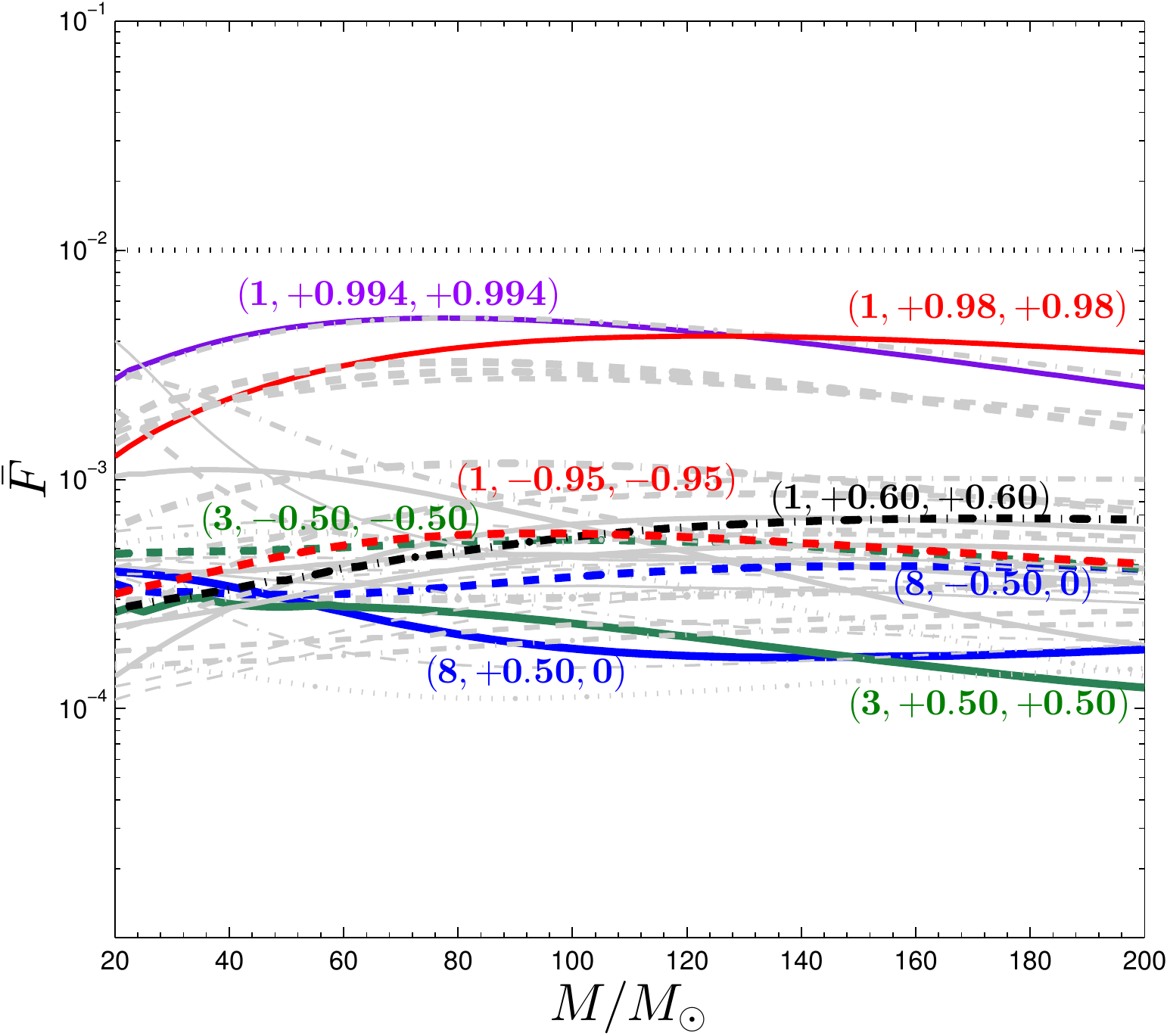}
\caption{\label{Fbar}Unfaithfulness of the \hbox{$\ell=m=2$} EOB waveforms with respect to the NR ones 
for all BBH datesets of Table~\ref{tab:configs}. The largest values of $\bar{F}$ corresponds 
to the quasi-extremal spinning configurations SXS:BBH:0177 and SXS:BBH:0178 in Table~\ref{tab:configs}. 
The labels of the configurations use the notation $(q,\chi_1,\chi_2)$; they are highlighted 
in color so as to ease a direct comparison with Fig.~1 of Ref.~\cite{Taracchini:2013rva}.}
\end{center}
\end{figure}
\begin{figure}[t]
\begin{center}
\includegraphics[width=0.48\textwidth]{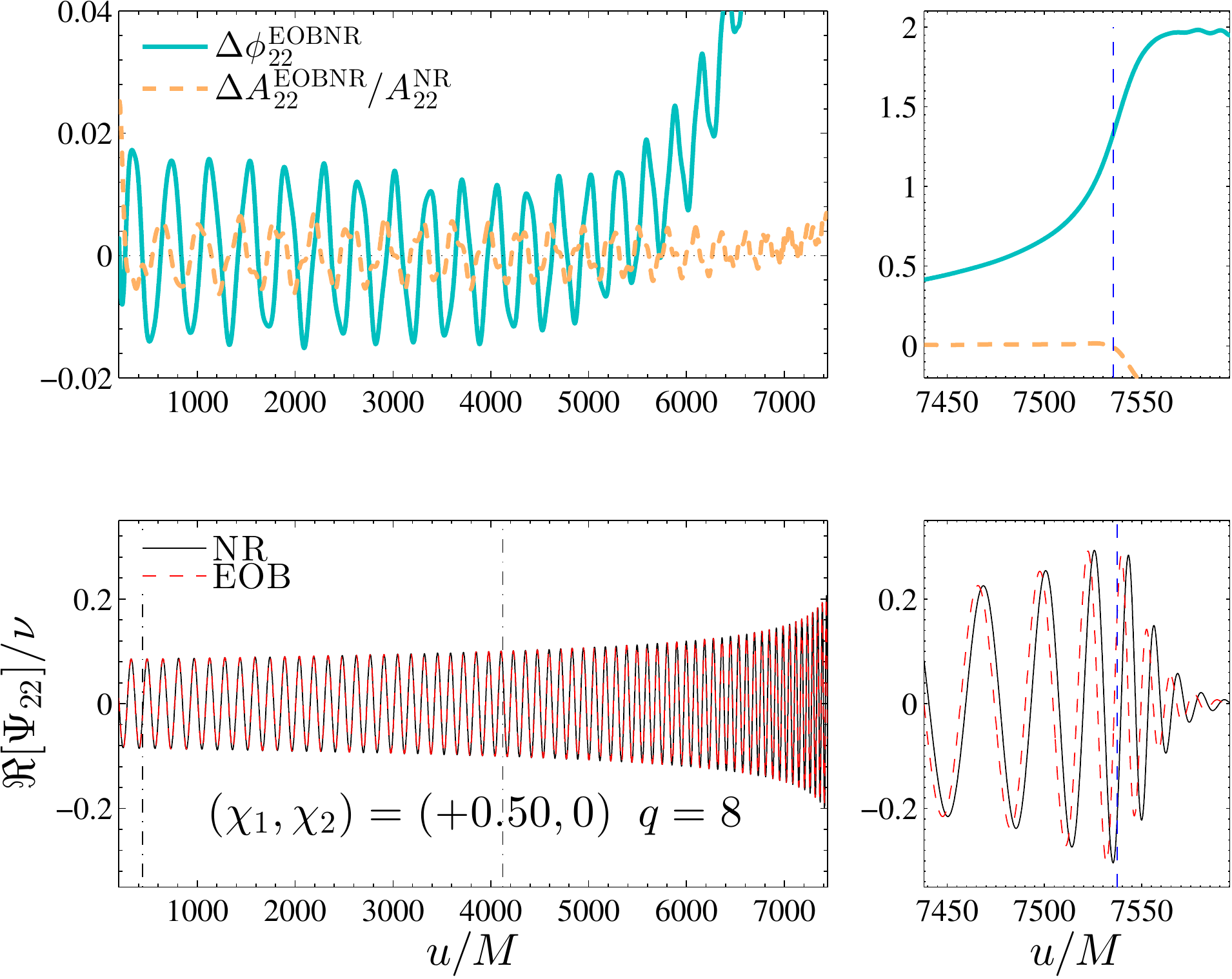}
\caption{\label{phasing_q805}EOBNR time-domain phasing comparison for SXS:BBH:0065 
configuration, $q=8$, $(\chi_1,\chi_2)=(+0.5,0)$. The EOBNR difference at merger
(dashed vertical line) is compatible with the corresponding NR uncertainty 
$\sim-3$~rad (see Table~\ref{tab:configs}). 
The eccentricity ($\sim 10^{-3}$) modulation in the phase difference is rather visible.}
\end{center}
\end{figure}
To further demonstrate the high-quality of the EOB model presented here, and to give 
a clearer physical meaning to the phase differences quoted above, we also measured
the agreement between the EOB waveforms and all the available NR ones by computing 
the EOB/NR unfaithfulness (as a function of the total mass $M$)
\be
\label{eq:barF}
\bar{F}(M)\equiv 1 - \max_{t_0,\phi_0}\dfrac{\langle h_{22}^{\rm EOB},h_{22}^{\rm NR} \rangle}{||h_{22}^{\rm EOB}||||h_{22}^{\rm NR}||} \ ,
\ee
where $t_0$ and $\phi_0$ are the initial time and phase, $||h||\equiv \sqrt{\langle h,h\rangle}$,
and the inner product between two waveforms is defined as 
$\langle h_1,h_2\rangle\equiv 4\Re \int_{f_{\rm min}^{\rm NR}(M)}^\infty \tilde{h}_1(f)\tilde{h}_2^*(f)/S_n(f)\, df$,
where $S_n(f)$ is the zero-detuned, high-power noise spectral density of advanced LIGO~\cite{shoemaker} 
and $f_{\rm min}^{\rm NR}(M)=\hat{f}^{\rm NR}_{\rm min}/M$
is the {\it starting frequency of the NR waveform} (after the junk radiation initial transient).
Both EOB and NR waveforms are tapered in the time-domain so as to reduce high-frequency oscillations 
in the corresponding Fourier transforms.
The procedure of tapering the waveforms is the same followed by Ref.~\cite{Taracchini:2013rva}, 
that introduced a different EOB model calibrated to NR, called SEOBNRv2, and performed the
same unfaithfulness analysis we are doing here.
Figure~\ref{Fbar} shows the so computed unfaithfulness as a function of the total mass of the binary for
all configurations we considered. The maximum of $\bar{F}(M)$ is also listed, for convenience, 
in the last column of Table~\ref{tab:configs}. One sees that for most of 
all considered configurations  the value of $\bar{F}$ always stays one order 
of magnitude below the reference value of $1\%$ 
(actually, most configurations range between $0.1\%$ and $0.01\%$) 
as the total mass of the binary ranges from 20 to 200$M_\odot$. 
Such a waveform quality implies a negligible loss in event rate due to the modeling 
uncertainty within the frequency range $f\geq f^{\rm NR}_{\rm min}(M)$. 
Note,however, that the NR waveforms do not cover the entire frequency band of the 
detector when $M\lesssim 100M_\odot$. For example, the longest NR waveform available, 
SXS:BBH:0002, has $f_{\rm min}^{\rm NR}(M)\approx 900 (M_\odot/M)$~Hz, so that
it starts at a frequency $\leq 10$~Hz when $M\geq 90~M_\odot$ (and  $\leq 20$~Hz when $M\geq 45~M_\odot$).
The corresponding value of the unfaithfulness for $M=90M_\odot$ is $\bar{F}=8.54\times 10^{-4}$.
For the highest spinning waveform, SXS:BBH:0178, we have  $f_{\rm min}^{\rm NR}(M)\approx 1300 (M_\odot/M)$~Hz
so that we need $M\geq 130M_\odot$ to reach $f_{\rm min}^{\rm NR}=10$~Hz.
However, several recent works have shown that the EOB formalism provided the best
available description of GW phasing even during early inspiral~\cite{Pan:2013tva,Szilagyi:2015rwa,Husa:2015iqa}.
In absence of longer NR waveforms able to cover the entire frequency range of the detector
for $M\geq 20M_\odot$, the values of $\bar{F}(M)$ displayed in Fig.~\ref{Fbar} do
not have a direct data-analysis meaning over the full plotted mass range 
$20M_\odot \leq M \leq 200 M_\odot$. We, however, expect, notably in view of 
Ref.~\cite{Szilagyi:2015rwa}, that the values of $\bar{F}$ will not significantly
degrade when using longer NR waveforms.

Figure~\ref{Fbar} highlights in color the same particular configurations that were
highlighted in Fig.~1 of Ref.~\cite{Taracchini:2013rva}, so as to prompt an easy and 
direct comparison. In addition, Fig.~\ref{Fbar} also shows, in purple, the 
configuration $(+0.994,+0.994)$ that was not available when 
Ref.~\cite{Taracchini:2013rva} was written.
Note that the worst global unfaithfulness, of the order of 
$\max{\bar F}\approx 0.005$, corresponds to the two quasi-extremal spinning 
cases $\chi_1=\chi_2=(+0.99,+0.994)$, see Table~\ref{tab:configs}.
A direct comparison with Fig.~1 of Ref.~\cite{Taracchini:2013rva}
allows us to make the following observations: (i) the configuration 
$(q,\chi_1,\chi_2)=(1,+0.6,+0.6)$, SXS:BBH:0152, delivers, within our EOB model, 
$\bar{F}(M)< 10^{-3}$, all over the total mass range considered. 
The corresponding curve in Fig.~1 of Ref.~\cite{Taracchini:2013rva} was starting 
around $\bar{F}\simeq 4.5\times 10^{-3}$ for $M=20M_\odot$, then increasing up to 
$\bar{F}\simeq 10^{-2}$ for $M\simeq 50M_\odot$ before decreasing again down 
to $2.5\times 10^{-3}$ for $M=200 M_\odot$. Similary, the other rather 
extreme case $(q,\chi_1,\chi_2)=(8,+0.5,0.0)$, SXS:BBH:0065, 
yields here an unfaithfulness of order $4\times 10^{-4}$ at 
$M=20~M_\odot$, that then {\it decreases} by a factor two
for larger total mass. By contrast, it is interesting to 
note that the EOB/NR unfaithfulness computed for the same 
configuration using SEOBNRv2 {\it increases} with $M$ 
from $\approx 0.3\%$ to $\approx 1\%$
(see Fig.~1 of Ref.~\cite{Taracchini:2013rva}). 

Note that the, numerical phase uncertainty on this dataset is as 
large as $-3$~rad at merger, so that $c_3$ was (conservatively) calibrated so as to yield 
a 1.3~rad EOB/NR phase difference at merger (see Fig.~\ref{phasing_q805}).
We advocate new simulations with reduced error bars for this configuration
to firm up (and possibly improve) the current EOB calibration.

Globally, the results collected in Fig.~\ref{Fbar} show that our model 
quantitatively improves upon existing results. We shall discuss more these 
and other aspects of our unfaithfulness comparison in the 
Conclusions (see in particular Fig.~\ref{histo} and related discussion there).

\begin{figure*}[t]
\begin{center}
\includegraphics[width=0.32\textwidth]{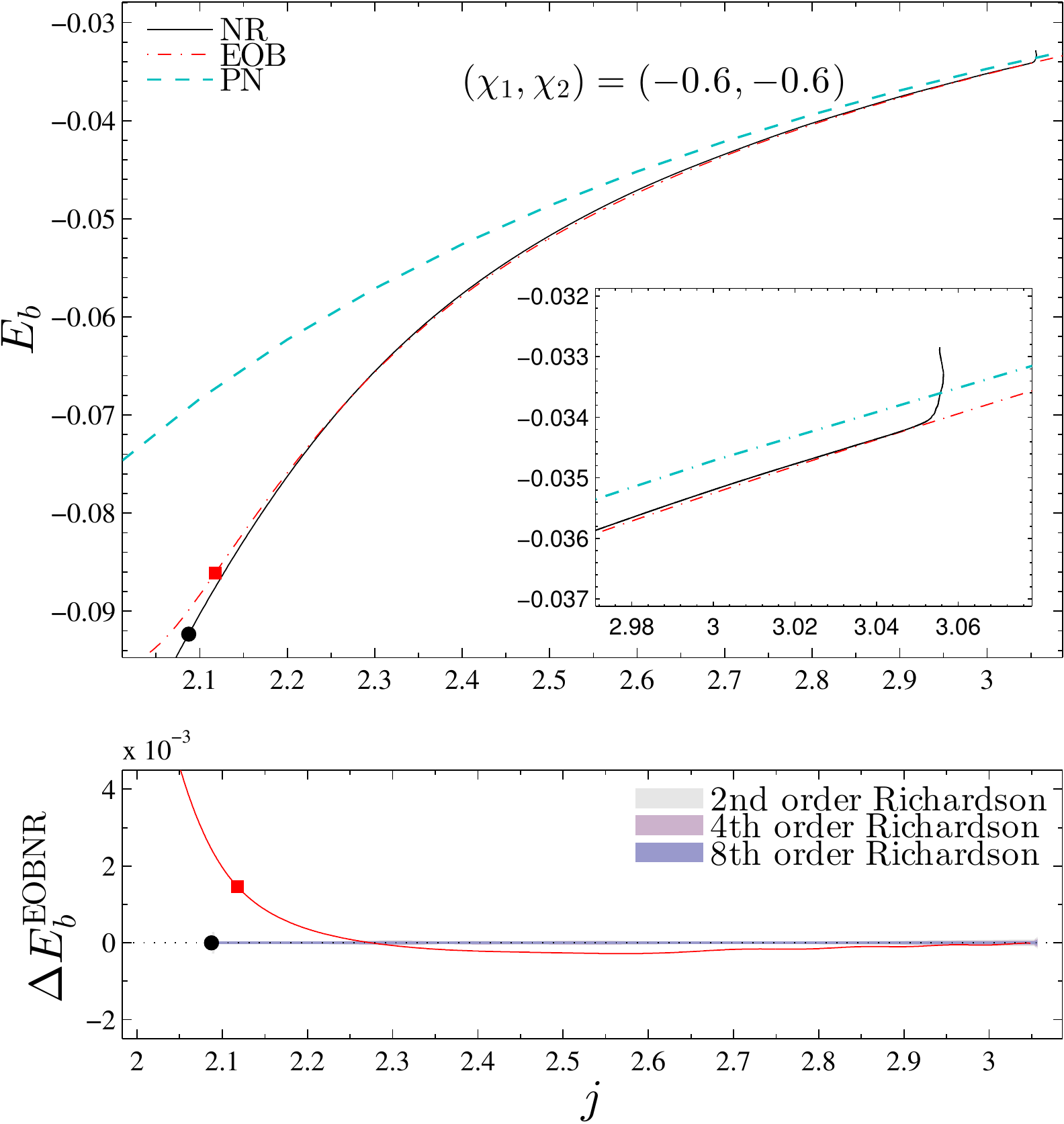}\;\;
\includegraphics[width=0.32\textwidth]{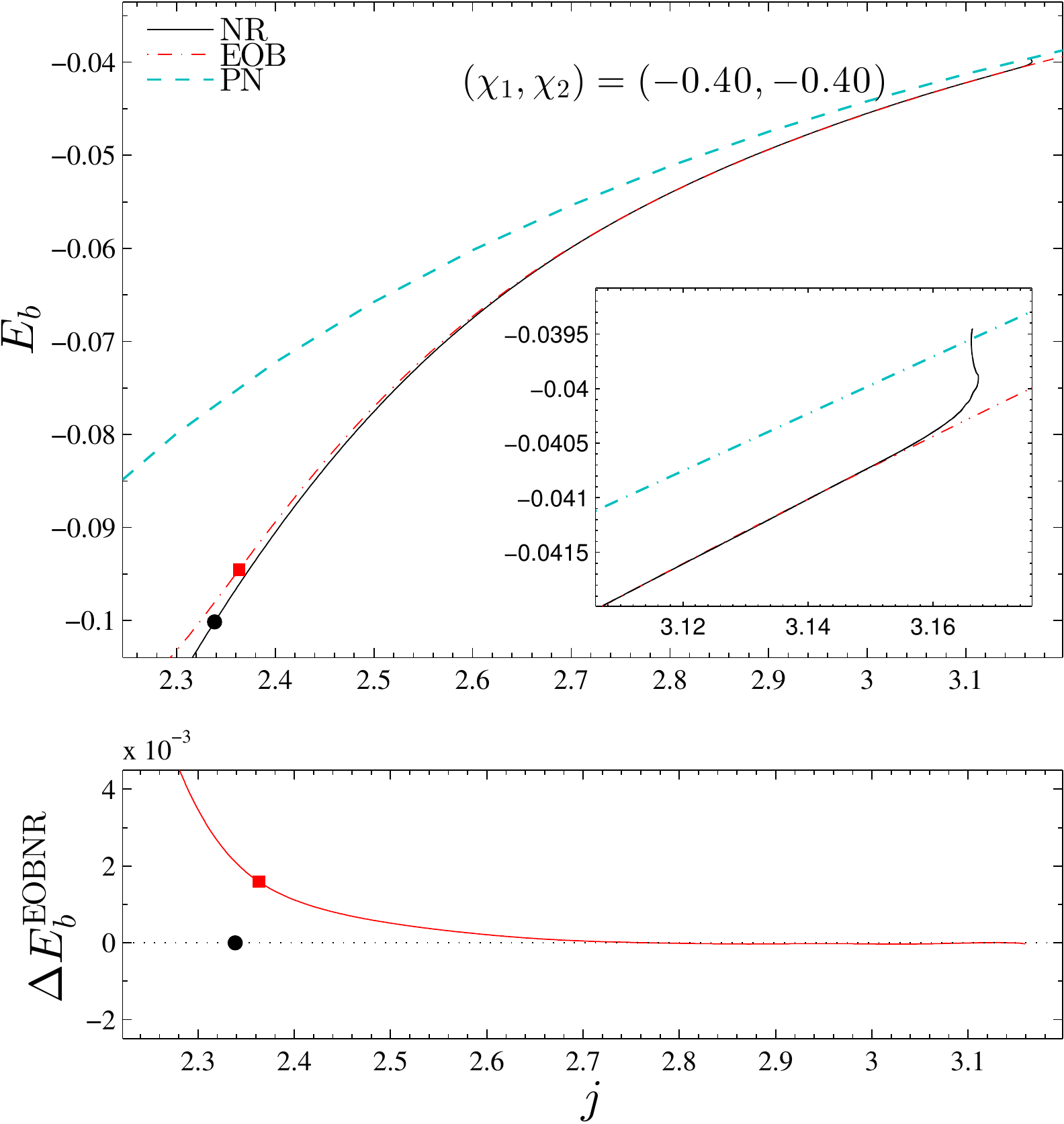}\;\;
\includegraphics[width=0.32\textwidth]{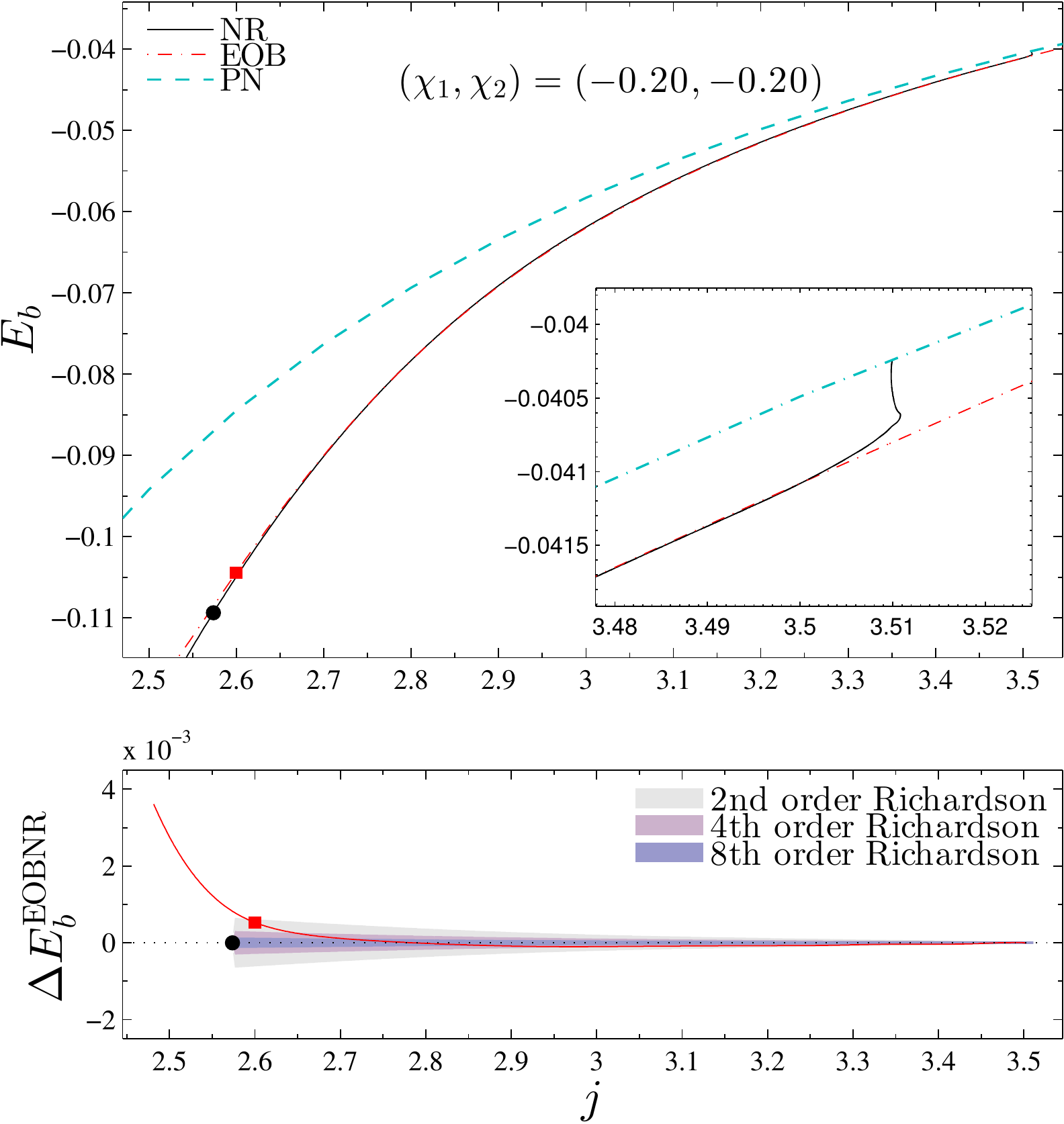}\\
\includegraphics[width=0.32\textwidth]{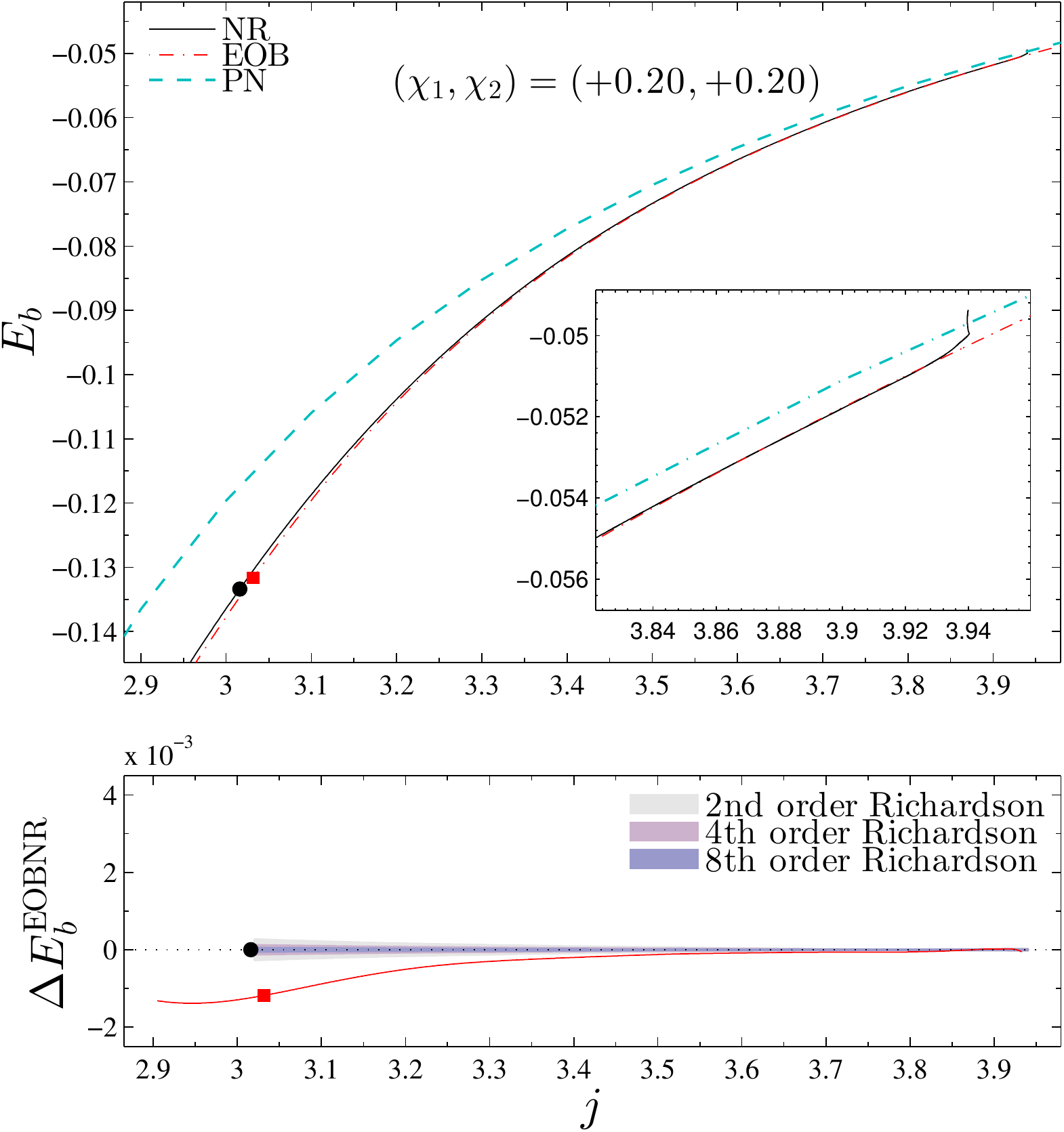}\;\;
\includegraphics[width=0.32\textwidth]{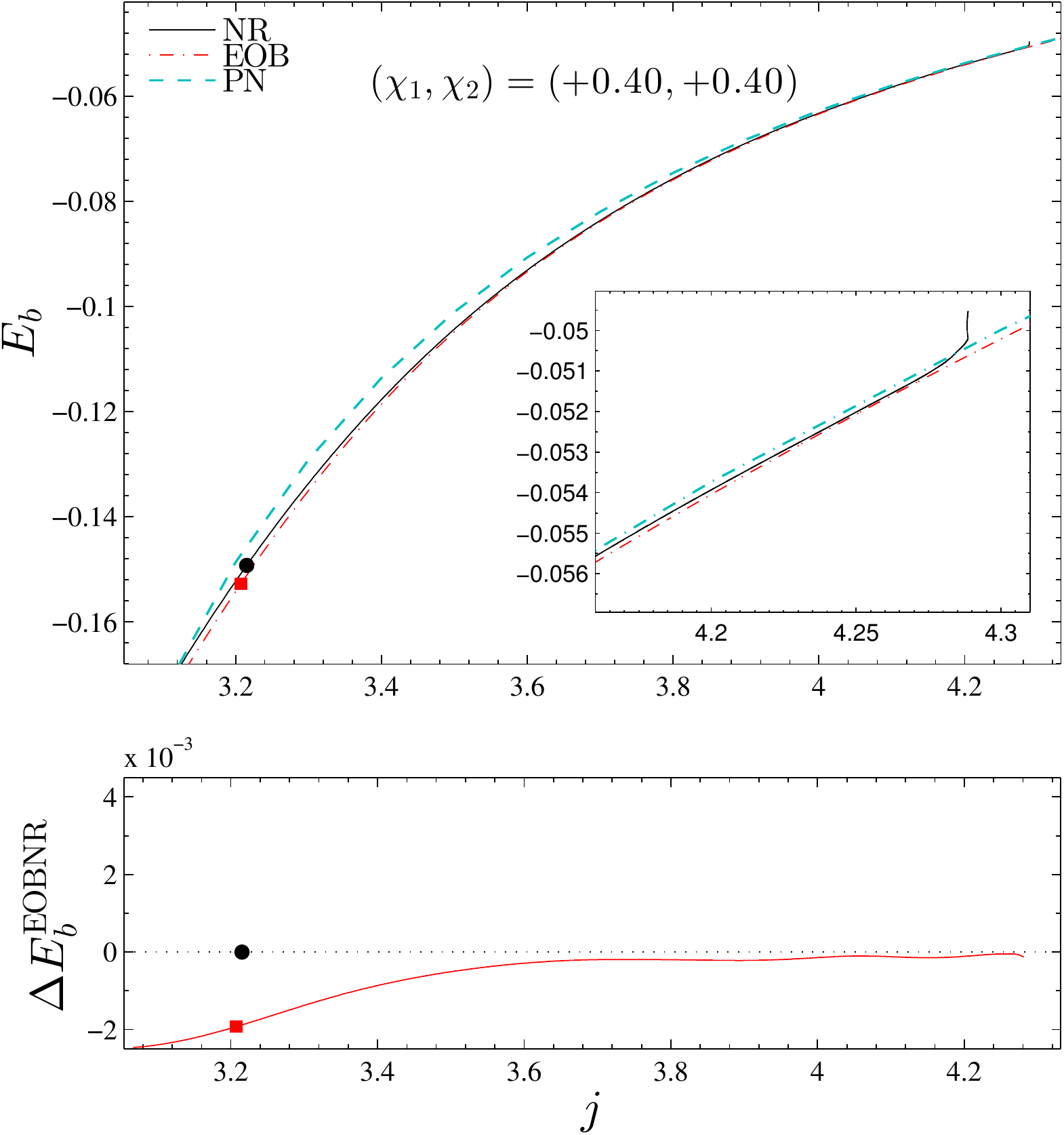}\;\;
\includegraphics[width=0.32\textwidth]{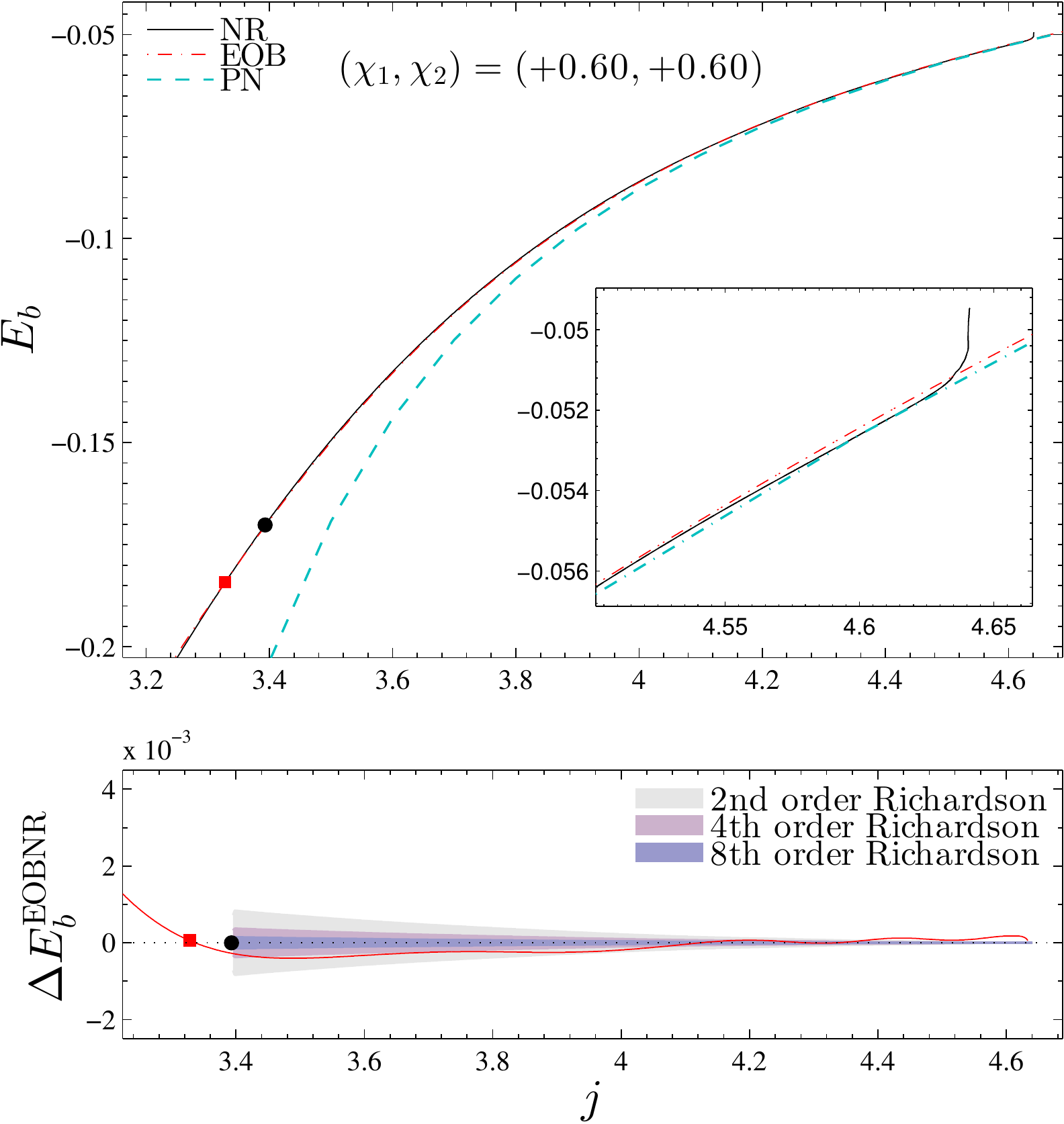}
\caption{\label{fig:Llama_ej}Energy versus angular momentum curves for spinning equal-mass binaries
with spin $\chi$ either aligned or anti-aligned to the orbital angular momentum, as obtained from Llama data. Good mutual
consistency EOB/NR is found. The shaded area indicates an error bar estimate obtained by Richardson
extrapolating three resolutions.}
\end{center}
\end{figure*}

\section{Energetics for spinning coalescences}
\label{sec:energetics}
\subsection{Energetics of spinning Llama data}
Let us finally discuss the energetics of spinning coalescences 
yielded by our newly calibrated EOB model.
We start doing this with Llama data and we will cross check our results with
SXS data in the next Section.
Figure~\ref{fig:Llama_ej} contrasts the NR and EOB $E_b(j)$ curves with 
$\chi_1=\chi_2=(\pm 0.2,\pm 0.4,\pm 0.6)$, with the EOB-NR difference $\Delta E_b(j)$ shown 
in each bottom subpanel. As before, the EOB (red) and NR (black) mergers 
are indicated by markers. One sees that the differences are of the order of
$10^{-4}$ (or less) during the inspiral, to grow up to approximately the $10^{-3}$
level around merger. One also notices that the disagreement between
NR and EOB merger quantities depends on the configuration.
In several representative cases we have indicated in Fig.~\ref{fig:Llama_ej}
an estimate of the NR uncertainty on the energetics.
The latter estimate was obtained by first taking the difference between the $E_b(j)$ curves 
computed from the highest ($\Delta x=0.512$) and second highest ($\Delta x=0.64$) 
resolution at our disposal and then Richardson extrapolating it assuming 
some convergence order. The three colored areas around the difference in the 
bottom panels of Fig.~\ref{fig:Llama_ej} display three different estimates
of this NR uncertainties, as obtained by: (i) conservatively assuming 2nd 
order convergence (lighter area); (ii) assuming 4th order convergence; and
(iii) assuming 8th order convergence. The most conservative estimate (light grey), 
may eventually dominate the error budget in the limit of infinite resolution
because of the presence of second-order finite difference operators in 
the numerical infrastructure. This gives bounds of the order $\pm 10^{-3}$ around merger.
Such a $10^{-3}$ level is compatible with the EOB/NR differences at merger 
that we find for all other configurations.
Other sources of uncertainty on the NR data such as: (i) the conversion 
from $\Psi^4_{\ell m}$ to $h_{\ell m}$ \cite{Reisswig:2010di}; (ii) the CCE error due to the choice
of the finite-radius worltube where the CCE is started~\cite{Reisswig:2009rx, Taylor:2013zia}; 
or (iii) the effect of higher multipoles, that contribute at the $10^{-4}$ level, 
are subdominant with respect to the resolution uncertainty.

\subsubsection{High spins: $\chi_1=\chi_2=\pm 0.8$ (Llama data)}
\label{sec:llama08}
\begin{figure}[t]
\begin{center}
\includegraphics[width=0.42\textwidth]{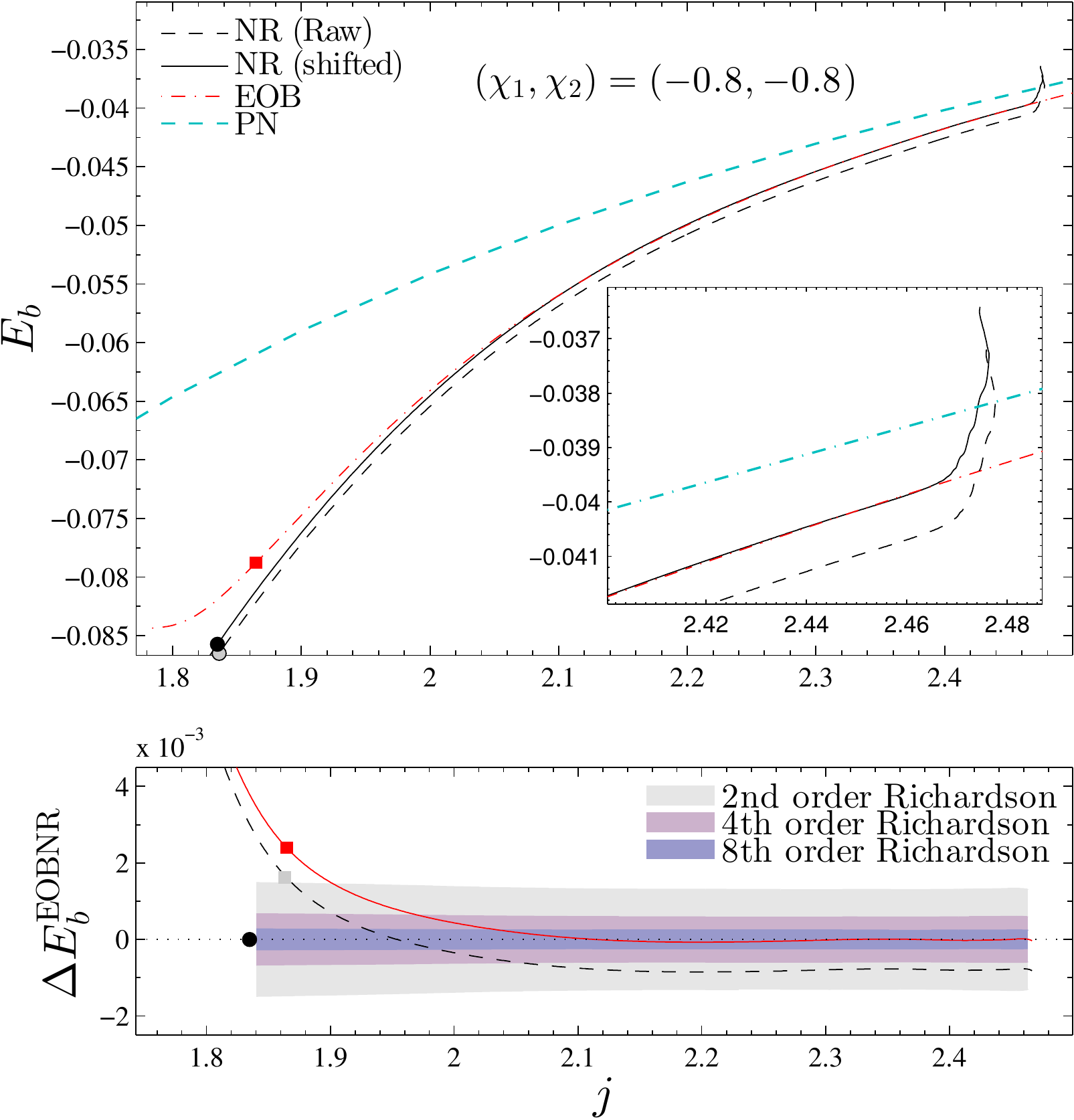}\\
\vspace{5mm}
\includegraphics[width=0.42\textwidth]{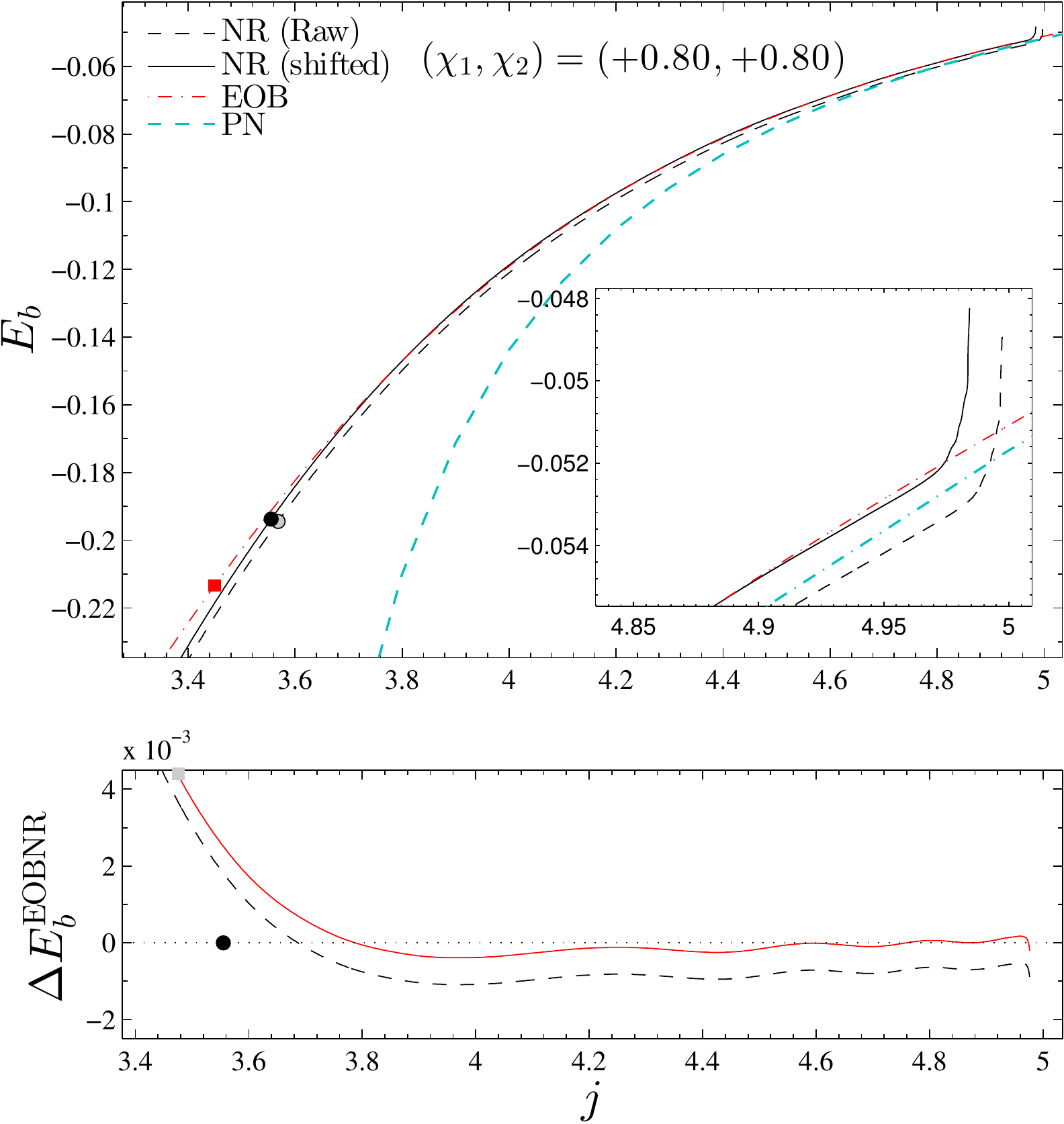}
\caption{\label{fig:Llama_ej_08}Same as Fig.~\ref{fig:Llama_ej} (still for Llama data), 
but for the  cases $\chi=\pm 0.8$. The black curves are obtained after shifting
the raw data (dashed line) with the vector $(\Delta j^0,\Delta E_b^0)$ so as to 
recover a good EOB/NR consistency for large values of $j$. The Taylor-expanded
PN curve is also added for comparison. The insets zoom on the initial (junk-related)
drop in the $E_b(j)$ curve.}
\end{center}
\end{figure}

\begin{figure}[t]
\begin{center}
\includegraphics[width=0.47\textwidth]{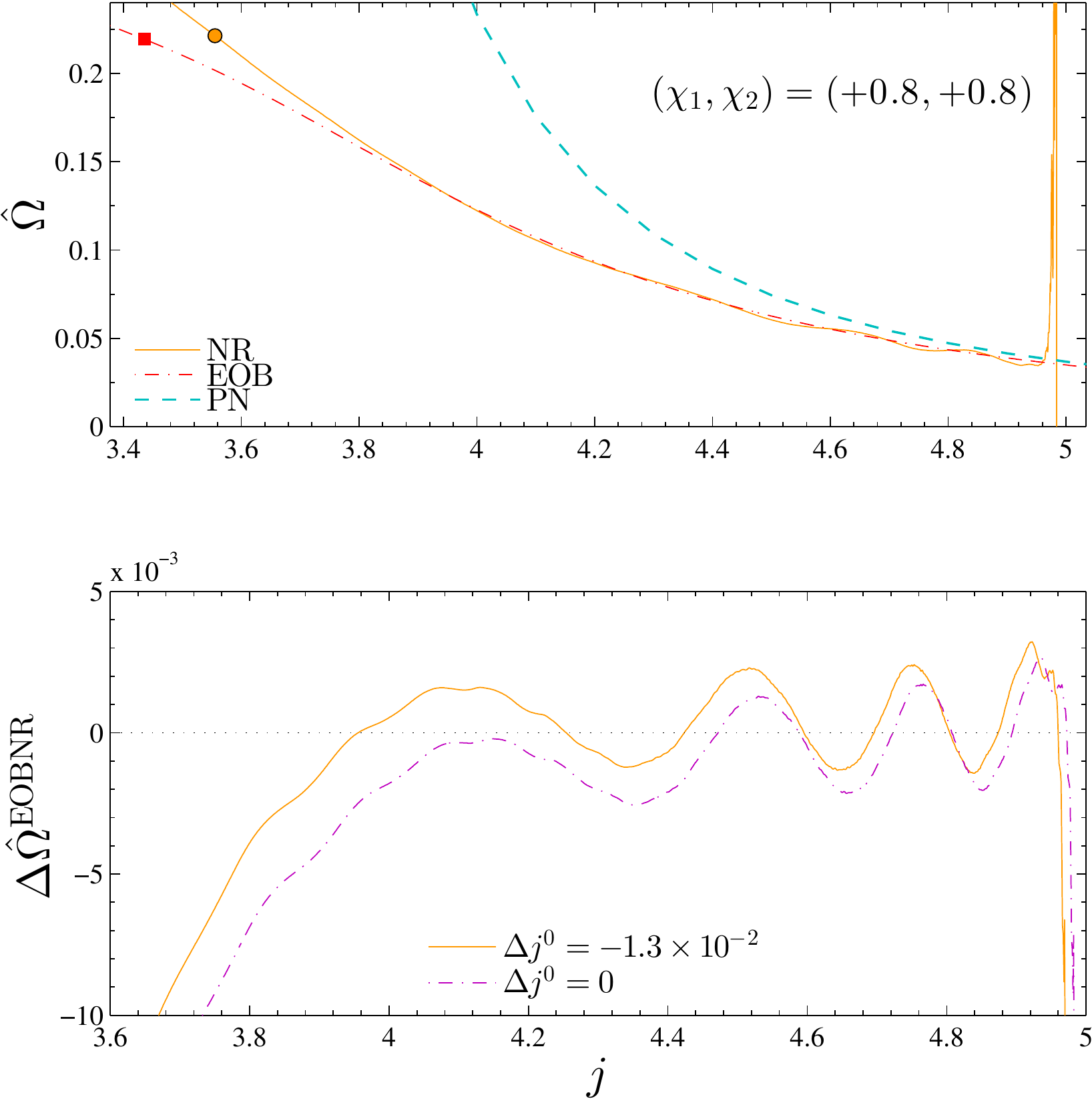}
\caption{\label{fig:Omg08} Comparison between the analytical and numerical dimensionless
orbital frequencies $\hat{\Omega}$ for the (Llama) case $\chi_1=\chi_2=+0.80$ in Fig.~\ref{fig:Llama_ej_08}.
The bottom panel shows how the difference $\Delta\hat{\Omega}^{\rm EOBNR}(j)=\hOmg^{\rm EOB}(j) -\hOmg^{\rm NR}(j)$
can be made to oscillate around zero by a proper choice of the angular momentum shift $\Delta j^0$.}
\end{center}
\end{figure}

The higher spin values $\chi\equiv \chi_1=\chi_2=\pm 0.8$ deserve a separate
discussion. First of all, inspecting the $E_{b}(j)$ curves one sees that, 
contrary to the previous cases, the junk radiation transient is such that
the  ``raw" NR curves (depicted as black, dashed lines in Fig.~\ref{fig:Llama_ej_08})
stand visibly below the EOB prediction (for both signs of spin), see insets of the figure. 
This phenomenon is analogous (though smaller in magnitude) 
to what we found in nonspinning SpEC data. 
As mentioned above, it suggests that the 
NR $E_{b} (j)$ curve should be corrected by an additional shift vector  
$(\Delta j^{0},\Delta E_{b}^{0})$ so to have consistency with the 
EOB prediction at large values of $j$. The phenomenon found here with 
Llama data is probably due to the fact that the higher multipoles ($\ell \geq 4$) 
of the junk-radiation are not resolved with sufficient accuracy 
at the spatial resolution we can afford (see below).
The error bars, computed as above (though only for one configuration) are also
rather large and essentially {\it constant} over the entire $j$-range. Assuming 
the conservative second-order convergence order, one sees that the observed energy 
difference is mostly compatible with the error bar, remaining rather flat up to $j\approx 2$. 
We then apply the same technique discussed above to determine  $(\Delta j^{0},\Delta E_{b}^{0})$,
i.e., we inspect the EOB/NR differences $\Delta \hOmg^{\rm EOBNR}(j)$ and  $\Delta E_{b}^{\rm EOBNR}(j)$ 
and determine the vectorial shift so as to have them as flat and as small as possible
in the first part of the inspiral (see Appendix for a fuller discussion).
We choose here $(\Delta j^{0},\Delta E_{b}^{0})=(-1.3\times 10^{-3},7.8\times 10^{-4})$ 
for $\chi=-0.8$ and $(\Delta j^{0},\Delta E_{b}^{0})=(-1.3\times 10^{-2},7\times 10^{-4})$ 
for $\chi=+0.8$. 
Figure~\ref{fig:Omg08} (that refers to the $\chi=+0.8$ case) illustrates the effect 
that a good choice of $\Delta j^0$ has in centering around a zero averaged value the (eccentricity driven) 
oscillation in $\Delta \hOmg^{\rm EOBNR}(j)$. 
Figure~\ref{fig:Llama_ej_08} (bottom panels) shows how the use of the vectorial shifts
substantially reduces the difference $\Delta E_b^{\rm EOBNR}$, making it compatible with
the 4th-order extrapolation error bar.
 
Let us finally comment on the possible origin of the excessive initial drop in 
the $E_b(j)$ curve (visible in the insets of Fig.~\ref{fig:Llama_ej_08})
yielded by the junk-radiation transient. We think it has to do with the 
under-resolution of the junk-radiation high-multipoles ($4 < \ell\leq 8$)
for the following reason: we noted that the energy loss during the junk-transient
yielded by the modes with $ 5\leq \ell \leq 8$ is 
approximately {\it as large as} that for  $ 2\leq \ell \leq 4$ for $\chi_1=\chi_2=\pm 0.8$;
on the contrary, for the other configurations one always observes a sort of convergence
of the junk radiation losses, with the higher multipoles contributing vertical drops 
that are progressively smaller than those due to the leading order modes. This makes
us suspect that if it were possible to improve the accuracy of the higher-modes
junk radiation one would be able to find the same straightforward (i.e., without shifts) 
EOB/NR consistency as for the case $-0.6\leq \chi \leq +0.6$. A more refined analysis of these issues
(either employing higher-resolution and/or starting the system at larger separations, 
so as to reduce the magnitude of the junk) is left to future work.

For the moment, to gain more confidence in our analytical model, as well as to gauge 
the evidence of systematic uncertainties in the NR data, we turn to explore the 
energetics of spinning binaries computed using SXS data.
\begin{figure*}[t]
\begin{center}
\includegraphics[width=0.32\textwidth]{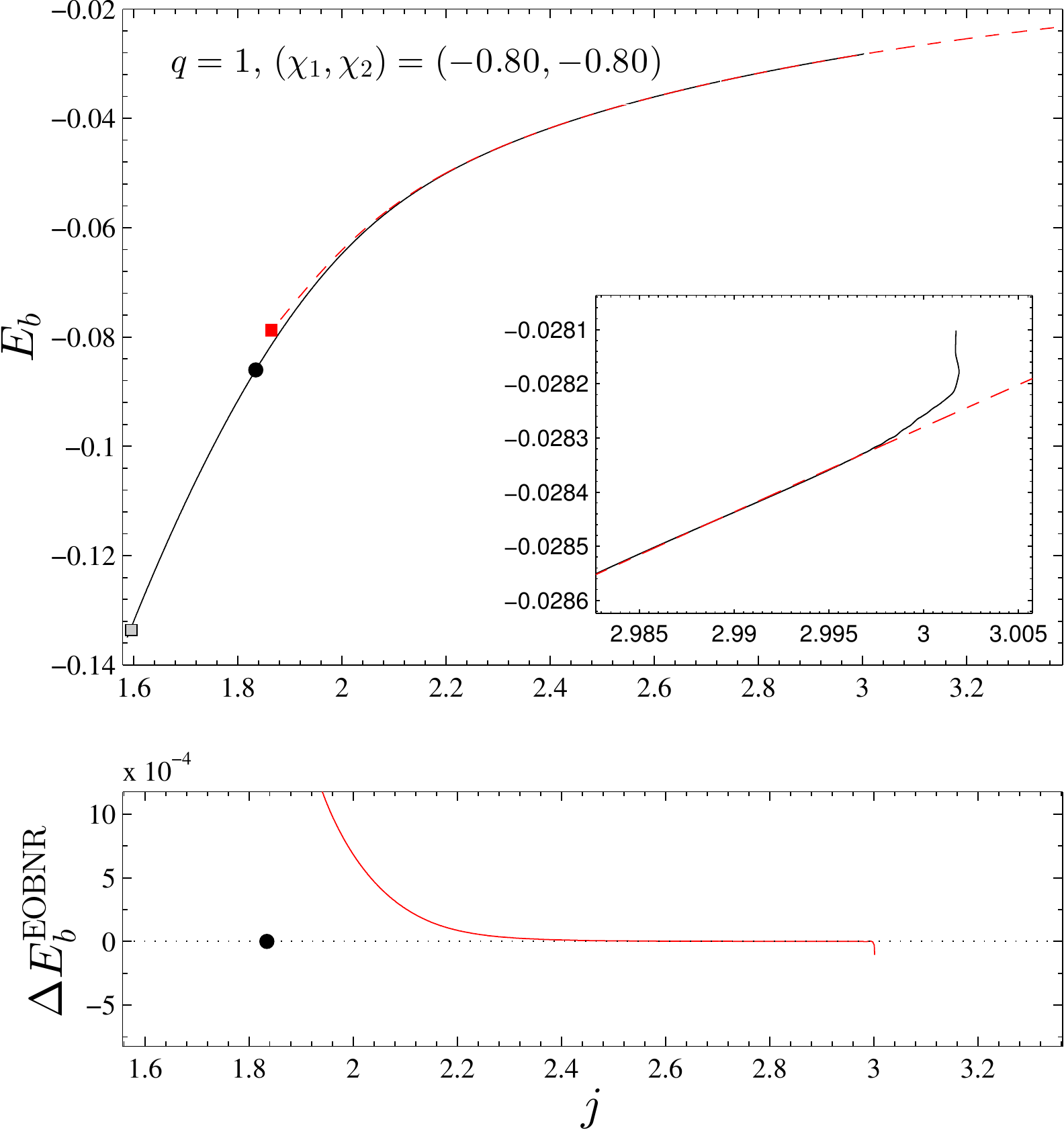}\;\;
\includegraphics[width=0.32\textwidth]{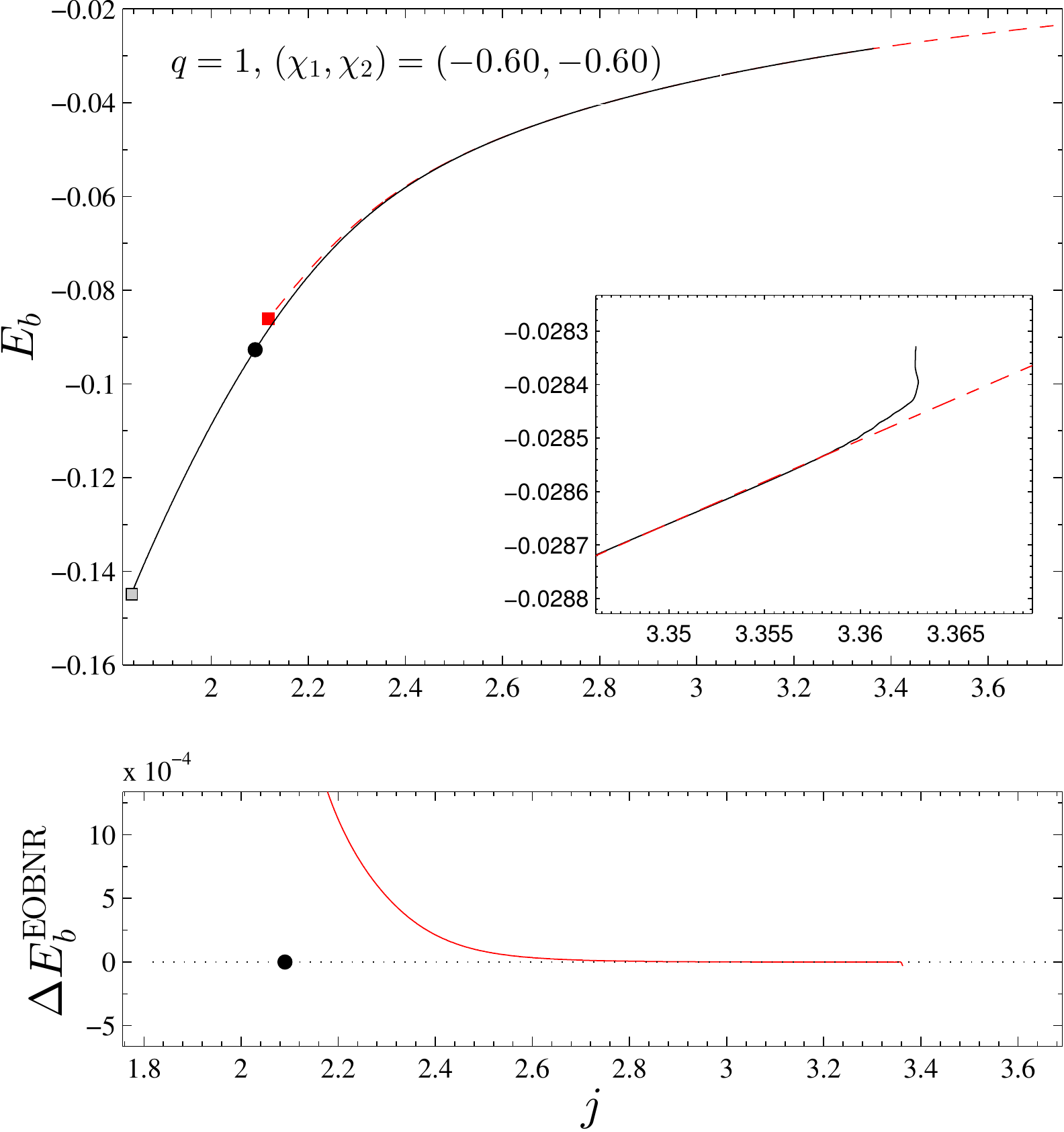}\;\;
\includegraphics[width=0.32\textwidth]{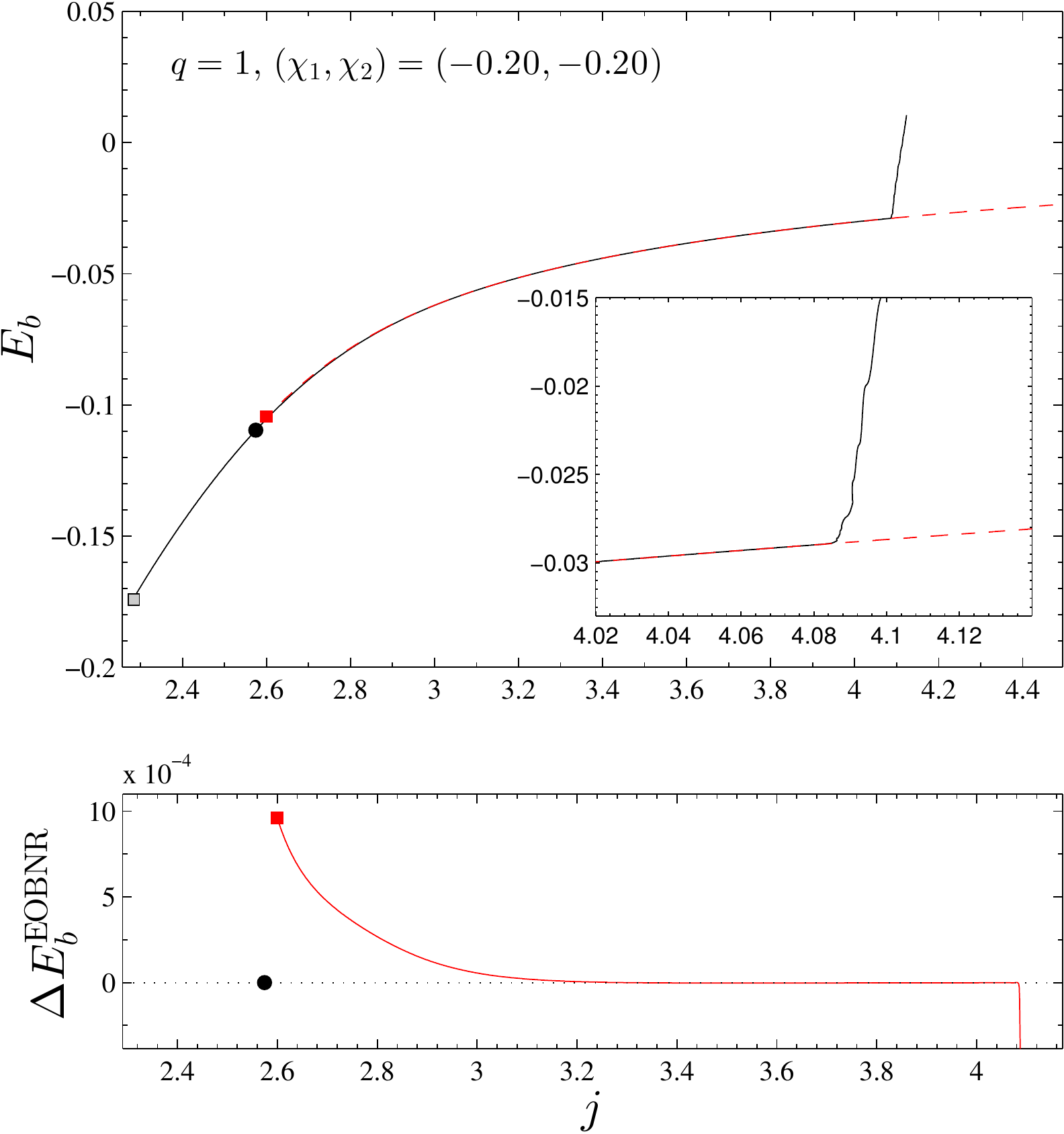}\\
\includegraphics[width=0.32\textwidth]{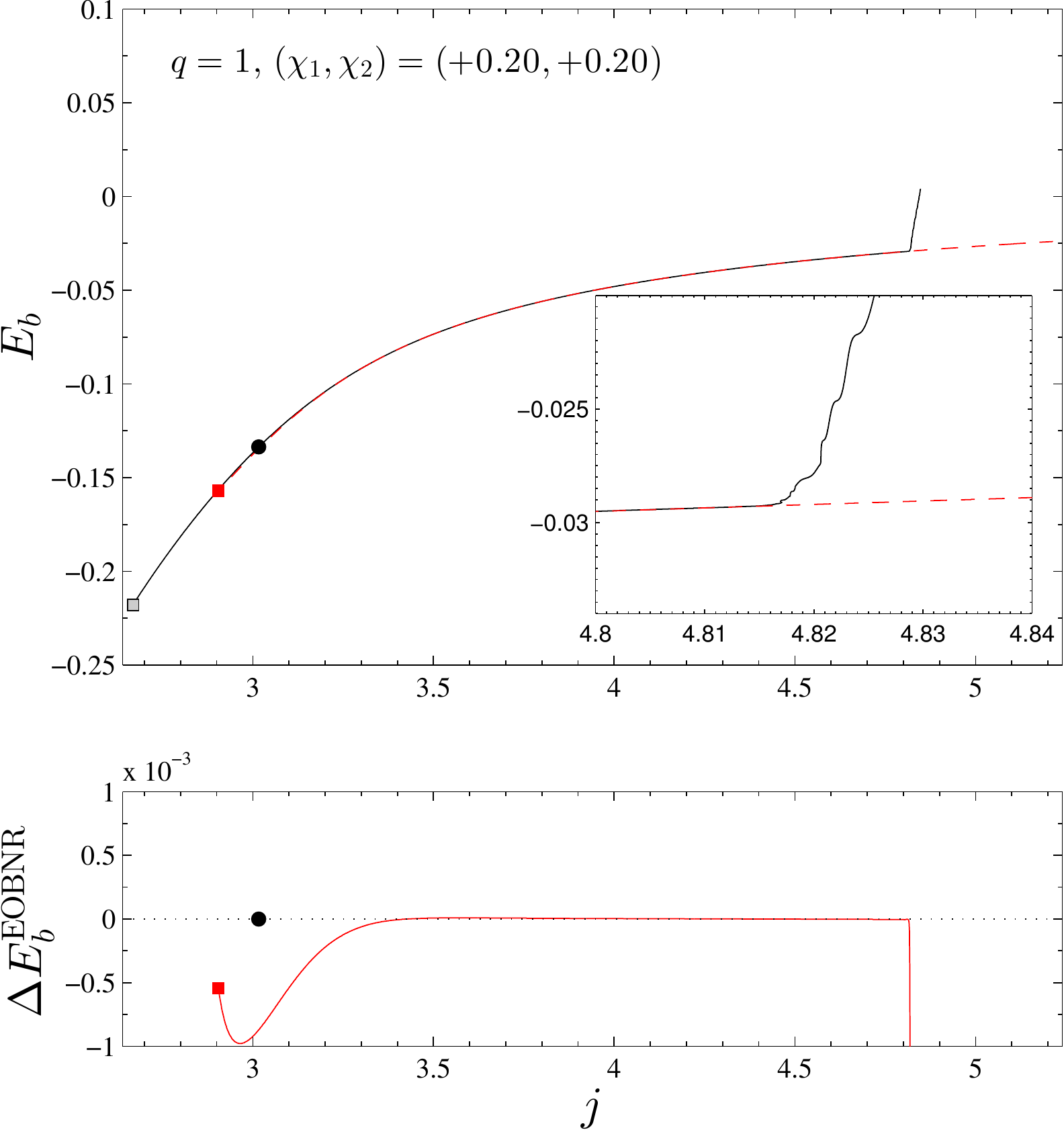}\;\;
\includegraphics[width=0.32\textwidth]{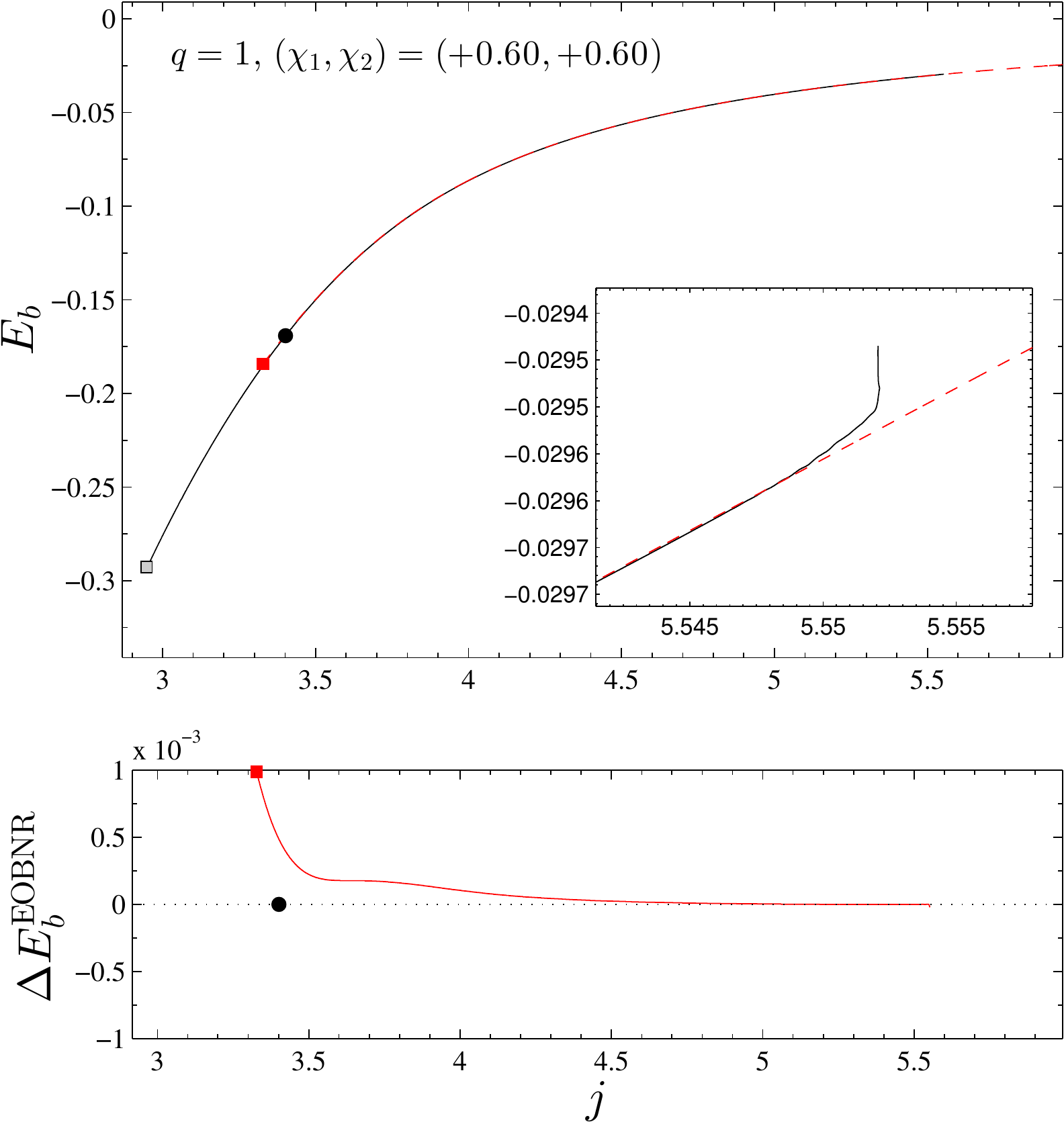}\;\;
\includegraphics[width=0.32\textwidth]{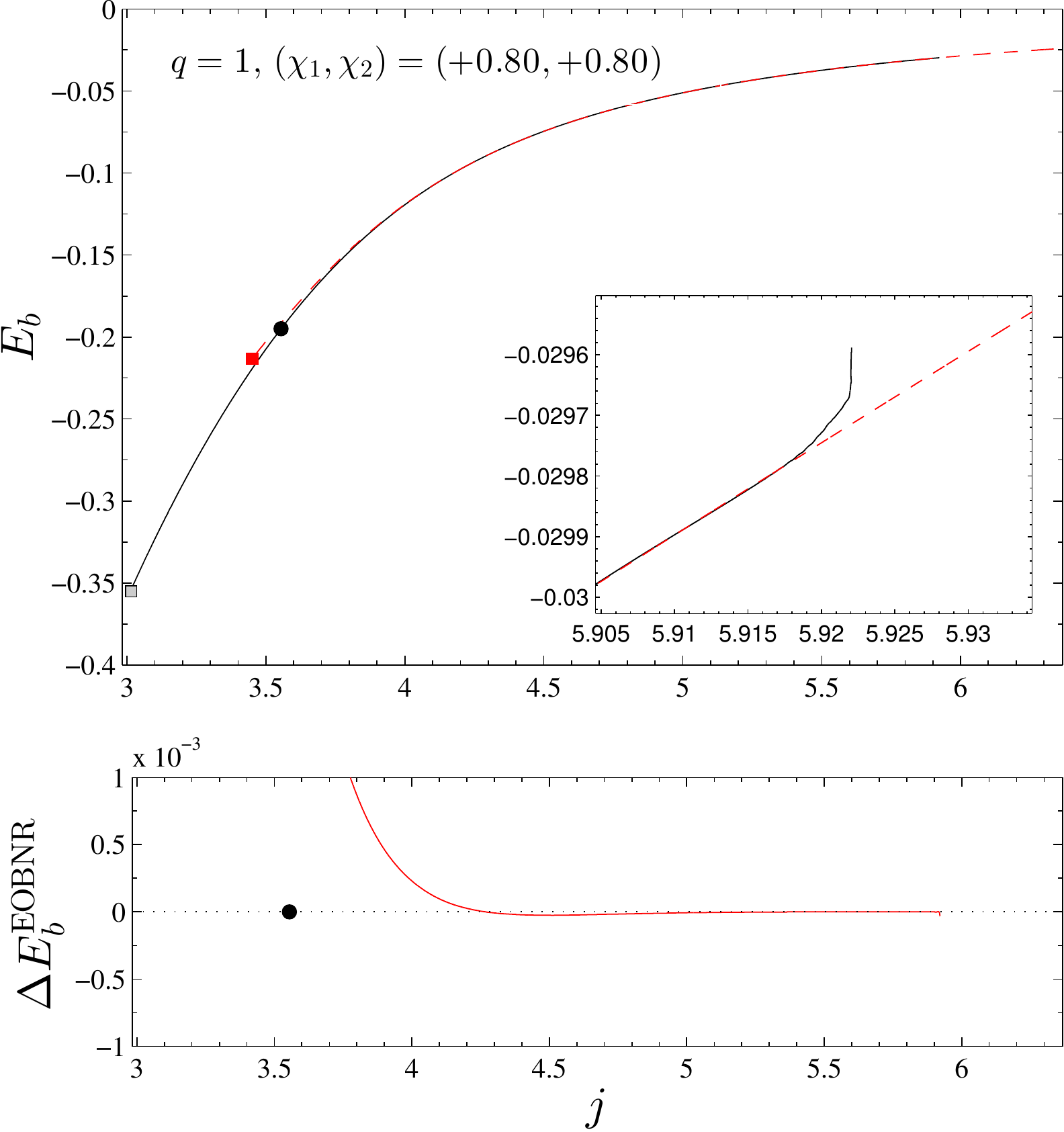}
\caption{\label{fig:sxs_ej}Energy versus angular momentum curves for equal-mass binaries
with spin $\chi$ either aligned or anti-aligned to the orbital angular momentum as obtained
from SXS dataset. This cross checks the above comparison with Llama data.}
\end{center}
\end{figure*}

\subsection{Energetics of SXS spinning data}

\subsubsection{Equal-mass, spinning binaries: $-0.8\leq \chi \leq +0.8$}
One of the crucial outcomes of the Llama/EOB comparison seen above is that, 
even when the separations are relatively small, the NR energetics,  
represented by $E_{b}(j)$,  quickly settles down to the EOB one after the initial junk radiation transient.
This was clearly the case for $-0.6\leq \chi \leq +0.6$. 
For $\chi=\pm 0.8$, we recovered approximately the same behavior after 
properly fixing a suitable vector shift $(\Delta j^{0},\Delta E_{b}^{0})$.
Such a ``convergence'' of the NR $E_{b}(j)$ curve to the EOB one after 
the initial transition, would, a priori, be expected to emerge 
even more clearly from any of the SXS dataset at our disposal, 
since the initial separation between the black holes is much larger,
so that the EOB curve should give an even better approximation to the NR one.
However, in practice, as detailed in the Appendix, the SXS junk radiation 
is often rather large and tends to produce unphysical effects in the $E_{b}(j)$ 
curve. In some cases, the ``raw'' $E_b(j)$ NR curve is shifted in a region 
of the $(j,E_b)$ plane which is clearly incompatible with the EOB prediction
even in the early inspiral (see in particular Fig.~\ref{burst_098} 
in the Appendix). As above, the problem is overcome by introducing a suitable 
shift vector $(\Delta j^{0},\Delta E_{b}^{0})$. More technical details, 
as well as the values of the shifts, are given in the Appendix. 

Figure~\ref{fig:sxs_ej} illustrates the $E_b(j)$ comparison between 
the EOB predictions and the SXS data. The figure displays just the restricted 
sample of SXS configurations that overlaps with the Llama ones considered 
above, i.e. $\chi=(\pm 0.8,\,\pm 0.6,\,\pm 0.2)$. One sees that the result 
is compatible with the previous one, though improved in the following 
aspects: (i) the NR curves extend from larger values of $j$ up to the final
state corresponding to the mass and angular momentum of the final black hole. 
The final state, as read from the {\tt metadata.txt} file in the SXS catalog,
is also indicated by the grey square marker in the plots; (ii) due to the improved 
accuracy of NR data, the differences between 
EOB and NR merger states (indicated by markers in the plots) are visibly smaller
(and more homogeneous) than with Llama data. Compare in particular the case
$\pm 0.2$.
Generally speaking, the EOB and NR curves are seen to progressively separate 
as the spin becomes negative and large. Still, the final merger states are 
closer when the spin is negative than when the spin is positive.
Such differences are typically larger than the numerical error, when 
available (shaded region in the plot): this suggests that the performance of
the EOB model should be improved close to merger when the 
BHs are spinning.

The insets in the panels of Fig.~\ref{fig:sxs_ej} are close-ups of 
the initial junk-radiation-transient parts. Note that the 
magnitude of the junk radiation transient
(and thus of the shift  $(\Delta j^{0},\Delta E_{b}^{0})$ that has to be applied) is very different depending on
the configuration. The cases $\chi=\pm 0.2$ are particularly remarkable. For them, 
rather large vectorial shifts need to be applied to reconcile the
NR curve with both the EOB at low frequencies and the final state.

\begin{figure}[t]
\begin{center}
\includegraphics[width=0.47\textwidth]{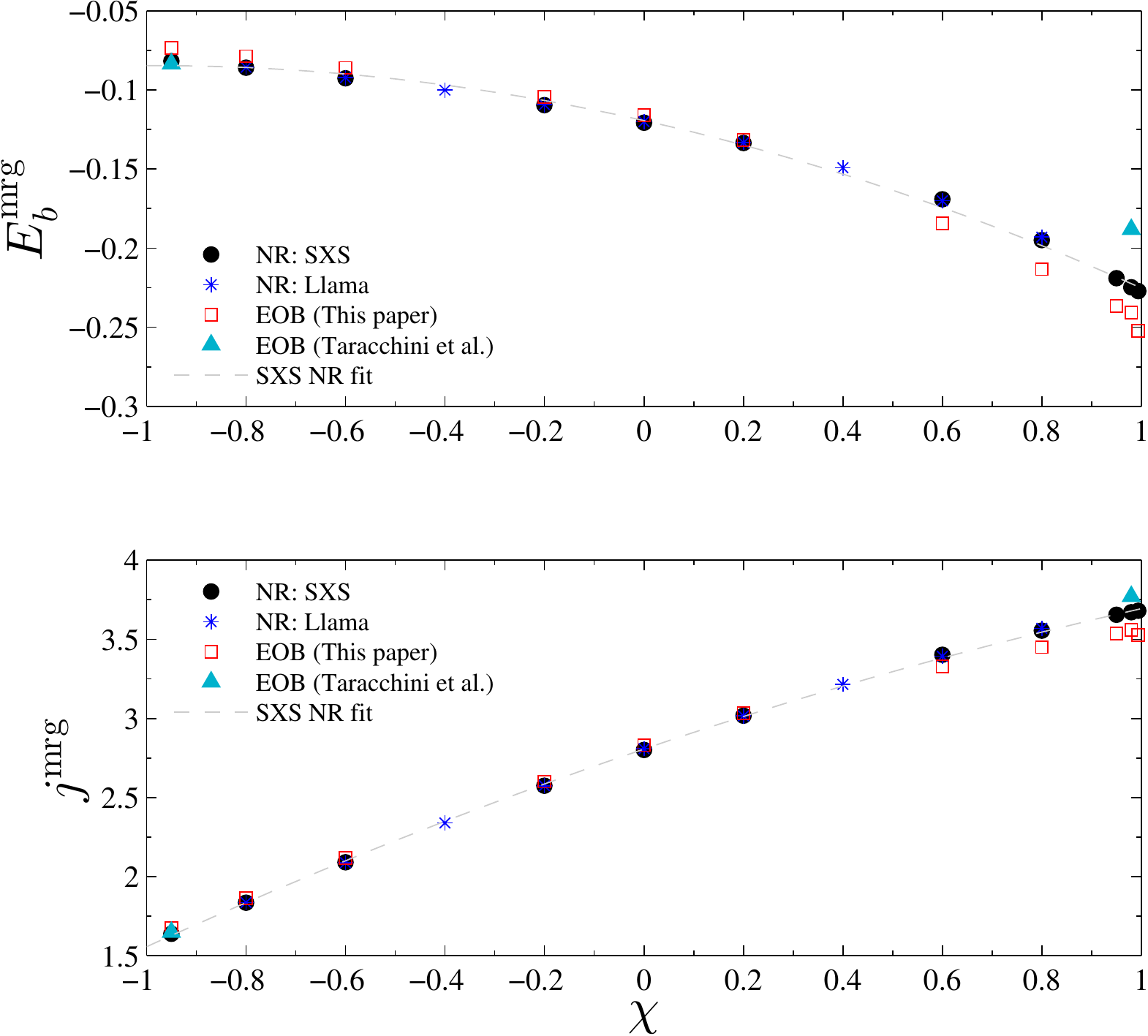}
\caption{\label{fig:ej_mrg_chi}Equal-mass, equal-spin case. Comparison between EOB and 
 NR merger quantities, binding energy (top) and angular momentum (bottom).}
\end{center}
\end{figure}

The difference between Llama and SXS merger states is rather small in absolute value 
($\sim 10^{-4}$), and compatible with the expected level of conservation of energy and 
angular momentum in the Llama code (see Table~2 of Ref.~\cite{Reisswig:2009rx}).
Note that the energy conservation in SXS data is on average more accurate by two orders of magnitude
(see Table~II of Ref.~\cite{Hemberger:2013hsa}).
The compatibility among NR merger data is highlighted in Fig.~\ref{fig:ej_mrg_chi}:
the figure plots the functions $E_b^{\rm mrg}(\chi)$ and $j^{\rm mrg}(\chi)$ 
(including also extremal spin values, see below): Llama and SXS data are 
hardly  distinguishable on this scale. The SXS NR merger states 
$E_b^{\rm mrg}(\chi)$ and $j^{\rm mrg}(\chi)$ shown in the picture
can be accurately fitted with quadratic polynomials in $\chi$,
\begin{align}  
E_{b,\rm SXS}^{\rm mrg}(\chi)&=-0.035546\chi^2 - 0.070311\chi - 0.119444,\\
j^{\rm mrg}_{\rm SXS}(\chi)&=-0.180948\chi^2 + 1.068176\chi + 2.805593.
\end{align}
The figure also summarizes  the performance of the EOB 
model at merger (the corresponding points are indicated
by red, empty, squares), highlighting that the EOB/NR agreement
between merger states, which is best when $\chi=0$, slightly worsens
when $\chi<0$ and worsens more markably when $\chi>0$, and especially
when  $\chi\to 1$. Such a disagreement calls 
for improvements in the analytical  EOB model that will be undertaken 
in future studies. We recall however that, despite such discrepancies 
in the energetics, the {\it phasing} provided by the EOB model discussed 
here is in excellent agreement with NR data, as illustrated by the 
dephasing and unfaithfulness data of Table~\ref{tab:configs}.

\subsubsection{Extremal spins and comparison with the results of Taracchini et al.~\cite{Taracchini:2013rva}}
%

\begin{figure}[t]
\begin{center}
\includegraphics[width=0.42\textwidth]{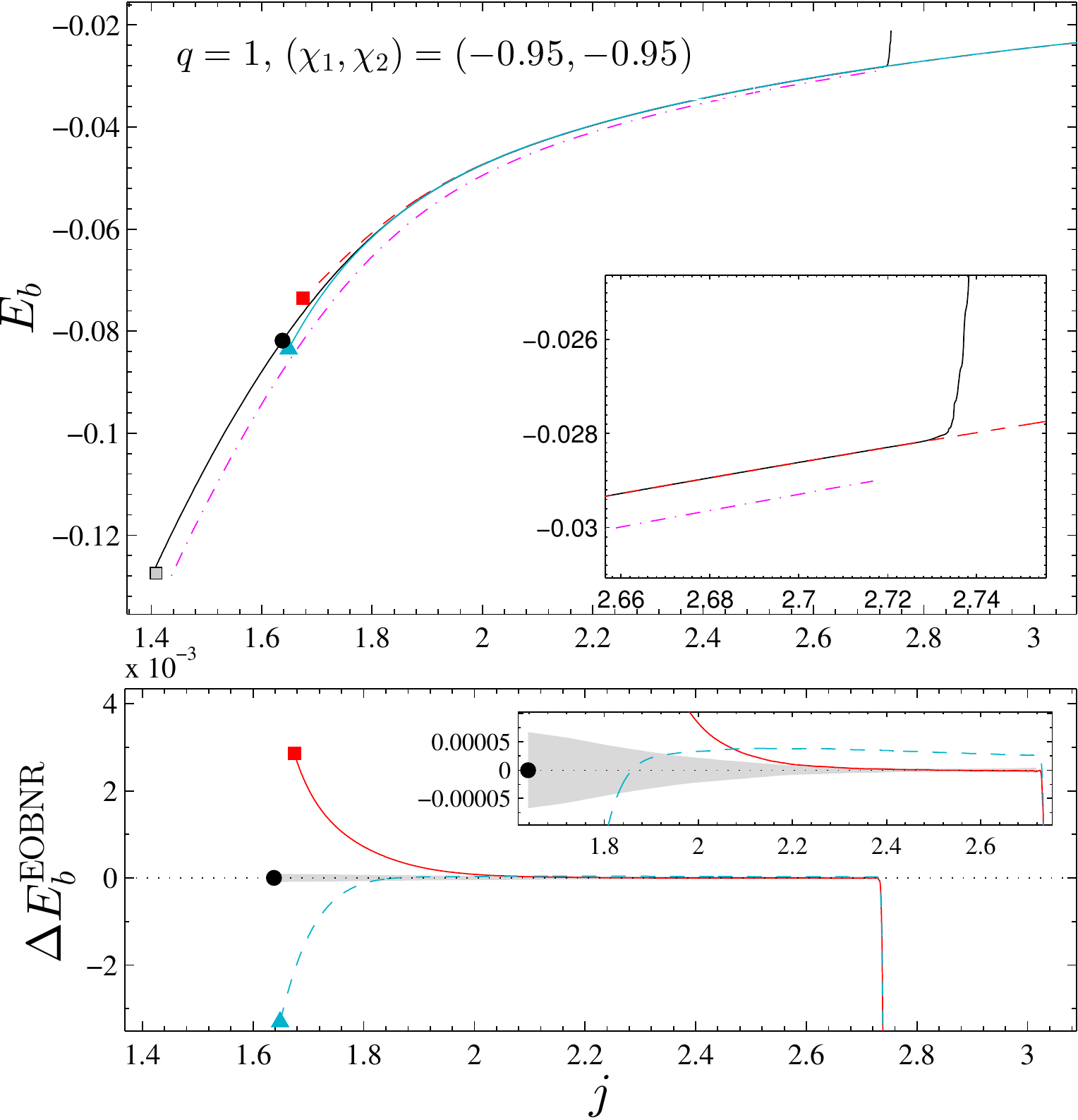}\\
\vspace{5mm}
\includegraphics[width=0.42\textwidth]{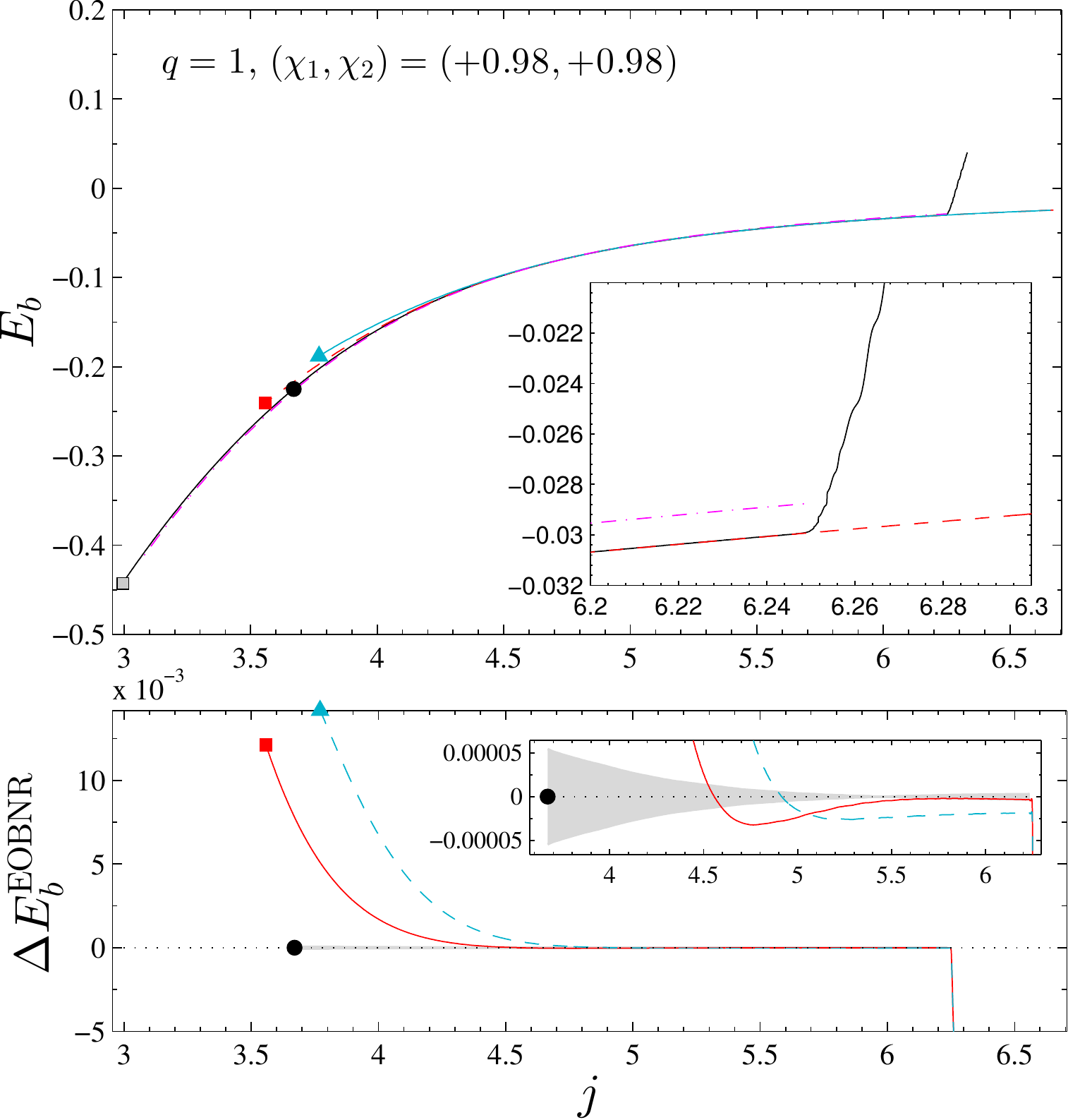}
\caption{\label{ej:sxs_ej_extramal} Energetics in the extremally spinning case: comparison with the EOB and NR 
data of Ref.~\cite{Taracchini:2013rva}. Markers indicate the merger for each dataset (except
the NR data of Ref.~\cite{Taracchini:2013rva}): NR (red); EOB model of Ref.~\cite{Taracchini:2013rva}
(lgiht-blue); our EOB model (red). The shaded area gives an estimate of the numerical uncertainty. 
Both EOB models are consistent among themselves and with the NR curves computed in this paper. 
For $\chi=+0.98$ our EOB model is closer to NR than the EOB model of~\cite{Taracchini:2013rva}, 
while the reverse is true for  $\chi=-0.95$.
}
\end{center}
\end{figure}

\begin{figure}[t]
\begin{center}
\includegraphics[width=0.47\textwidth]{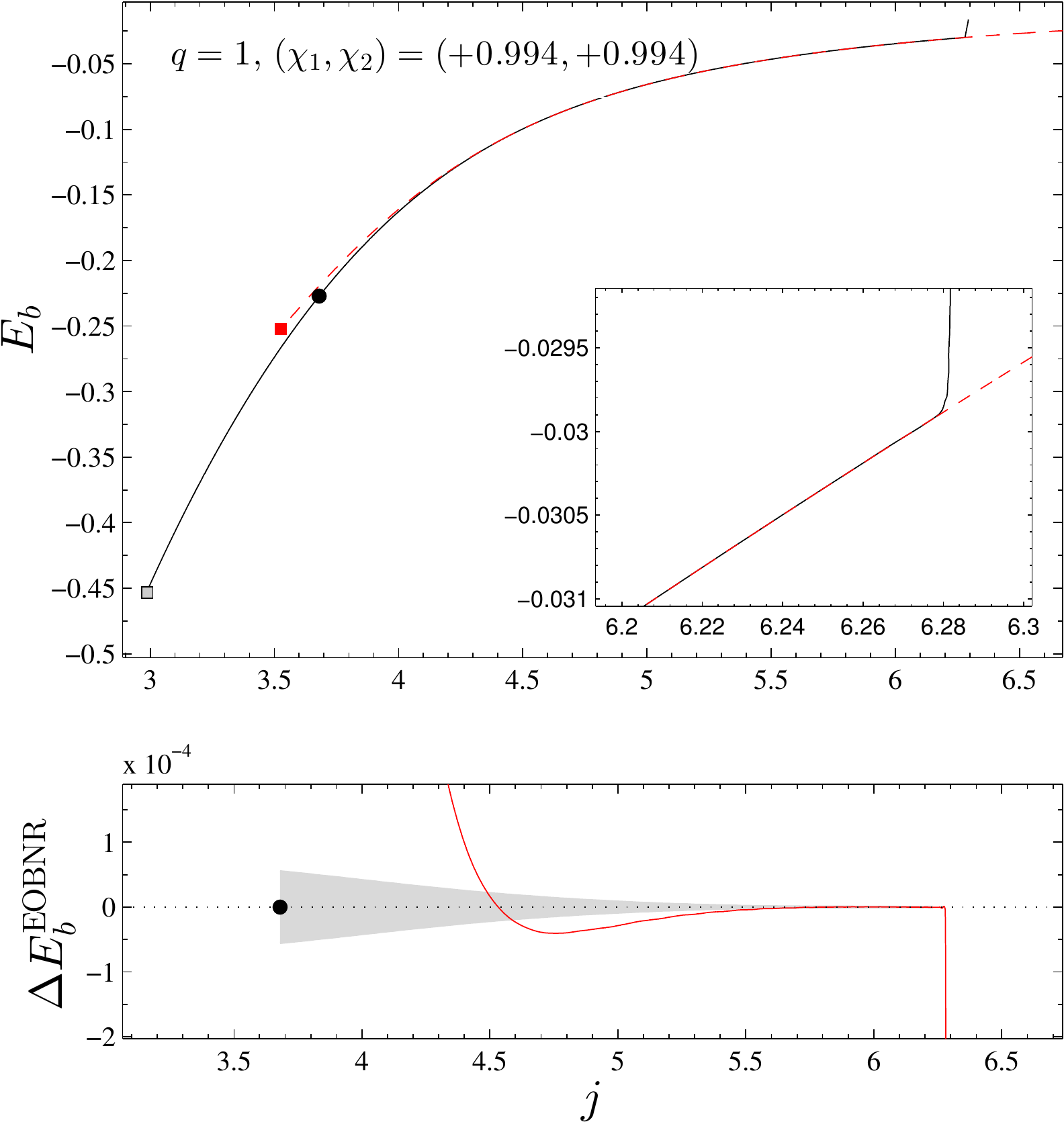}
\caption{\label{ej0994} EOB/NR comparison for $q=1$, $\chi_1=\chi_2=+0.994$ (bottom), 
the highest spin simulated so far. The NR uncertainty (shaded region) is 
obtained by taking the difference between the data at the two highest 
resolutions after determining the corresponding vectorial shifts $(\Delta j^0,\Delta E_b^0)$.}
\end{center}
\end{figure}

Let us finally analyze in detail some (equal-mass, equal-spins) 
quasi-extremally spinning binaries that are present in the 
SXS catalog, $\chi = -0.95$, $\chi=+0.98$ and $\chi=+0.994$,
the highest spin value simulated so far.
We recall that before the SXS catalog became
public, data with $\chi=-0.95$ and $\chi=+0.98$ were used by Taracchini 
et al.~\cite{Taracchini:2013rva} to  test the phasing and 
energetics of their EOBNR model. We note that the EOB model of 
Ref.~\cite{Taracchini:2013rva} differs in many aspects from 
the one we are presenting here. Since we have presented a new 
computation  of $E_b(j)$  taking advantage  of an improved 
understanding of various subtleties,
we think it is pedagogically useful to compare and contrast our new NR-based 
results with both the EOB and the NR curves of Ref.~\cite{Taracchini:2013rva}.
The latter data were kindly given to us by A.~Taracchini and A.~Buonanno. 
The result of our comparison is illustrated in Fig.~\ref{ej:sxs_ej_extramal}.
Each top panel of Fig.~\ref{ej:sxs_ej_extramal} contrasts four curves: 
(i) our newly computed NR curve, using a suitable vectorial shift, (black); 
(ii) our new EOB curve (red, dashed); (iii) the EOB curve of 
Ref.~\cite{Taracchini:2013rva} (light blue); 
(iv) the NR curve of Ref.~\cite{Taracchini:2013rva} (dot-dashed, magenta).
In addition, in the bottom panel of Fig.~\ref{ej:sxs_ej_extramal}, we show
the corresponding differences together with the NR numerical uncertainties,
depicted as a shaded grey region. Such NR uncertainties were computed 
by the following procedure: (i) for the two highest resolutions available,
we determined the corresponding $E_b(j)$ curves and the related vectorial
shifts $(\Delta j_0,\Delta E_b^0)$; (ii) we took the difference between the
so computed highest and second highest resolution data, $\Delta E_b^{\rm NRNR}$; 
(iii) the shaded region corresponds to $\pm \Delta E_b^{\rm NRNR}$.
The uncertainty we obtain in this way looks rather small: it is $\sim 10^{-6}$
for large values of $j$, to grow up to a mere $\sim 10^{-4}$ towards merger. 

A careful inspection of this figure tells us that:
\begin{itemize}
\item[(i)] The energetics yielded by our EOB model is consistent with 
extremally spinning NR data, and exhibits the continuation of the trend 
we found above for  $\vert \chi \vert \leq 0.8$ .
For both spins, our EOB curves are slightly {\it above} the NR ones close to EOB merger.
\item[(ii)] Regarding the comparison with the EOB model of Ref.~\cite{Taracchini:2013rva} we note
first that the two EOB's are perfectly consistent among themselves (and with the NR data)
during most of the early inspiral. The two EOB models start differing among themselves 
(and with NR) when getting close to merger. For $\chi=-0.95$ the EOB model 
of Ref.~\cite{Taracchini:2013rva} gives a better approximation to the final state 
than our model; on the contrary, for $\chi=+0.98$ our model performs quantitatively 
better near merger. Note that, in spite of these differences in energetics, the 
waveforms delivered by both EOB models agree  well with NR data 
within the NR uncertainty~\cite{Taracchini:2013rva}.
\item[(iii)]
On the same plot, we superpose the NR $E_b(j)$ curves presented in Ref.~\cite{Taracchini:2013rva}
(dashed lines; red online).
The curves exhibit significant differences with both EOB models already in the early inspiral.
Moreover their end point (after merger and ringdown) differs from the actual final mass and 
angular momentum of the BH (as extracted from the SXS file {\tt metadata.txt} and displayed
in the figure).  By contrast we note that  we determined the additional, needed, final shifts 
taking care that this last point (computed with high accuracy) approximately ($~10^{-4}$ level) 
coincides with the final point of the $E_b(j)$ curve. No details were given in  
Ref.~\cite{Taracchini:2013rva} on how their $E_b(j)$ curve was computed. 
\item[(iv)]
As a side remark, we note that the red and light-blue curves in each insets of 
Fig.~\ref{ej:sxs_ej_extramal} differ at the $\sim 10^{-5}$ level. Since both
curves were obtained subtracting the corresponding EOB to the same NR curve, 
this $10^{-5}$ level essentially quantifies the analytical difference, in 
the early inspiral, between the conservative dynamics of our EOB model and
that of SEOBNRv2.

\end{itemize}

To conclude, Fig.~\ref{ej0994} illustrates the EOB/NR comparison for the highest
value of the spin available, $\chi=+0.994$. The numerical uncertainty,
computed as outlined above, is again rather small, as it does not exceed the $10^{-4}$ 
level towards merger. The performance of the EOB in this case is essentially analogous
to the $\chi=+0.98$ case illustrated above, though with a slightly larger displacement
between the EOB and NR merger states. 

\subsubsection{Unequal-mass, spinning binaries}

We conclude this survey of EOB/NR energetics comparison by also discussing 
explicitly a few representative configurations with both unequal masses 
and unequal spins.
\begin{figure*}[t]
\begin{center}
\includegraphics[width=0.32\textwidth]{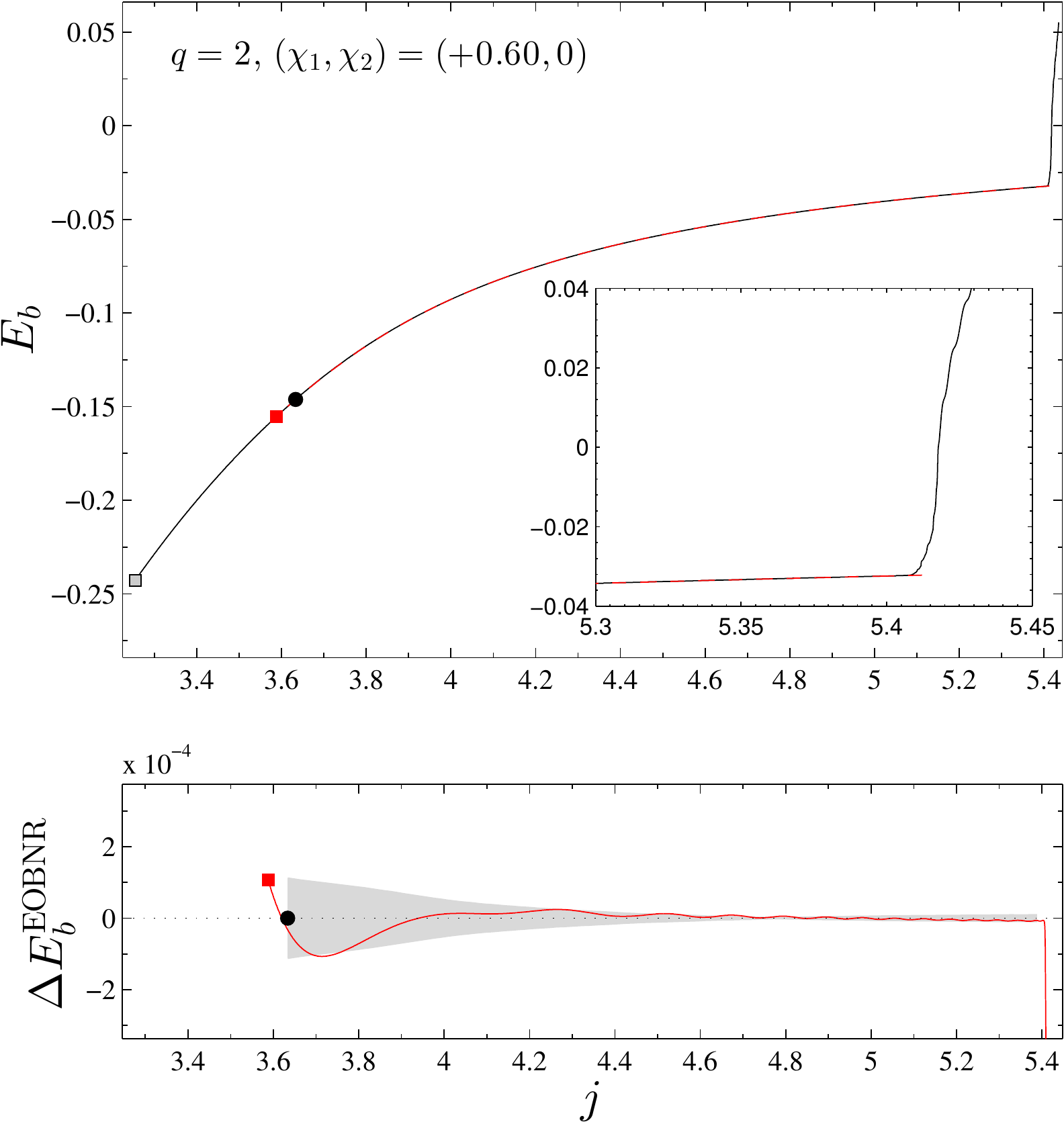}\;
\includegraphics[width=0.32\textwidth]{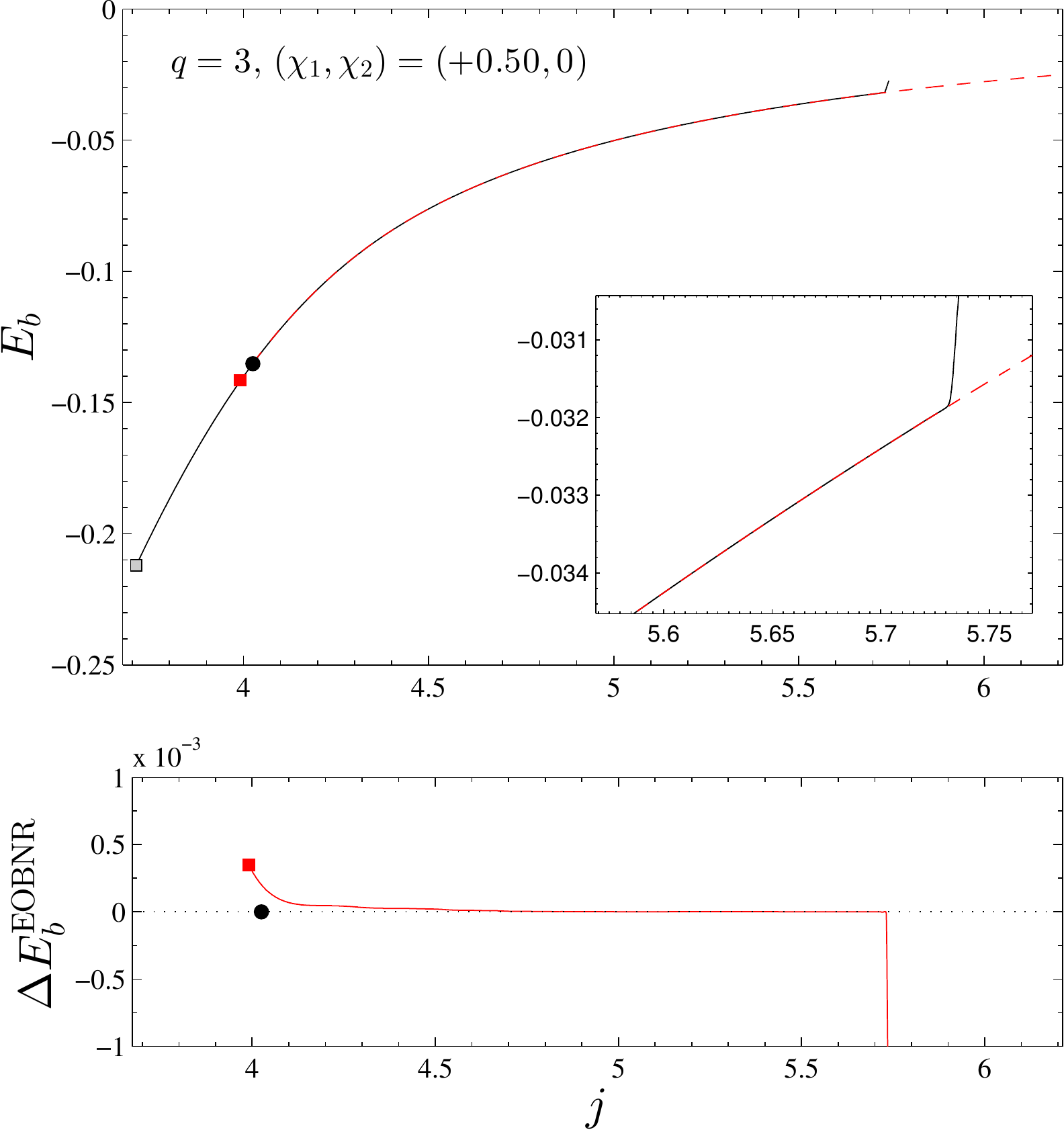}\;
\includegraphics[width=0.32\textwidth]{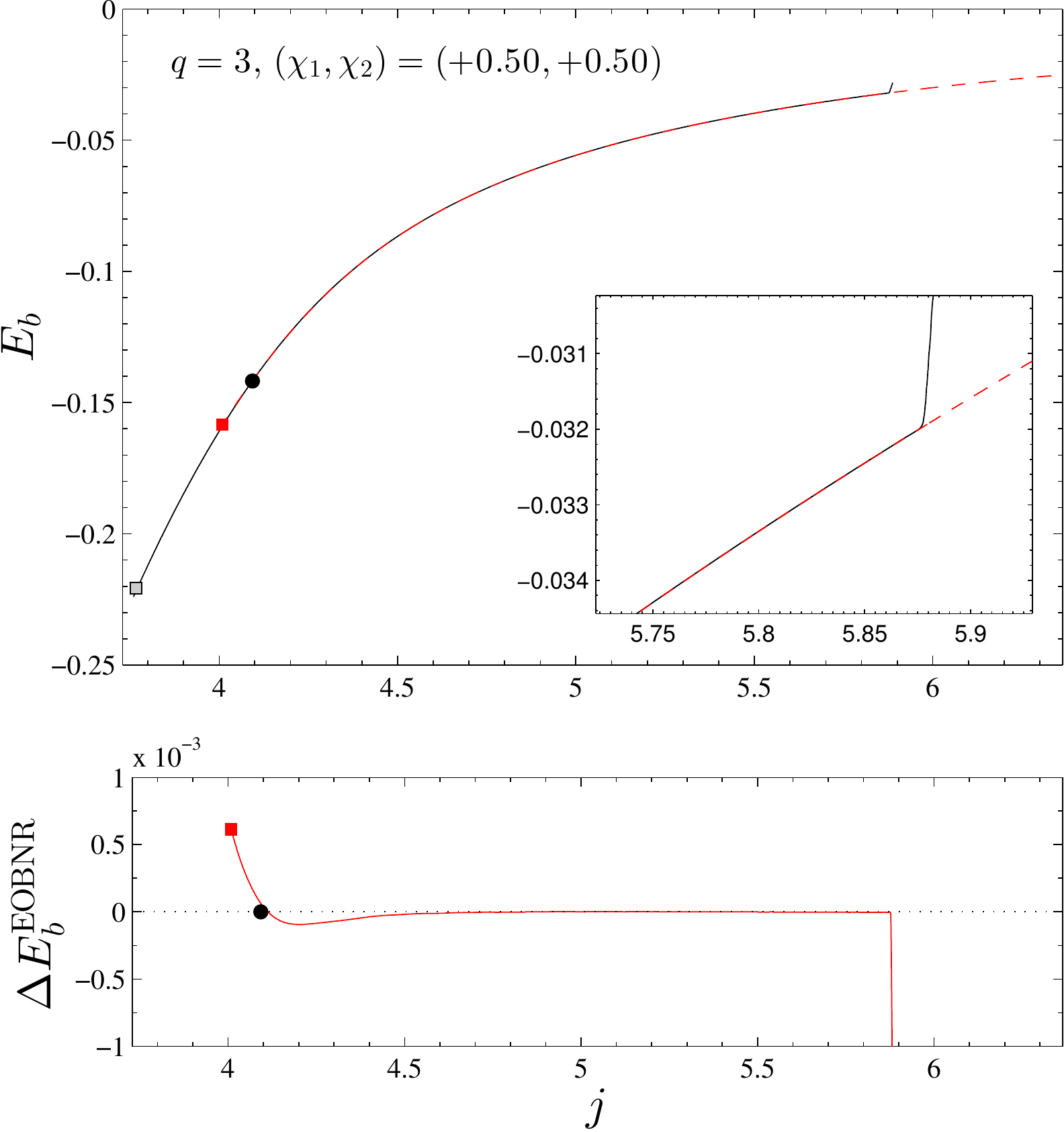}\\
\includegraphics[width=0.32\textwidth]{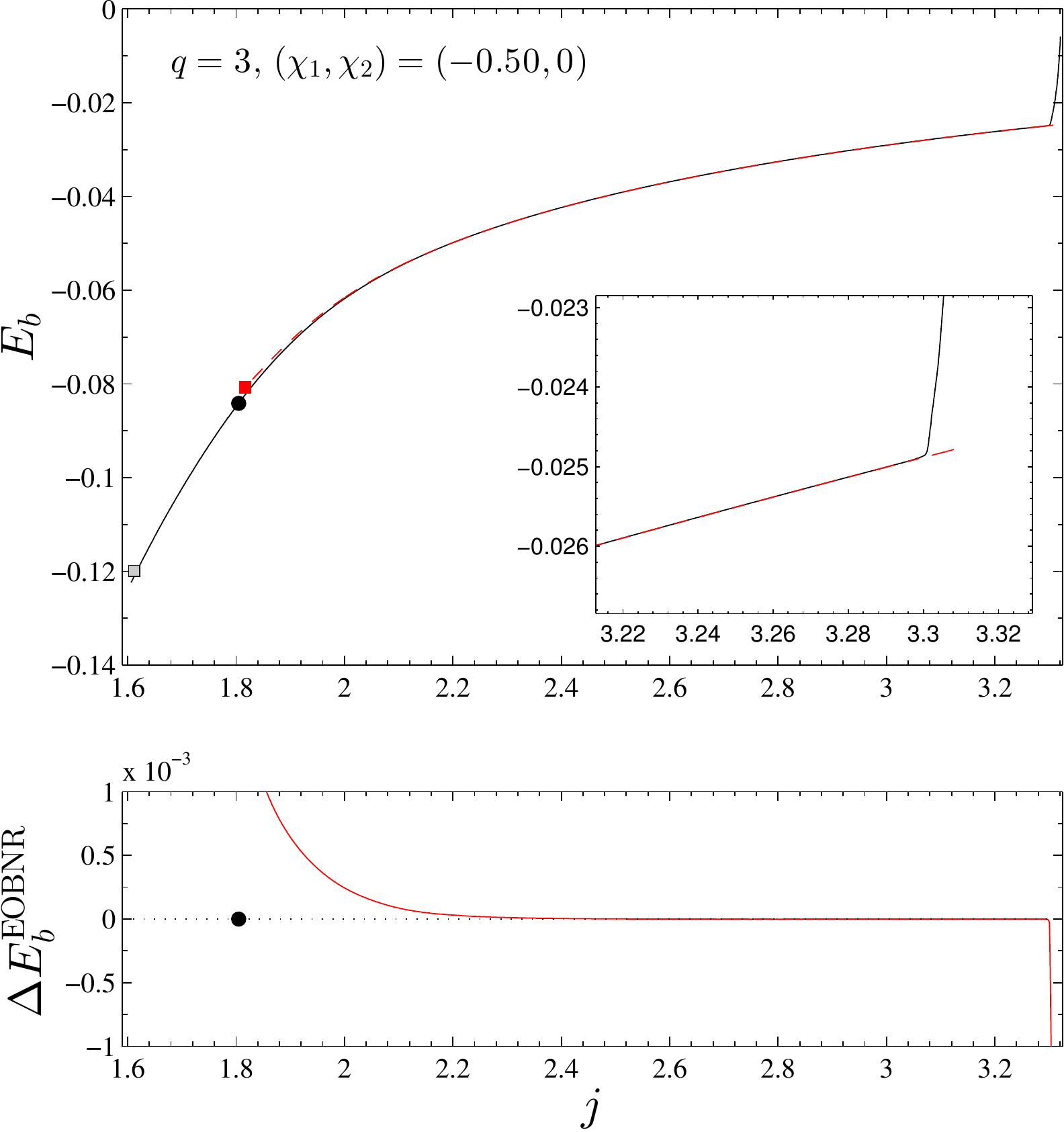}\;
\includegraphics[width=0.32\textwidth]{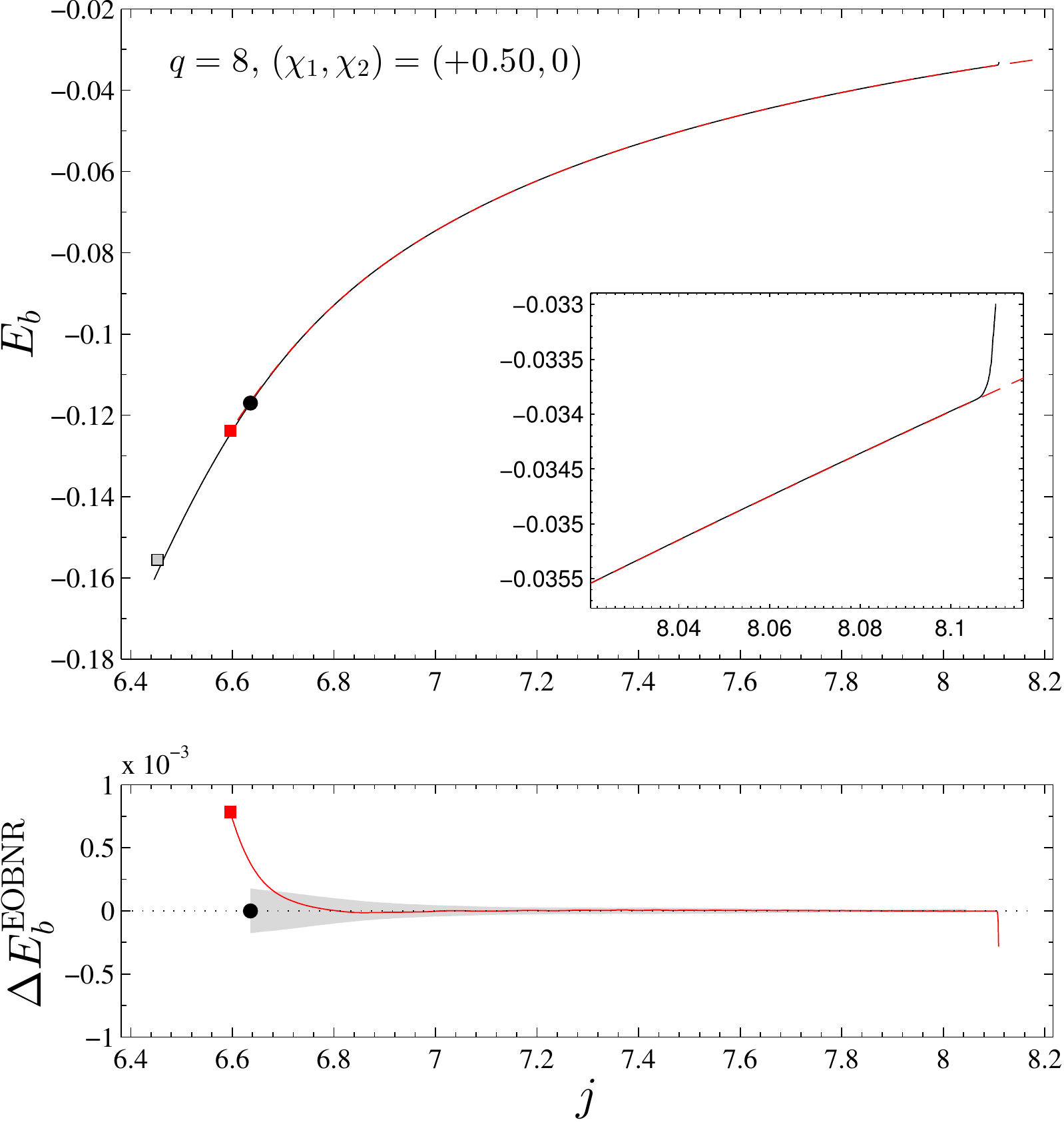}\;
\includegraphics[width=0.32\textwidth]{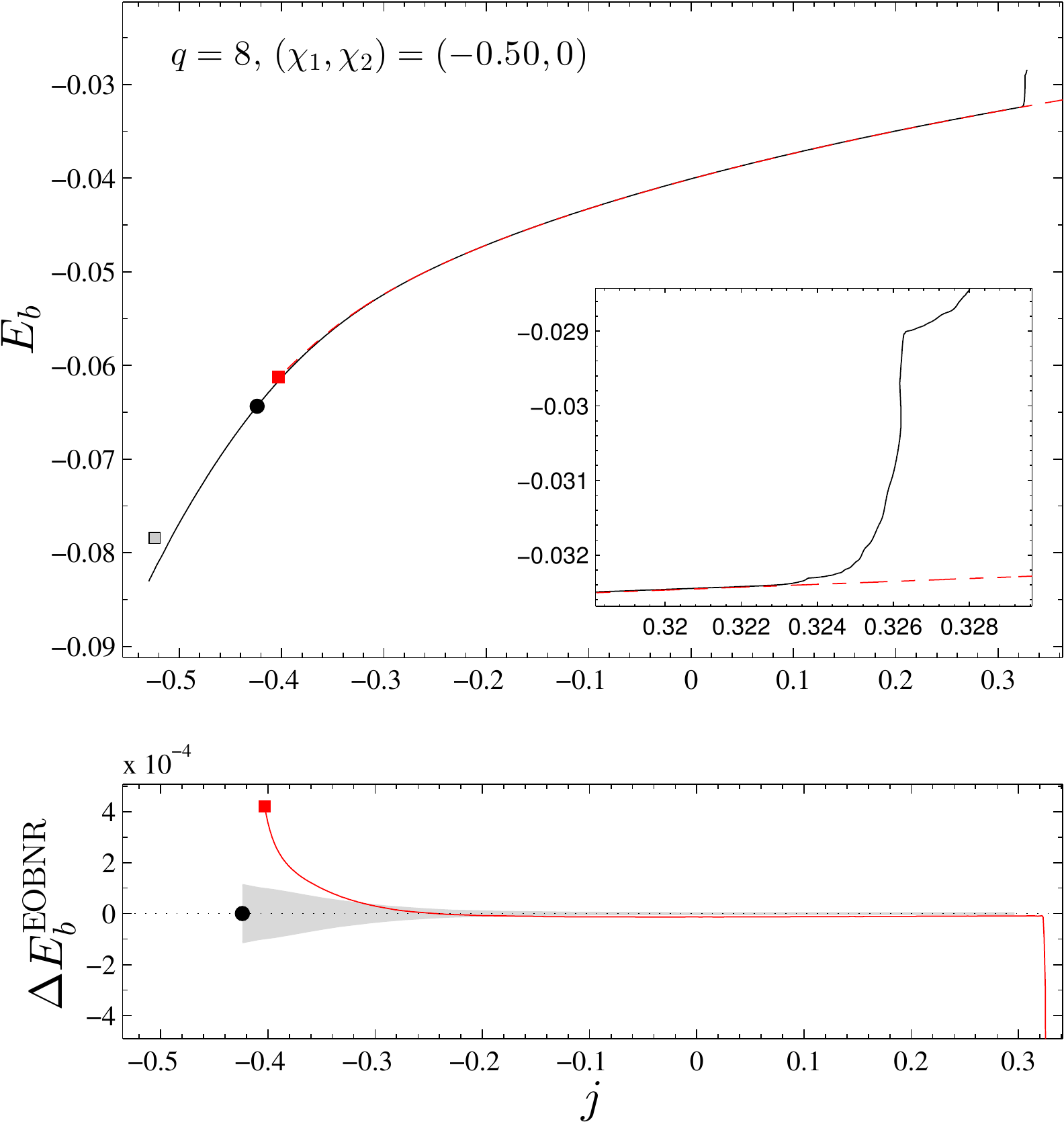}
\caption{\label{ej:unequal_mass}Energetics comparison for a selected sample of 
unequal-mass, unequal-spin binaries. The EOB/NR compatibility is essentially 
comparable to the equal-mass results shown in 
Figs.~\ref{fig:sxs_ej},~\ref{ej:sxs_ej_extramal}, \ref{ej0994}. }
\end{center}
\end{figure*}
The comparison is illustrated in Fig.~\ref{ej:unequal_mass}, which collects six
representative configurations, notably the high-mass-ratio ones $(8,\pm0.50,0)$.
We show here explicitly those cases where only one of the two black holes is spinning,
since in these cases we have enough resolutions to compute error bars. The two cases
$(3,-0.50,-0.50)$ and $(3,+0.50,+0.50)$ yield equivalent results.
The numerical uncertainties, computed as in the previous section, are represented
by light-gray shaded areas: notably, these error bars are compatible with the
extremally spinning cases discussed above, i.e. about $2\times 10^{-4}$ towards
merger (we expect to get similar results also for the other configurations).
Generally speaking, we see here the same features we were finding in the equal-mass,
equal-spin cases shown in Figs.~\ref{fig:sxs_ej},~\ref{ej:sxs_ej_extramal} and~\ref{ej0994}.
In particular, one sees that: (i) when the BH spin is positive, the EOB model
predicts values of the merger states that are slightly smaller than what they should
be. This occurs independently of the mass ratio. Similarly, (ii), when the BH spin
is negative, the EOB predicts values that are slightly, but significantly, higher than the NR prediction.
If both black holes are spinning, these effects remain qualitatively the same,
though quantitatively are slightly magnified.
In other words, the $E_b(j)$ analysis seems to indicate that the current EOB model
is slightly overestimating the magnitude of the spin-orbit interaction. Note however 
that the $E_{b}(j)$ diagnostics is here magnifying a difference that corresponds
to a very small part (near merger) of the time-domain waveform. 
Let us only note here that this shows the usefulness of the $E_{b}(j)$ diagnostics 
for understanding subtle physical effects in the spin-orbit interaction that cannot 
be appreciated by just inspecting the phasing.
 
\section{Preliminary comparison of the two existing spinning EOB models}
\label{sec:analytics}

Let us finally highlight some differences between the analytic structure of
our EOB model and that of Ref.~\cite{Taracchini:2013rva} by comparing some
crucial quantities that enter the construction of the models.

\begin{figure}[t]
\begin{center}
\includegraphics[width=0.47\textwidth]{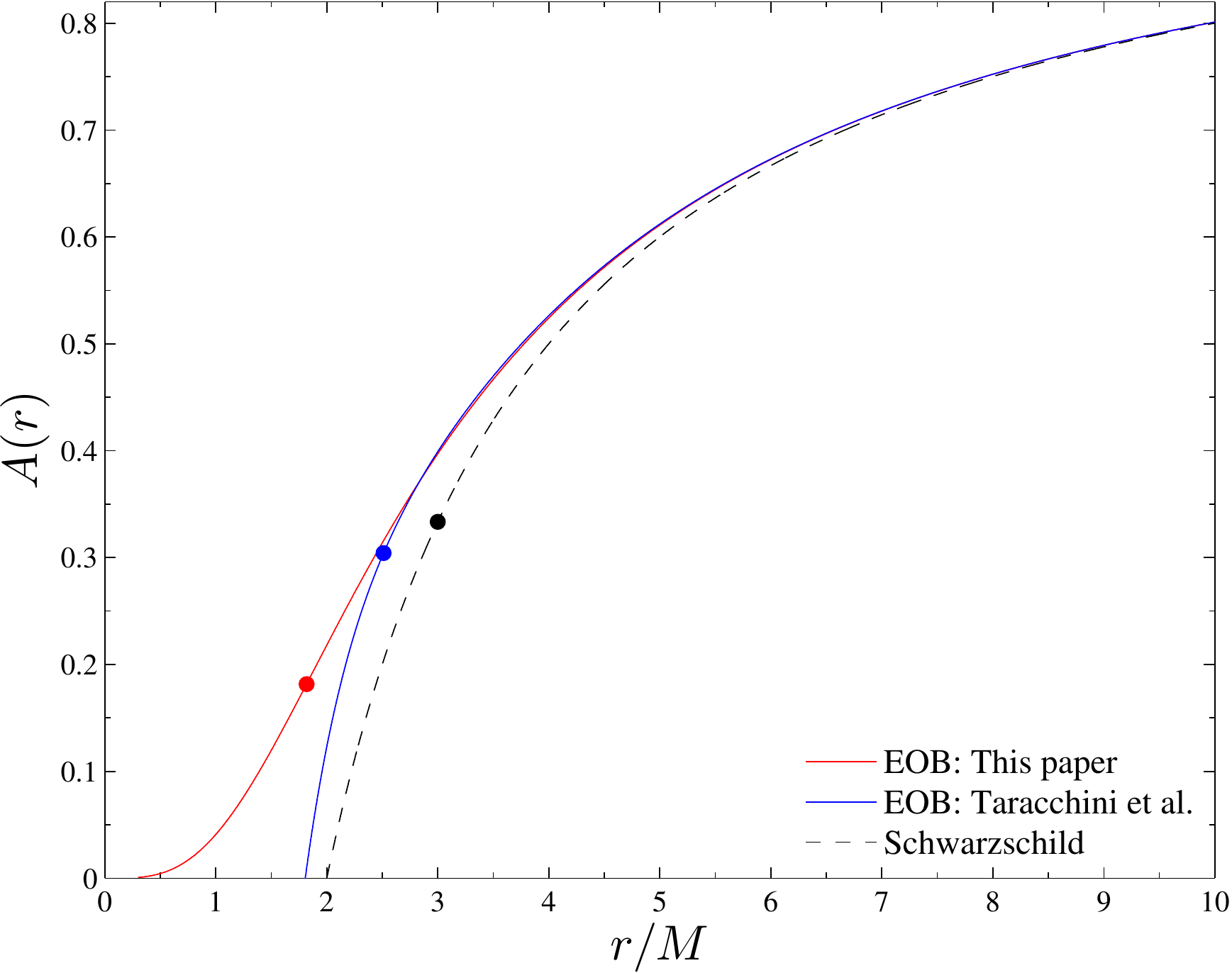}
\caption{\label{Afun_q1}Comparison between EOB radial potentials $A(r)$: the Schwarzschild case
is contrasted with the $q=1$, nonspinning function of Taracchini et al.~\cite{Taracchini:2013rva} 
and the Pad\'e-resummed, NR completed one of this paper, Eq.~\eqref{eq:Anonspinning}, with $a_6^c(\nu)$ 
given by Eq.~\eqref{eq:a6c_calibrate}. The marker indicates the location of the ``light-ring'',
i.e. the peak of the effective ``photon-potential'' $A(r)/r^2$ (located at $r=3M$ for Schwarzschild).
}
\end{center}
\end{figure}

As a first example, we focus on the $q=1$, nonspinning case.
Figure~\ref{Afun_q1} compares three $A(r)$ curves: our NR tuned  $A_{\rm orb}(r)$ curve, with $a_6^c$
given by  Eq.~\eqref{eq:a6c_calibrate}; the $A(r)$ of Ref.~\cite{Taracchini:2013rva},
calibrated to the same sample of nonspinning NR waveform data we used
here; and the Schwarzschild $A(r)$. The markers on the plot indicate the
location of the (adiabatic) ``light-ring'', defined as the peak of
the effective ``photon-potential'' $A(r)/r^2$ (corresponding to $r=3M$
for the Schwarzschild metric).  
Let us first note a qualitative difference: our choice of resumming the Taylor-expanded
$A$ function with a Pad\'e $(1,5)$ approximant entails that our function 
$A_{\rm orb}(r)$ vanishes at $r=0$. This is qualitatively different from 
$A^{\rm Tar}(r)$, whose (non Pad\'e) resummation is chosen so as to
impose the existence of a horizon at a nonzero value of $r$. As a consequence,
our radial potential is more repulsive than  $A^{\rm Tar}(r)$ when $r \lesssim 2$.

From the quantitative point of view, let us note that our $A_{\rm orb}(r)$ and  $A^{\rm Tar}(r)$ 
are rather close up to the lightring of $A^{\rm Tar}(r)$. Their difference (up to their
light-ring location, blue dot) ranges between $-3\times 10^{-3}$
and approximately $7\times 10^{-3}$ at the lightring point. 
Although such differences look small, let us emphasize that, as discussed
in Ref.~\cite{Damour:2012ky}, these differences are large enough to make the two 
potentials belong to {\it different} ``equivalence classes", i.e. to yield significantly
different conservative dynamics. Indeed,  Ref.~\cite{Damour:2012ky} found 
that one needed differences smaller than the $10^{-4}$ level to lead to 
indistinguishable dynamics.

\begin{figure}[t]
\begin{center}
\includegraphics[width=0.47\textwidth]{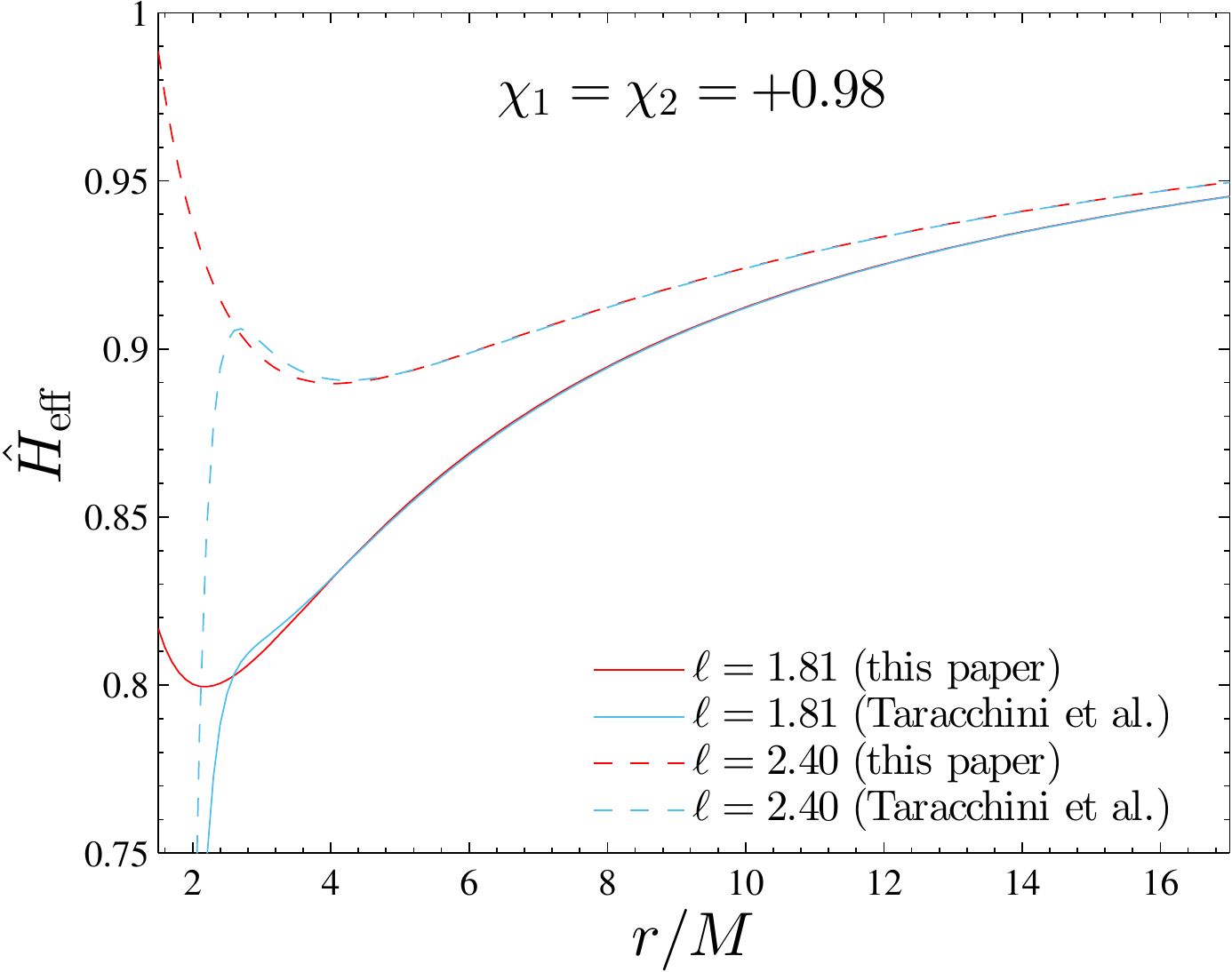}
\caption{\label{Heff_chi098} Comparison between effective Hamiltonians $\hat{H}_{\rm eff}(r,\ell,\chi)$
obtained by using our EOB spinning model and the one of Ref.~\cite{Taracchini:2013rva} for $\chi=+0.98$
and for two values of the dimensionless orbital angular momentum $\ell$. The value $\ell=1.81$ corresponds
to the merger location of the blue curve in Fig.~\ref{Afun_q1}. For $\ell=1.81$ our effective
potential is such that the system is still inspiralling.}
\end{center}
\end{figure}

As a second example of comparison between our EOB dynamics and that of 
Ref.~\cite{Taracchini:2013rva} we compare in Fig.~\ref{Heff_chi098} the dimensionless effective
potential  $\hat{H}_{\rm eff}(r,\ell,\chi)$ (including spin-orbit and spin-spin interactions,
but with $p_{r_*} =0$), for $\chi=\chi_1=\chi_2= + 0.98$ and two representative values of the 
dimensionless orbital angular momentum $\ell$. For the value $\ell=2.40$, both effective 
potentials exhibit a local minimum, corresponding to a stable circular orbit. 
On the other hand, for  $\ell=1.81$  (which corresponds
to the merger location of the blue curve in Fig.~\ref{Afun_q1}) the model of Taracchini et al. no longer allows for
a stable circular orbit, while ours still does. In other words, our dynamics is still inspiralling, while that
of  Ref.~\cite{Taracchini:2013rva} is already ``plunging".  Note that for  $\chi=\chi_1=\chi_2= + 0.98$
we have seen above that our energetics is closer to the NR one. It should, however, be kept in mind
that we are here comparing two conservative dynamics while the real EOB dynamics is also
(especially near merger) strongly modified by radiation reaction effects.
%

\section{Conclusions}
\label{end}

\begin{figure}[t]
\begin{center}
\includegraphics[width=0.47\textwidth]{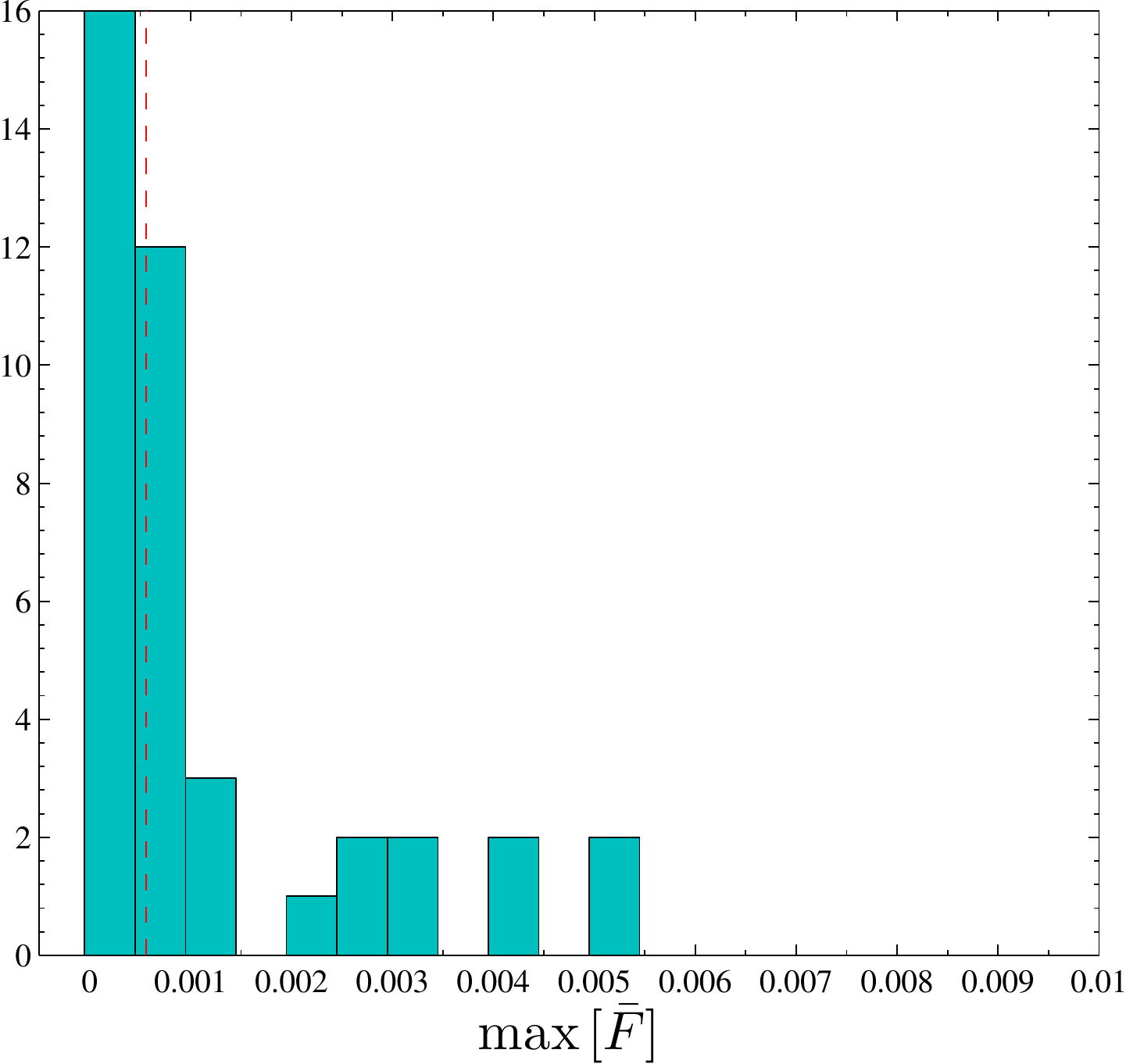}
\caption{\label{histo}Maximum value of the EOBNR unfaithfulness $\bar{F}$ 
(listed in Table~\ref{tab:configs}) for all SXS dataset considered. 
The vertical line indicates the median of the histogram, corresponding 
to 0.00056105.}
\end{center}
\end{figure}

In this paper we have presented a new NR completion of the spinning, 
nonprecessing, EOB model introduced in Ref.~\cite{Damour:2014sva}. 

Our new EOB model uses only {\it two} flexibility\footnote{Let us recall that the idea of using the ``flexibility'' of the EOB
formalism to determine effective values of yet uncomputed higher-order analytical functional parameters (such as $a_6(\nu)$) by
fitting EOB predictions to NR data was introduced, and first implemented, in Ref.~\cite{Damour:2002qh}.} parameters, 
one for the nonspinning sector $a_6^c(\nu)$, and one for the spinning sector $c_3(\ha_1,\ha_2;\,\nu)$. 
We have calibrated the functions $a_6^c(\nu)$ and $c_3(\ha_1,\ha_2;\,\nu)$ by
comparing EOB predictions to a large sample ($\sim 50$) of NR data waveforms
computed with two different numerical codes (Llama and SpEC).
This calibration is done (for every step) through a simple one-parameter 
search by requiring that the EOB/NR time-domain phase difference (after alignment)
stays within NR uncertainty up to merger. This is achieved in two steps:
the function $a_6^c(\nu)$ is determined first by using nonspinning data;
then the single remaining flexibility parameter $c_3(\nu;\, \ha_1,\ha_2)$ is 
determined by using spinning data. 

The EOB/NR comparison is done both for phasing and energetics  (gauge-invariant relation $E_b(j)$  
between binding energy and angular momentum).

Our main results are:
\begin{itemize}
\item[(i)] The \hbox{$\ell=m=2$} GW phasing performance of our new spinning EOB model (measured both 
by the time-domain EOB/NR phase difference and through faithfulness computations) is 
at least as good as that of the most recent version of the spinning EOBNR model developed 
by the  group of A.~Buonanno~\cite{Taracchini:2012ig,Pan:2013tva,Taracchini:2013rva}.
To quantitatively assess the quality of our unfaithfulness results, we display in Fig.~\ref{histo}
a histogram of the maximized-over-mass values of the unfaithfulness\footnote{We recall that those
unfaithfulnesses are only integrated over the NR frequency range.}, $\Fmax\equiv \max_M(\bar{F}(M))$, 
over the entire sample of SXS waveforms we use. 
Note that the histogram is skewed  towards very small values 
of $\Fmax$, with a median value equal  to 0.00056105 
(see vertical line on the figure).
It is interesting to note that $\Fmax\lesssim 0.003$ for all data except for 
the case $(q,\chi_1,\chi_2)=(1,-0.50,0)$ and the near-extremal spin $q=1$ data sets 
with $\chi=\chi_1=\chi_2=(+0.98,+0.99,+0.994)$. For the latter datasets,
not that $\chi=+0.99$ and $\chi=+0.994)$ both yield $\Fmax\approx 0.005$, while 
$\chi=0.98$ yields $\Fmax\approx 0.0042$.

When comparing the mass dependence of $\bar{F}$ in our Fig.~\ref{Fbar} with the corresponding
quantity displayed in Fig.~1 of Ref.~\cite{Taracchini:2013rva} one notices the
following facts: (i) our $\bar{F}(M)$ is, for all values except a few, maximal around 
$M=20M_\odot$ and then monotonically decreases as $M$ increases. By contrast, 
the mass dependence of the $\bar{F}(M)$ of Ref.~\cite{Taracchini:2013rva} 
is quite different: either it increases monotonically from $20M_\odot$ to $200M_\odot$ 
or it has a local maximum. We attribute the mass behavior of our $\bar{F}(M)$ 
to the new, more accurate attachment of the ringdown part~\cite{Damour:2014yha} which we 
have implemented in our model; (ii) In Fig.~1 of~\cite{Taracchini:2013rva} we spot 
{\it two} outliers with $\Fmax\approx 0.01$: one is SXS:BBH:0065, with
$q=8$ $(\chi_1,\chi_2)=(+0.5,0)$; the other outlier of Ref.~\cite{Taracchini:2013rva} 
is SXS:BBH:0152, with $q=1$, $\chi_1=\chi_2=+0.6$. We stress that our 
corresponding values $\bar{F}(M)$ remain below $10^{-3}$, as displayed in Fig.~\ref{Fbar}.

\item [(ii)] We have presented the first systematic, accurate computation 
of $E_b(j)$ curves from SXS and Llama data of nonprecessing, spinning 
(or nonspinning) configurations. Our results significantly improve 
previous attempts to do so and give us a reliable target for 
EOB calibration, that complements the usual phasing analysis.

For the nonspinning sector of the model, the inspection of the NR $E_b(j)$ curves 
led us to decide that the choice ${\cal F}_{r_*}=0$ (which was already used in 
early EOB works~\cite{Buonanno:2000ef,Damour:2009kr}) leads to a good EOBNR 
agreement between the corresponding energetics, at the $10^{-4}$ level up to merger.  
We hope to further study (both analytically and numerically) the role of the radial  
part of radiation reaction, and its influence on Schott terms.
The results presented in Figs.~\ref{fig:ejq8} and ~\ref{fig:diff} have shown the potential
of SXS data for very accurate investigations of energetics in coalescing 
black hole binaries.

The $E_b(j)$ NR curves were computed using two different sets of numerical data, one by the
Llama code and the other by the SpEC code. We solved several subtle issues related to the 
computation of $E_b(j)$ from SXS data.
In particular, for the cases of extremal spins $\chi_1=\chi_2=-0.95$ and $\chi_1=\chi_2=+0.98$, 
we showed that the corresponding NR curves presented in Ref.~\cite{Taracchini:2013rva} were 
inconsistent both with the NR final state (as provided in the SXS files itself) and 
with early inspiral EOB predictions. This allowed us to meaningfully compare, for the first time, extremal-spin
NR data both to our EOB model and to the one of Ref.~\cite{Taracchini:2013rva}.
The main outcomes of this comparison are: (i) for $\chi=+0.98$, the energetics predicted
by our model model is closer, than that of the model of Ref.~\cite{Taracchini:2013rva},
to the NR one, during the late inspiral and near merger; (ii) for $\chi=-0.95$, the 
opposite is true, with the model of Ref.~\cite{Taracchini:2013rva} being closer 
than ours to the NR values near merger.

\item[(iii)] The EOB  model we are presenting here is analytically simpler than 
the one of Ref.~\cite{Taracchini:2013rva}, while it performs at a comparable accuracy 
(which is demonstrably better in some situations, e.g. the case of SXS:BBH:0152 data). 
The structural simplicity of our spinning EOB model is such that it is easy to 
recalibrate it if the needs occurs, e.g. in the case that further linear-in-$\nu$ 
corrections to the gyro-gravitomagnetic ratios
$(G_S,G_{S_*})$, as defined in Paper~I, are introduced. Different analytical expressions would
imply a different determination of the function  $c_3(\ha,\ha;\,\nu)$.
In addition, it is straighforward to incorporate tidal effects in the model simply by augmenting
the pure orbital $A$ function with the tidal potential recently determined in 
Ref.~\cite{Bernuzzi:2014owa} so as to build a BNS (or BHNS) spinning EOB model. 
The performance of such a tidal, spinning, EOBmodel, as well as its eventual calibration, 
will have to be carefully assessed against state-of-the-art NR simulations of spinning
NS binaries~\cite{Bernuzzi:2013rza}.

\item[(iv)] The preliminary comparison given here between the performances of our model 
and the EOB model of Ref.~\cite{Taracchini:2013rva}, was purely based on information
that was either given to us or that was extracted from the literature. 
We also emphasized some of the structural differences between these two spinning EOB models:
(i) in Fig.~\ref{Afun_q1} we constrasted the two different EOB $A(r)$ radial potentials for
$q=1$ nonspinning black holes; and (ii) in Fig.~\ref{Heff_chi098} we contrasted
the two effective Hamiltonians $\hat{H}_{\rm eff}(r)$ for $q=1$ and $\chi=+0.98$. 
The differences are nonnegligible and are probably partly responsible of the 
different performances of the two EOB models 
versus the NR $E_b(j)$ curve for $\chi=+0.98$.
We hope that it will be possible to perform soon a more detailed comparison between 
the two spinning EOB models, notably by directly comparing waveforms.
We think that a synergetic effort dedicated to transfering the best features of 
each EOB model into a new model for spinning binaries will be crucial for 
the forthcoming gravitational wave astronomy.

\end{itemize}

\acknowledgments 
We thank: Andrea Taracchini, Alessandra Buonanno and the other authors 
of Ref.~\cite{Taracchini:2013rva} for sharing with us the NR and EOB
$E_b(j)$ curves for $\chi=-0.95$ and $\chi=+0.98$; Sascha Husa for 
assistance in computing low eccentricity initial data; and Simone
Balmelli for help with generating the data of Fig.~\ref{Heff_chi098}.
We finally thank Michael Boyle for having computed for us the data
of Fig.~\ref{fig:0152_com} so as to approximately remove the drift of
the center of mass.
D.~P. was supported by grants 
CSD2007-00042 and FPA2010-16495 of the Spanish Ministry of Science. 
CR acknowledges support by NASA through
Einstein Postdoctoral Fellowship grant number PF2-130099 awarded by
the Chandra X-ray center, which is operated by the Smithsonian
Astrophysical Observatory for NASA under contract NAS8-03060. 
Computations were performed using resources of the 
NASA High-End Computing (HEC) Program through the 
NASA Advanced Supercomputing (NAS) Division at Ames Research Center 
and NSF XSEDE (allocation TG-MCA02N014 and TG-PHY100033).
Additional computations were performed on the Caltech computer 
cluster ``Zwicky'' (NSF MRI award No.\ PHY-0960291).

\appendix*

\section{Computing $E_b(j)$ using SXS waveforms}
In this Appendix we discuss in detail the procedure we adopt to compute the
function $E_b(j)$ using SXS NR data. First, in Sec.~\ref{sec:a1} we contrast
SXS with Llama data and highlight some (minor) unsatisfactory features of 
the SXS waveforms. The actual $E_b(j)$ computation is detailed in Sec.~\ref{sec:a2} below.

\subsection{Unphysical effects in the $q=1$, SXS multipolar waveforms}
\label{sec:a1}
\begin{figure}[t]
\begin{center}
\includegraphics[width=0.47\textwidth]{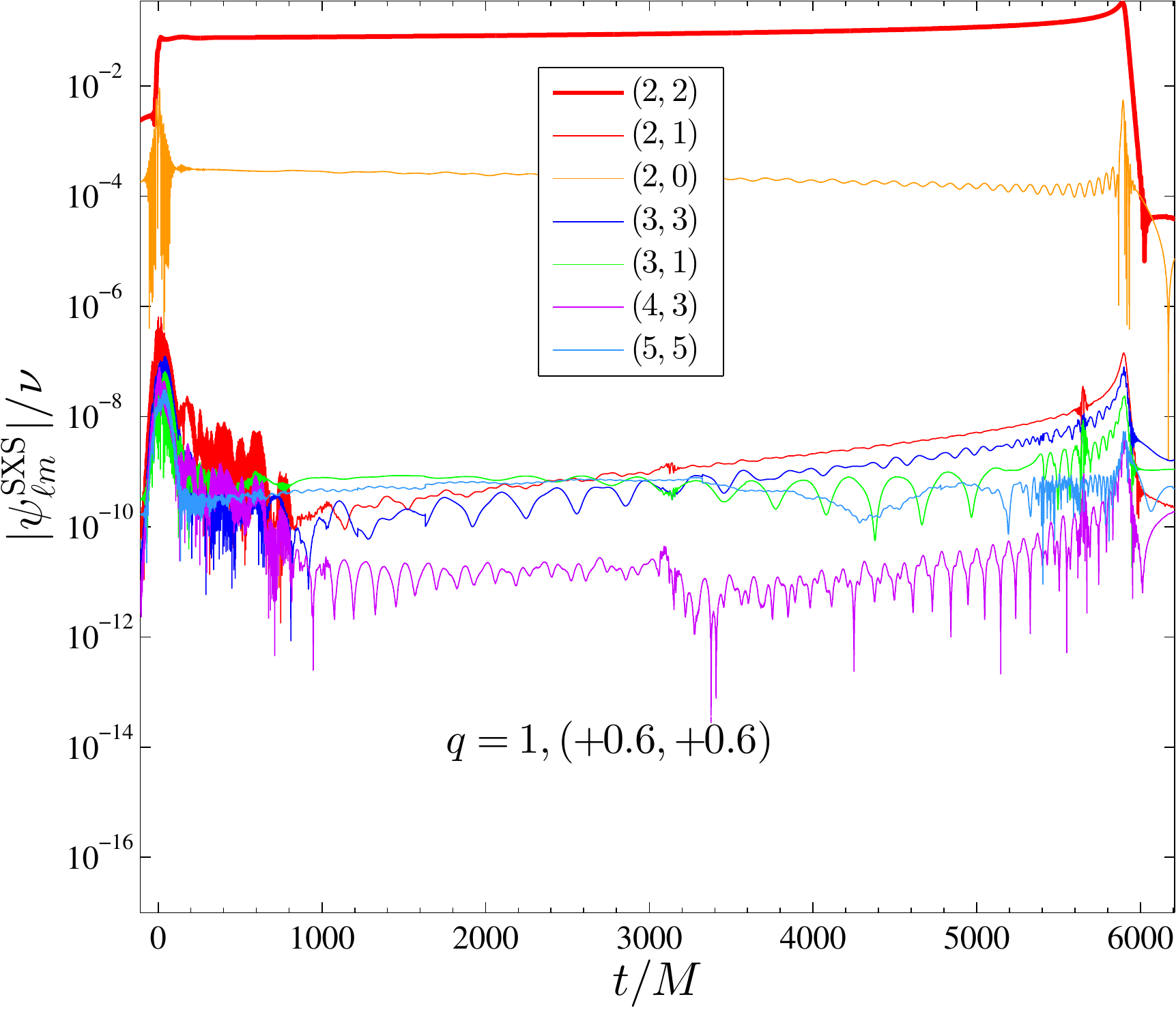}\\
\vspace{5mm}
\includegraphics[width=0.47\textwidth]{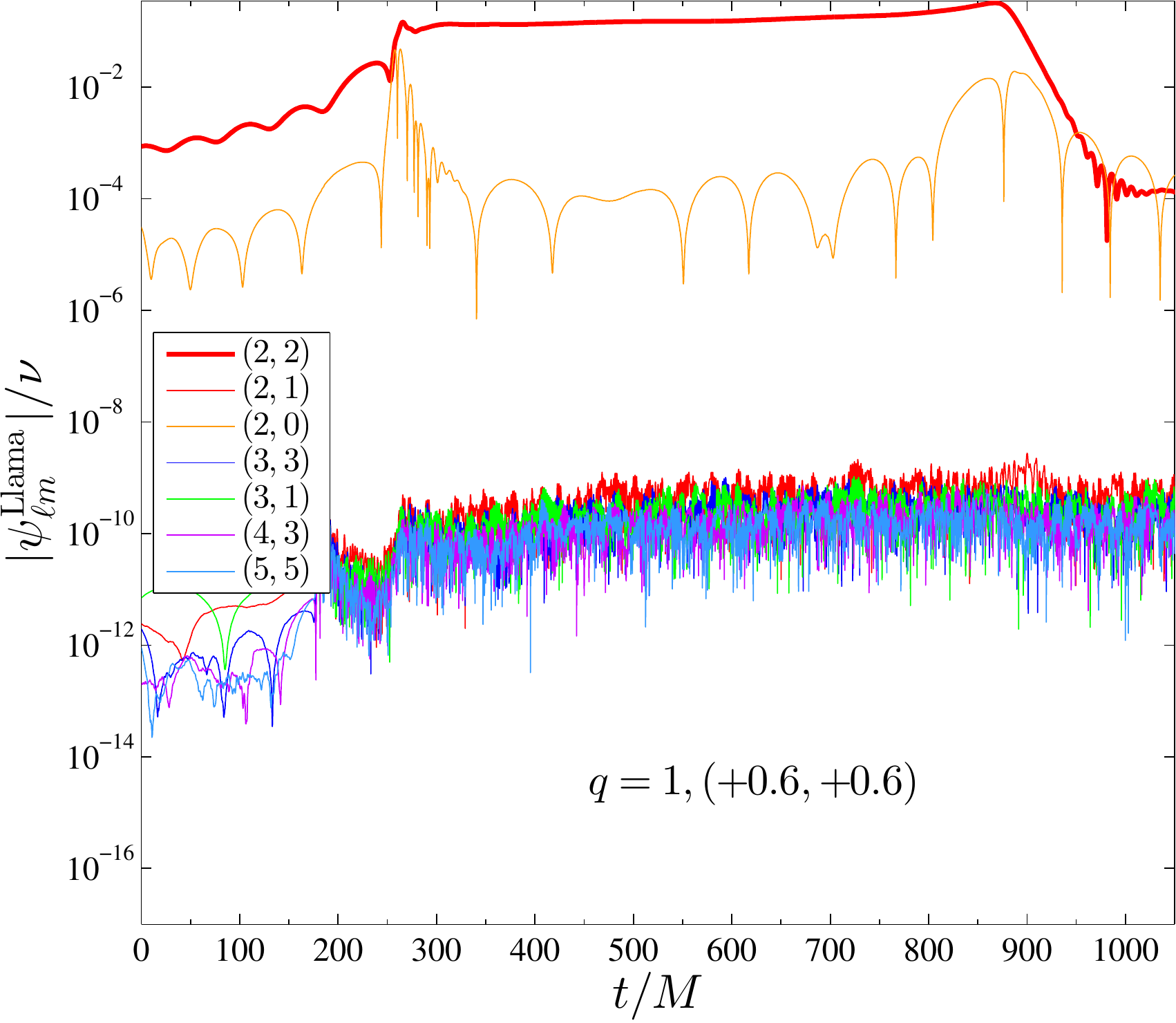}
\caption{\label{ej:psilm_Llama_SXS} Analysis of the subdominant 
multipoles of $q=1$ SXS waveforms (top panel). The figure compares the amplitudes 
of various subdominant multipoles to the $\ell=m=2$ one. The multipolar amplitudes
with $m=\text{odd}$  grow in time likewise the dominant one.
This effect (that is found in all $q=1$ data) is (partly) related to the
motion of the center of mass and to residual tiny asymmetries in the intial
data, as discussed in the text. It is, however, absent in the corresponding 
Llama waveform data (bottom panel).}
\end{center}
\end{figure}

For an equal-mass binary, with equal (aligned or anti-aligned) spins, 
the fact that the system is symmetric by exchange of the two black holes
ensures that all multipoles with $m=\text{odd}$ have to vanish. 
We verified that this is essentially the case for all Llama simulations at our 
disposal: in the time plots of these multipoles, only uncorrelated noise is found.
On the contrary, a careful inspection of SXS data led us to the discovery
that the $m=\text{odd}$ multipoles have  amplitudes, which are
small but which are not just uncorrelated noise.
Rather they qualitatively show some sort of chirping structure 
present in the actual physical modes with $m=\text{even}$. 
This is illustrated in Fig.~\ref{ej:psilm_Llama_SXS} for the
case $q=1$, $\chi_1=\chi_2=+0.6$, corresponding to dataset SXS:BBH:0152
of the SXS catalog. The figure contrasts 
the {\tt SpEC} modes (top panel) with the corresponding Llama modes 
(bottom panel). One immediately sees the striking qualitative difference 
between the two independent simulations of the same physical system.
Though, the amplitude of these modes is so small that it does not have
any practical influence on the $E_b(j)$ 
computation\footnote{We have analyzed what happens if the $m=\text{odd}$ modes are
put to zero. We find that the resulting small difference (mainly coming 
from the $m=\text{odd}$ junk radiation) can be absorbed by modifying
the vectorial 
shift $(\Delta j^0,\Delta E_b^0)$ mentioned in the main text and discussed
below. In practice, all our $E(j)$ curves are computed using the full
multipolar information at our disposal including multipoles up to $\ell=m=8$.}, 
it is interesting to briefly point out the impact of two physical effects 
on shaping the multipolar structure of the modes: (i) the motion of the center 
of mass (CoM) of the system and (ii) the (tiny) asymmetry of the (nominally symmetric) 
initial binary configuration.

\begin{figure}[t]
\begin{center}
\includegraphics[width=0.47\textwidth]{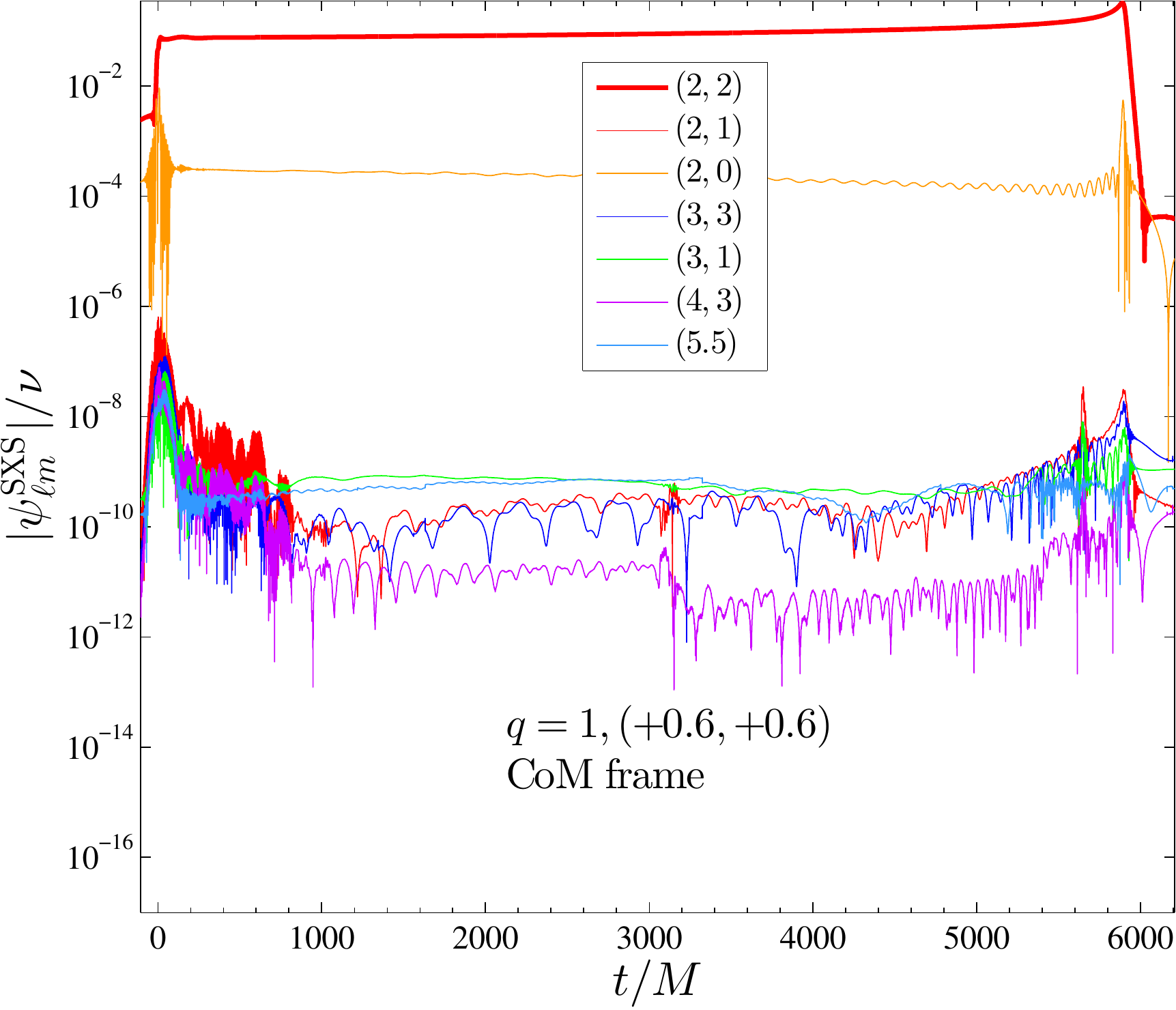}
\caption{\label{fig:0152_com}The same SXS multipolar waveform of 
 Fig.~\ref{ej:psilm_Llama_SXS} after M.~Boyle removed the drift of the CoM using the procedure 
of Ref.~\cite{Boyle:2015nqa}. The subdominant multipolar amplitudes look flatter than in 
Fig.~\ref{ej:psilm_Llama_SXS} and qualitatively closer to the Llama ones.}
\end{center}
\end{figure}

The effect of the motion of the CoM due to residual linear momentum in the 
initial data has been investigated in some very recent 
work~\cite{Boyle:2015nqa,Ossokine:2015yla}. Notably, Ref.~\cite{Ossokine:2015yla} 
introduced a method to eliminate such residual momentum from the initial data, 
so that future SXS waveforms will be amended of such systematic uncertainty.
However, as emphasized by Boyle~\cite{Boyle:2015nqa}, the
waveforms currently present in the SXS catalog {\it are not} expressed in
the CoM frame (as are the EOB ones) and this introduces systematic uncertainties
that should be removed. As a matter of fact, Ref.~\cite{Boyle:2015nqa} (that 
appeared while this paper was under evaluation) proposed a way to 
remove the effect of the CoM drift from the waveforms. 
We asked Micheal Boyle to kindly apply his removal procedure to 
the SXS:BBH:0152 configuration we are discussing here. 
His result is shown (with his permission) in Fig.~\ref{fig:0152_com}. 
One sees that the secular growth of the mode amplitudes 
is visibly reduced (though it is not completely removed, especially 
towards merger). This is particularly visible for the 
$(2,1)$ mode, where the trend visible in Fig.~\ref{ej:psilm_Llama_SXS} 
is replaced by random-like oscillations.
Globally, the amplitude of the modes is rather constant, with less structure
than before, and visually more consistent with the outcome of the Llama 
simulation. Improvements in the determination of the CoM frame of a NR 
simulation may flatten the $m=\text{odd}$ modes even further. 

\begin{figure}[t]
\begin{center}
\includegraphics[width=0.47\textwidth]{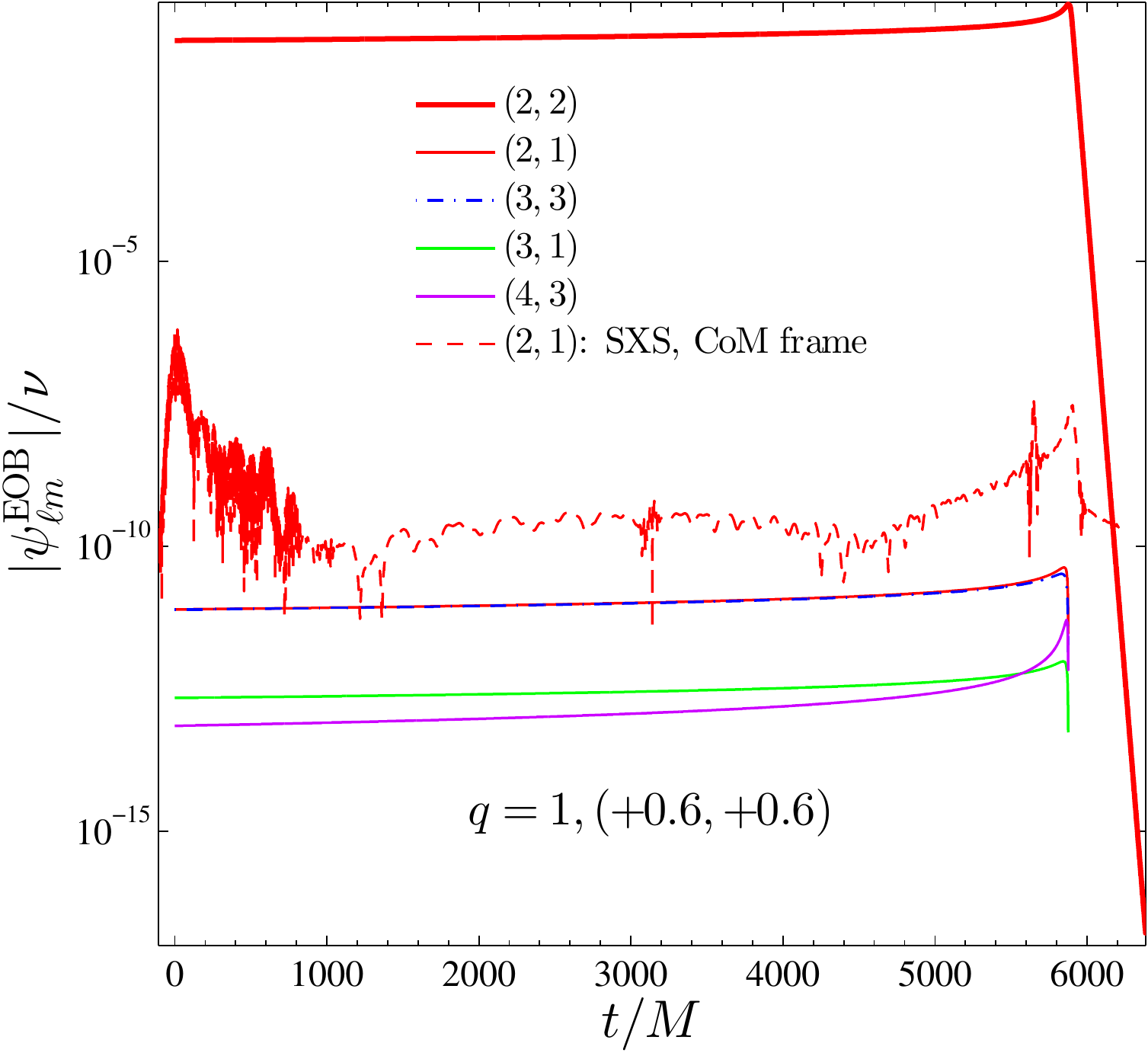}
\caption{\label{fig:eob_multipoles}Investigating the effect of tiny asymmetry in
the initial data at the level of EOB waveform. The $(2,1)$ mode of Fig.~\ref{fig:0152_com} is also included
for visual comparison. The asymmetry in the initial data is not sufficient to fully explain the magnitude of the
subdominant modes in the data of Fig.~\ref{fig:0152_com}.}
\end{center}
\end{figure}

Let us now turn to estimating the effect on the $m=\text{odd}$ multipoles
due to the tiny asymmetry present in the NR initial data.
Inspecting the file {\tt metadata.txt} of dataset SXS:BBH:0152, one finds
that the initial massed and initial spins are actually slightly different; 
i.e., one has $m_1=0.499999998745987$, $m_2=0.499999997467155$ and
$\chi_1 =  0.599963311735964$, $\chi_2=0.599963312827750$.
These values of the initial masses yield $\nu=0.25$ at machine precision,
though $X_1-X_2=(m_1-m_2)/M=1.278831984752316\times 10^{-9}$.
We then estimated the impact of these tiny asymmetries by running an 
EOB simulation with precisely these spin parameters and this value of $X_1-X_2$
entering the subdominant odd-parity modes. The multipolar amplitudes of the EOB
waveform computed with these choices are exhibited in Fig.~\ref{fig:eob_multipoles}: 
one sees that during inspiral, where the analytical (EOB, resummed) waveform is 
highly reliable, the amplitude of subdominant modes is from 1 (for the $\psi_{21}$ mode) 
to approximately 2 (for the $\psi_{43}$ mode) orders of magnitude smaller 
than the corresponding SXS multipoles in the CoM frame.
In Fig.~\ref{fig:eob_multipoles} we also added the SXS $(2,1)$ mode, in
the CoM frame, to ease the comparison.
This finding suggests that the magnitude of the subdominant multipolar amplitudes 
we see in Fig.~\ref{fig:0152_com} is only partially due to the asymmetry in 
the initial data and that other effects are present.
In any case, since the analysis we did here confirms that the effects of the 
asymmetries are practically negligible, we are entitled to ignore them and 
consider exactly symmetric initial data for the EOB evolutions.

Finally, let us note in passing that $|\psi_{21}^{\rm EOB}|\approx |\psi_{33}^{\rm EOB}|$
up to merger, likewise the corresponding SXS multipoles in the CoM frame 
in Fig.~\ref{fig:0152_com}. This result contrasts with the top panel of 
Fig.~\ref{ej:psilm_Llama_SXS}, where the $(2,1)$ and $(3,3)$ modes were not 
coinciding, and gives further evidence of the need of removing the effects of 
the CoM motion motion from the SXS waveforms.
Although the numerical effects we are pointing out here are small and practically 
negligible for the equal-mass configurations, they may not be as small in other 
situations (e.g., as pointed out in Ref.~\cite{Boyle:2015nqa} for the SXS:BBH:0004 
dataset, $(q,\chi_1,\chi_2)=(1,-0.50,0)$. In principle, we think that a careful analysis 
of the (hopefully very small) impact of the CoM drifts on all the EOB/NR comparisons
discussed in this paper, both for phasing and energetics, will be needed in the future.

\subsection{Computation of $E_b(j)$ from SXS waveform data}
\label{sec:a2}

\begin{figure}[t]
\begin{center}
\hspace{4mm}
\includegraphics[width=0.42\textwidth]{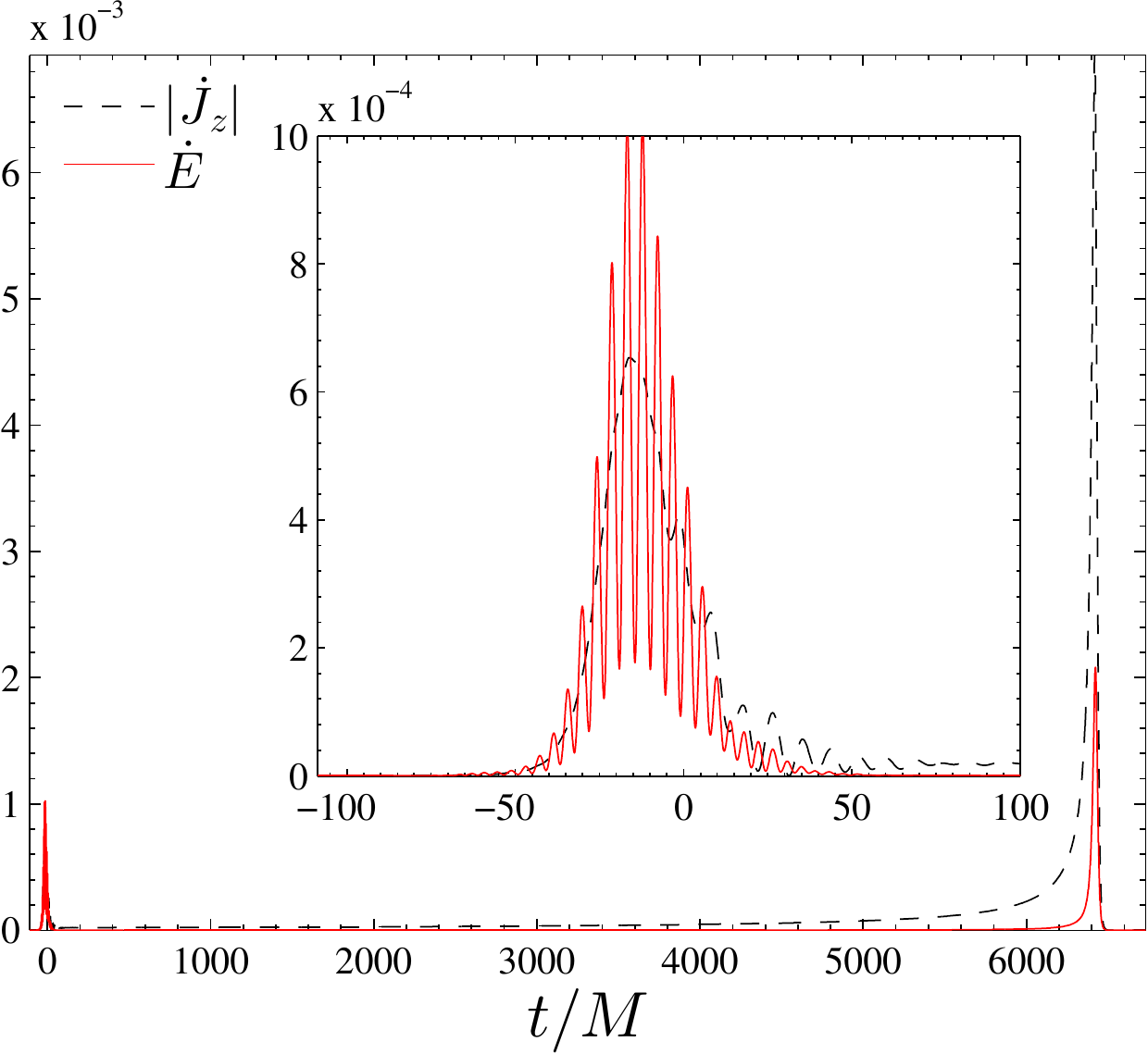}\\
\vspace{5mm}
\includegraphics[width=0.45\textwidth]{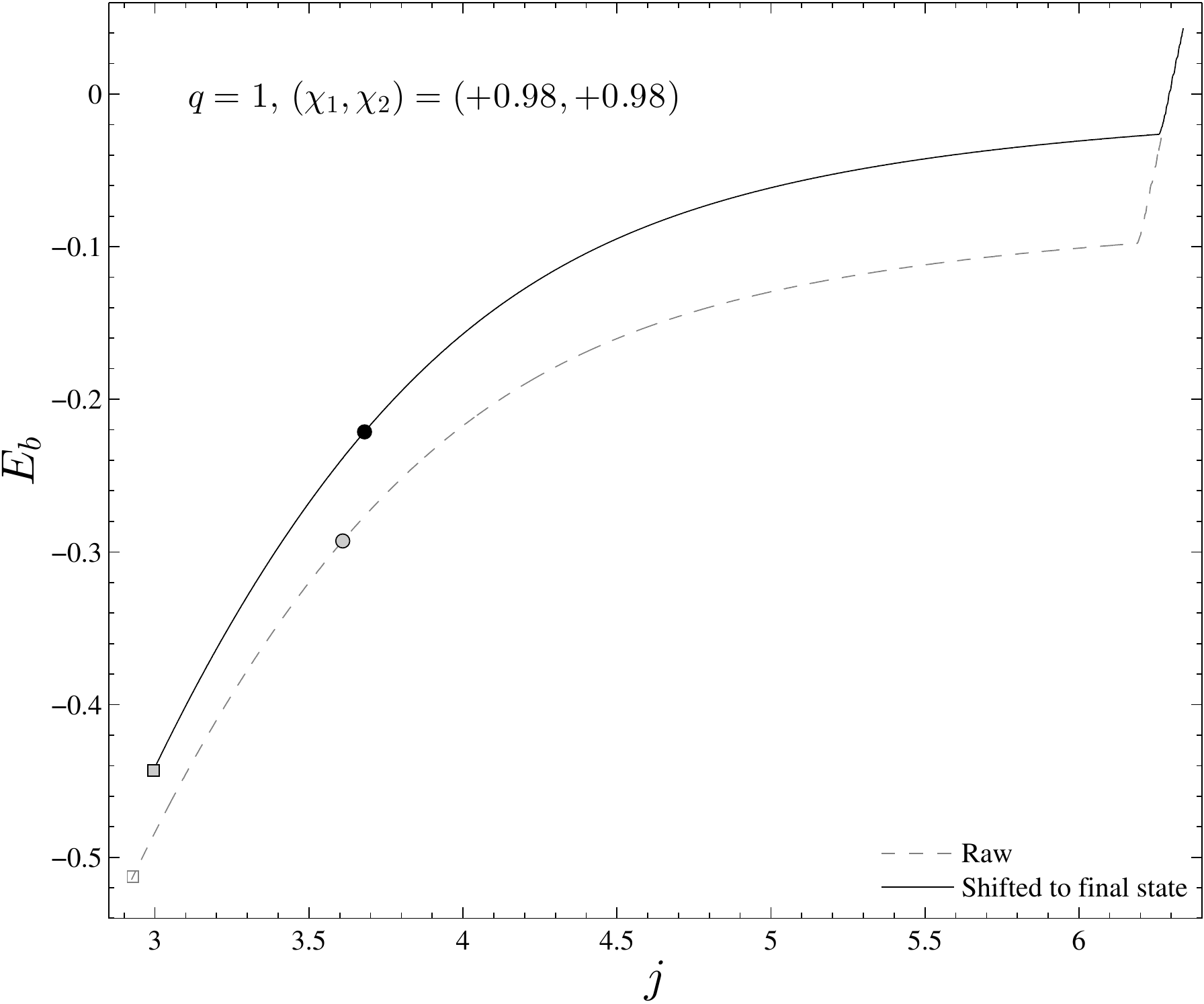}
\caption{\label{burst_098}Computation of $E_b(j)$ for $\chi_1=\chi_2=+0.98$.
The huge burst of junk radiation (top panel) introduces a huge, unphysical, vectorial shift
of the raw curve (grey, dashed), whose end point is strongly displaced from the 
actual final state $(j_f,M_f)$ (grey square on the plot) given in the {\tt metadata.txt} 
file of the SXS catalog. The solid curve is obtained by applying a shift vector 
$(\Delta j^0,\Delta E^0_b)$ to the raw curve so that the final point of 
the shifted $E_b(j)$ coincides with $(j_f,M_f)$. The filled circles mark the merger
location. See text for further details.}
\end{center}
\end{figure}

\begin{figure}[t]
\begin{center}
\hspace{4mm}
\includegraphics[width=0.43\textwidth]{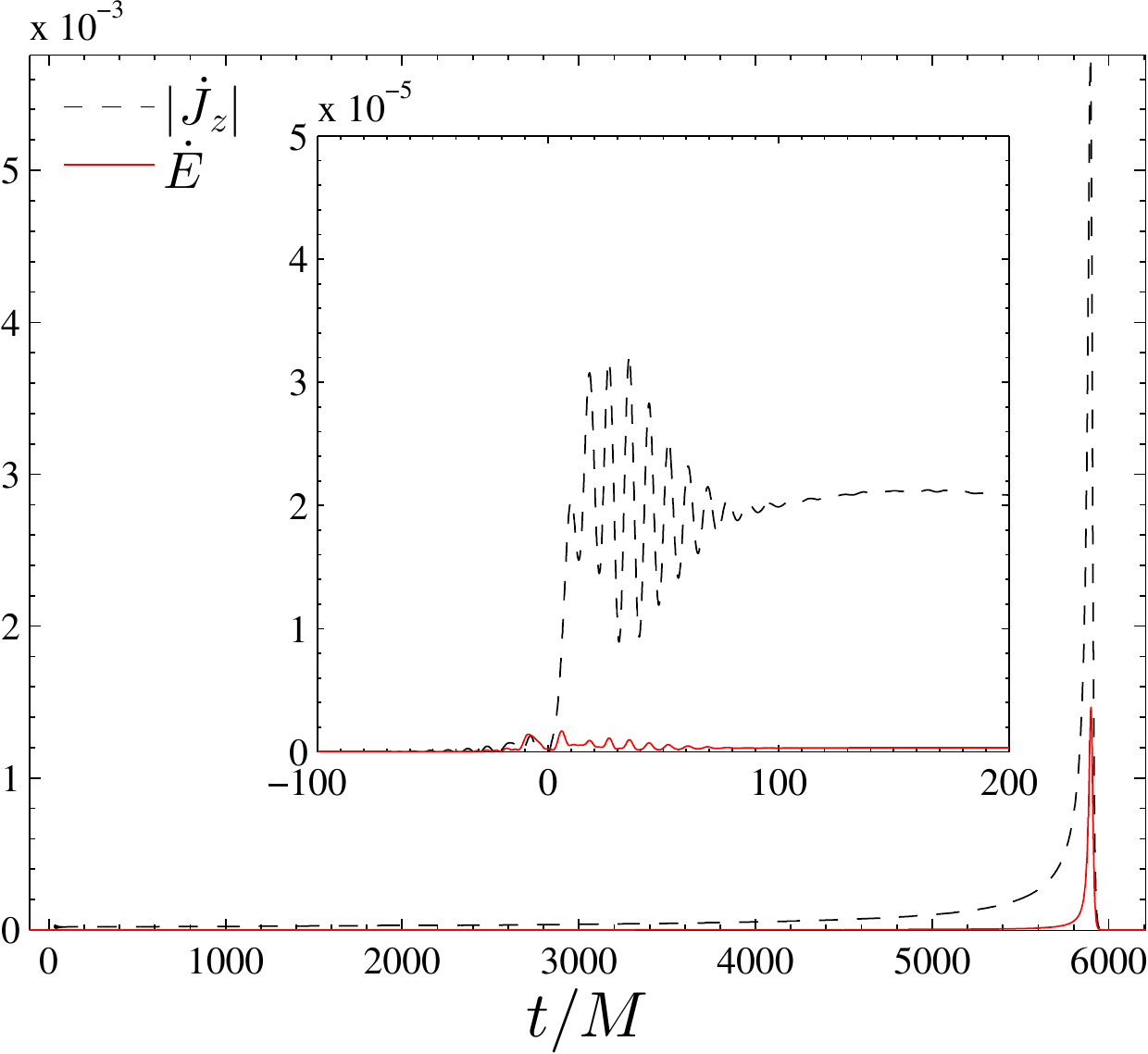}\\
\vspace{4mm}
\includegraphics[width=0.46\textwidth]{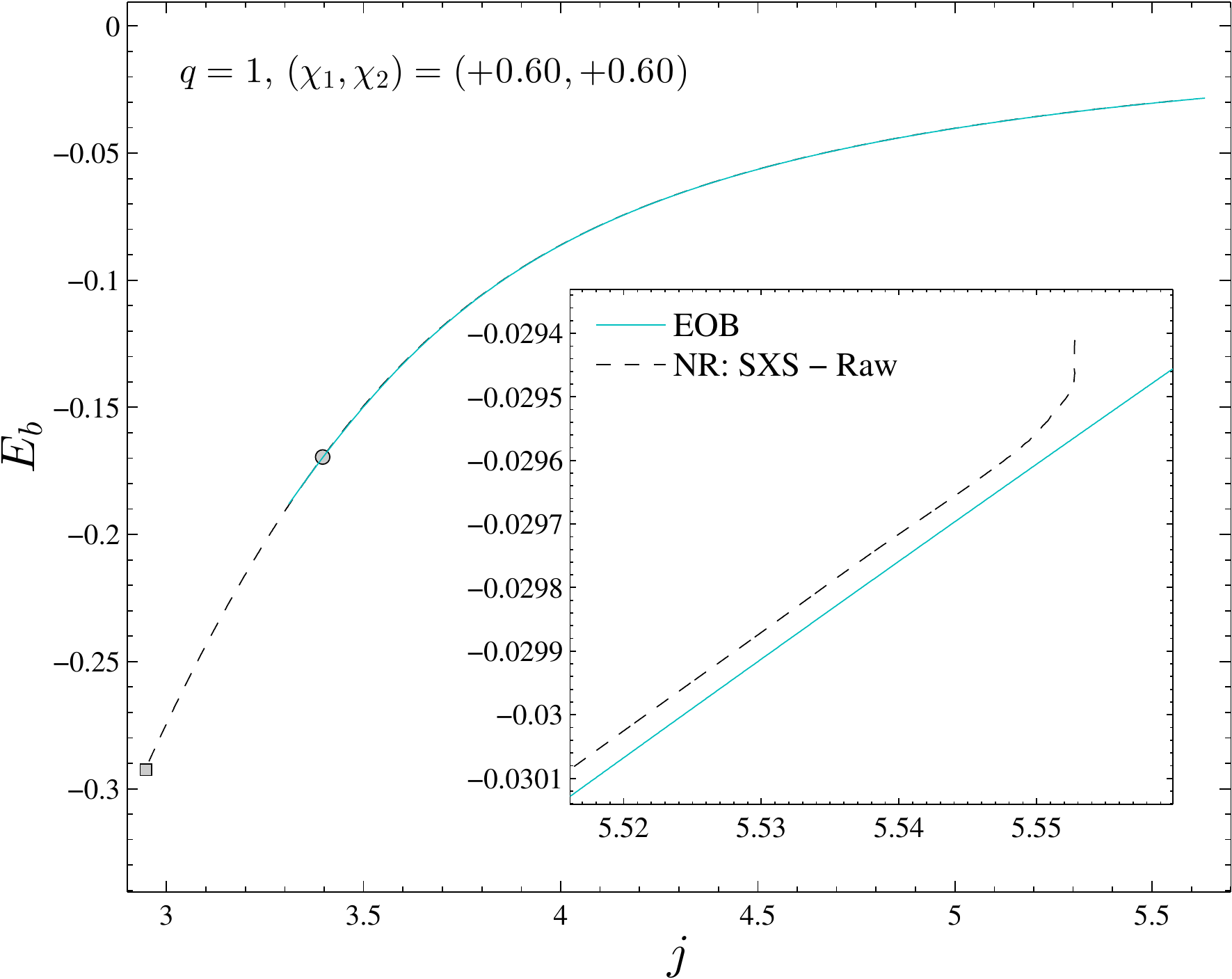}
\caption{\label{burst_060} Different phenomenology for $\chi_1=\chi_2=+0.60$. The initial
burst of radiation is much more moderate than the $\chi=+0.98$ case so that the direct, 
raw computation of $E_b(j)$ looks already visually consistent with the final state 
and the EOB curve. Additional fine tuning of $(\Delta j^0,\Delta E^0_b)$ is still 
allowed and it is then performed to further minimize the EOB-NR difference 
(within the NR uncertainty) for large values of $j$. See text for details.}
\end{center}
\end{figure}

\begin{figure}[t]
\begin{center}
\includegraphics[width=0.4\textwidth]{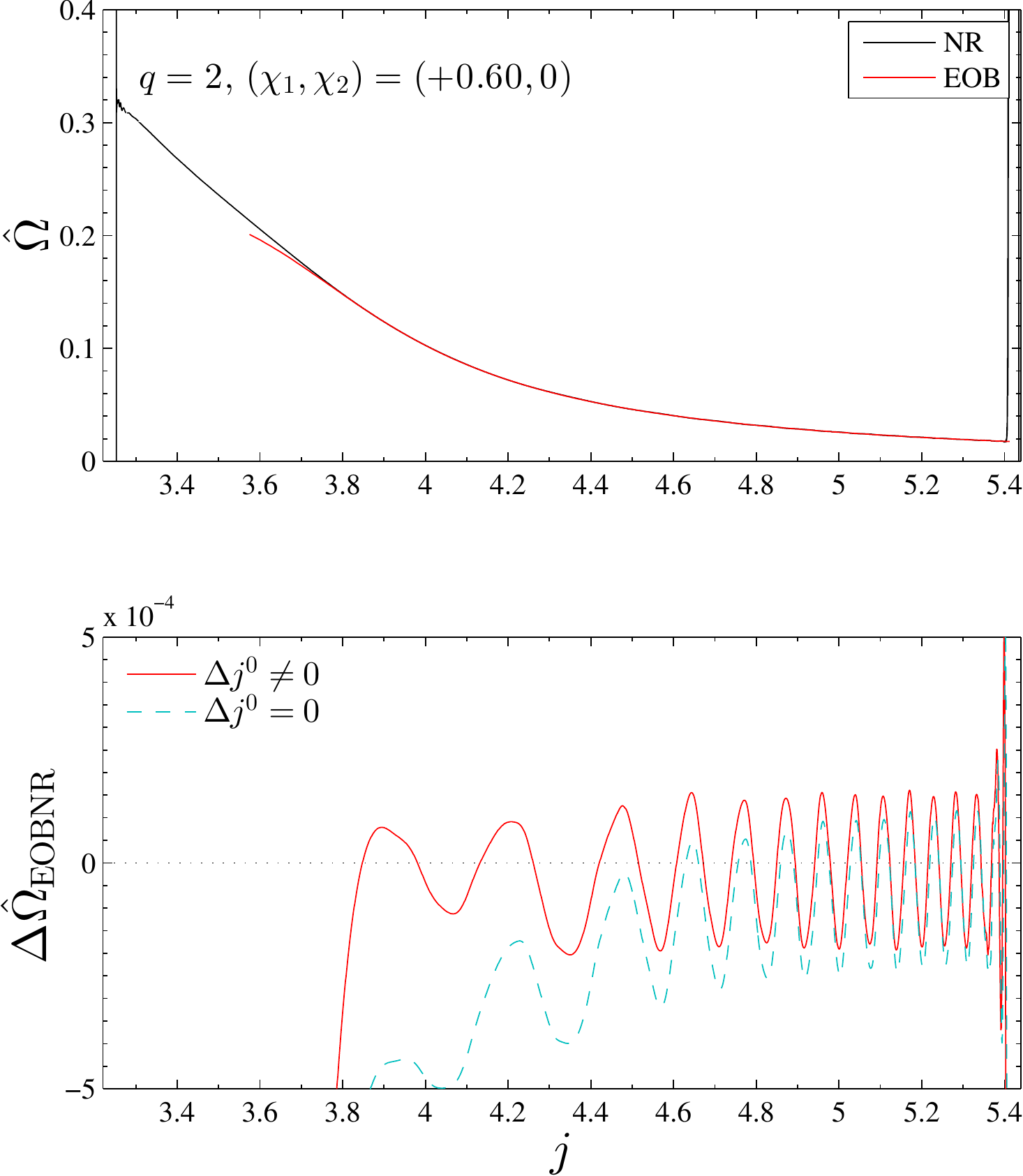}\\
\vspace{5mm}
\includegraphics[width=0.4\textwidth]{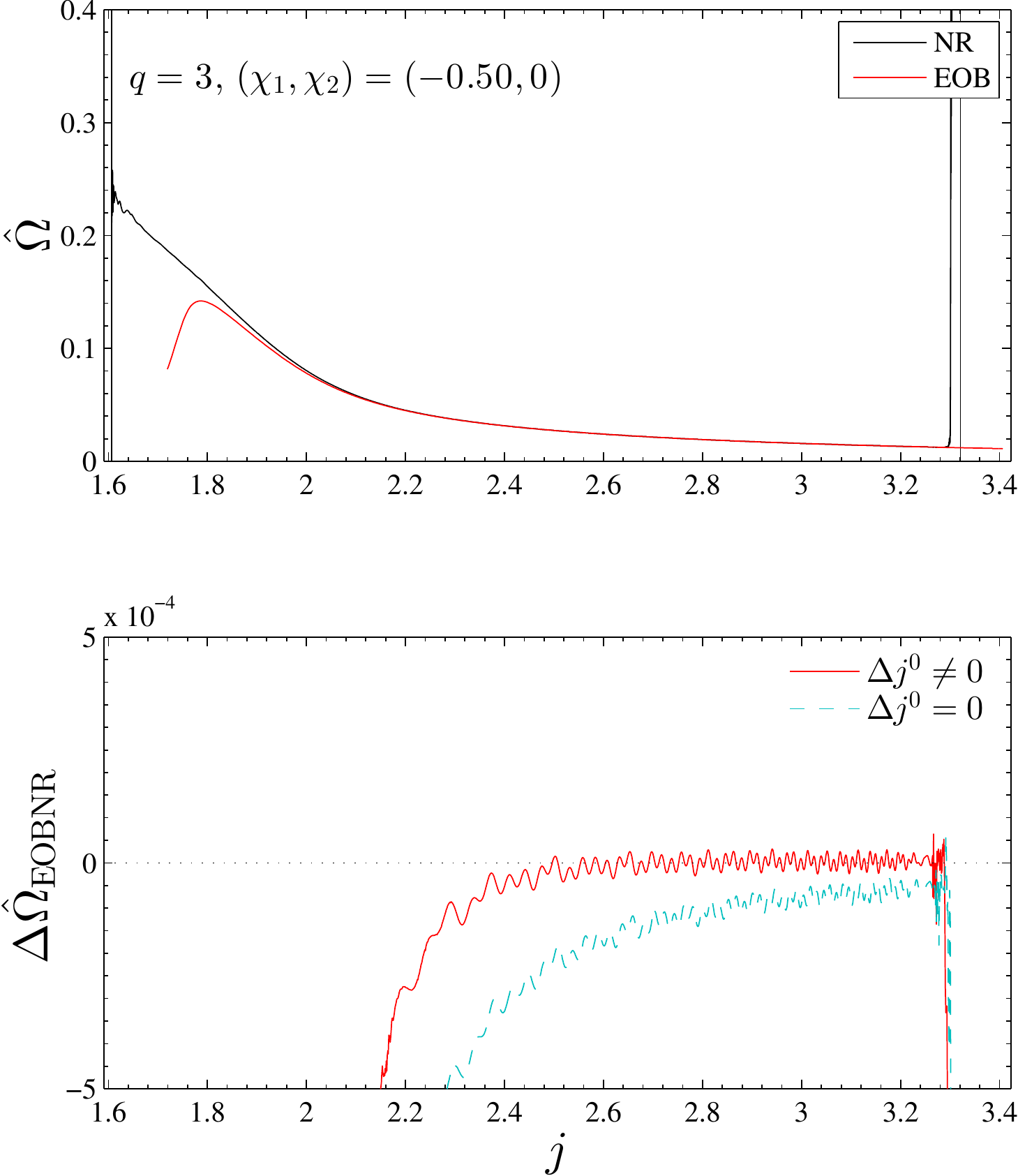}
\caption{\label{fig:Omg} Choosing $\Delta j^{0}$ by inspecting the EOB/NR differences 
between dimensionless orbital frequencies $\hOmg$ for two representative datasets.
The bottom subpanel of each configuration illustrates  how $\Delta \hOmg^{\rm EOBNR}$ 
is flattened (and the eccentricity-driven oscillation averages zero) on a large 
$j$-interval once a good choice of $\Delta j^{0}$ is made.}
\end{center}
\end{figure}

Let us now turn to explaining how $E_b(j)$ curves are computed in practice. First of all, we need to
identify the initial state of the system, i.e. the initial $(J_{\rm ADM}^0,M_{\rm ADM}^0)$.
For each simulation, this information is provided in the file {\tt metadata.txt} that 
is found in any of the {\tt Lev*} directories that can be downloaded at
{\tt http://www.black-holes.org/waveforms/} for any given configuration.
Then, the next step is to take all multipoles up to $\ell=8$ and use them
to compute the fluxes of energy and angular momentum as functions of time. 
In inspecting them, one quickly  realizes that all simulations basically 
split into two equivalence classes depending on the properties of the initial 
transient in the fluxes driven by the junk radiation: in class (A), one sees 
a huge burst of energy and angular momentum radiated during approximately the 
first $150~M$ of evolution. This behavior is displayed in
the top panel of Fig.~\ref{burst_098}, that refers to the $\chi_1=\chi_2=+0.98$ case.
Once the total losses at time $t$ are obtained (by integrating the fluxes up to time $t$),
the presence of such a big burst of radiation results in a huge shift in the $(j,E_b)$ 
plane away from $(J_{\rm ADM}^0,M_{\rm ADM}^0)$. This, raw, curve is represented as a gray,
dashed line in Fig.~\ref{burst_098}. The location of the merger is indicated by a 
filled, grey, circle. 
The unphysical nature of the initial, junk-related, shift is indicated by the fact 
that the last point of the $E_b(j)$ curve, which should coincide, to a good approximation, 
with the mass and angular momentum of the final black hole $(j_f,M_f)$, significantly differs from them.

The value of $(j_f,M_f)$ is among the data given in the SXS catalog, 
and is notably included in the {\tt metadata.txt} file. In Fig.~\ref{burst_098}, 
the actual $(j_f,M_f)$ is represented by a filled, grey, square on the plot, while the end point
of the raw $E_b(j)$ curve (indicated by an empty square) is noticeably displaced 
from $(j_f,M_f)$ due to the effect of the junk radiation. Such displacement should be
corrected, as we shall discuss below. In the second class of data, that we call (B), 
such an unphysical initial burst of GW radiation is much smaller and the raw computation
of the $E_b(j)$ curve gives a result that, without any additional shift, is  
visually consistent with the known final NR state.
This is illustrated for the case $\chi_1=\chi_2=+0.60$ in Fig.~\ref{burst_060}. 
Still (as highlighted in the inset of the figure) the NR curve, when zooming on 
the difference in the early part (see inset), exhibits a significant early 
deviation from the EOB, contrary to what was happening for 
the $\chi_1=\chi_2=+0.60$ case with Llama data, a simulation 
starting at a smaller value of $j$.

Finally, in both cases, (A) or (B), we need to apply a shift vector 
($\Delta j^0,\Delta E_b^0)$, which will be larger in case (A). 
In a first step (which is mostly useful for case (A)) we apply a first shift 
$(\Delta j^{f},\Delta E_b^{f})$ allowing the final point of the $E_b(j)$
curve to essentially (within $\lesssim 10^{-4}$) coincide with $(j_f,M_f)$. 
In a second step, we refine the determination of the vectorial shift 
by performing a second, smaller, shift, ($\Delta j^0,\Delta E_b^0)$.

\begin{table}[t]
\caption{The shift vector $(\Delta j^0,\Delta E_b^0)$ that is applied to the raw $E_b(j)$ 
         curves computed for an illustrative subset of the configuration listed in Table~\ref{tab:configs}.
         The numbers listed refer to curves where the final state  $(j_f,M_f)$ was already 
         imposed on the raw data, except for those datasets marked with an asterisk, $^*$.}
\centering
\begin{ruledtabular}
\begin{tabular}{lcccccccc}
  Name & $q$ & $\chi_1$  & $\chi_2$ & $\Delta E_b^0$ & $\Delta j^0$ \\
\hline 
SXS:BBH:0066       & 1   &    0      &   0       &    $-5.583\text{e-03}$    & $-2.13\text{e-02}$ \\
SXS:BBH:0169       & 2   &    0      &   0       &    $+1.224\text{e-05}$    & $-5.0\text{e-04}$  \\
SXS:BBH:0030       & 3   &    0      &   0       &    $-1.669\text{e-03}$     & $-9.0\text{e-03}$ \\
SXS:BBH:0167       & 4   &    0      &   0       &    $+1.917\text{e-04}$     & $+7.0\text{e-04}$ \\
SXS:BBH:0056       & 5   &    0      &   0       &    $-2.571\text{e-03}$    & $-1.25\text{e-02}$ \\
SXS:BBH:0166       & 6   &    0      &   0       &    $+1.942\text{e-04}$    & $+7.0\text{e-04}$  \\
SXS:BBH:0063       & 8   &    0      &   0       &    $-7.7709\text{e-04}$   & $-4.0\text{e-03}$  \\
SXS:BBH:0185       & 9.989 &  0      &   0       &    $+4.053\text{e-04}$    & $-2.4\text{e-03}$  \\
SXS:BBH:0156       & 1   &  $-0.95$  &   $-0.95$ &    $-9.496\text{e-05}$    & $-6.0\text{e-03}$  \\
SXS:BBH:0154$^*$   & 1   &  $-0.80$  &   $-0.80$ &    $-3.018\text{e-04}$    & $-0.9\text{e-03}$  \\
SXS:BBH:0151$^*$   & 1   &  $-0.60$  &   $-0.60$ &    $-6.507\text{e-05}$    & $-5.7\text{e-04}$  \\
SXS:BBH:0149       & 1   &  $-0.20$  &   $-0.20$ &    $-4.460\text{e-04}$    & $-3.8\text{e-03}$  \\
SXS:BBH:0150       & 1   &  $+0.20$  &   $+0.20$ &    $-4.200\text{e-04}$    & $-2.8\text{e-03}$  \\
SXS:BBH:0152$^*$   & 1   &  $+0.60$  &   $+0.60$ &    $-7.393\text{e-05}$    & $-0.7\text{e-03}$  \\
SXS:BBH:0155$^*$   & 1   &  $+0.80$  &   $+0.80$ &    $-2.536\text{e-04}$    & $-0.5\text{e-03}$               \\
SXS:BBH:0172       & 1   &  $+0.98$  &   $+0.98$ &    $-3.640\text{e-03}$    & $-1.05\text{e-02}$  \\
SXS:BBH:0178       & 1   &  $+0.994$ &   $+0.994$&    $-2.657\text{e-03}$    & $-8.8\text{e-03}$  \\
SXS:BBH:0162       & 2   &  $+0.60$  &   $0$     &    $-5.240\text{e-04}$    &  $-2.3\text{e-03}$   \\        
SXS:BBH:0036       & 3   &  $-0.50$  &   $ 0$    &    $-2.434\text{e-03}$    & $-5.5\text{e-03}$    \\
SXS:BBH:0031       & 3   &  $+0.50$  &   $ 0$    &    $-2.326\text{e-03}$    & $-7.1\text{e-03}$    \\
SXS:BBH:0064       & 8   &  $-0.50$  &   $ 0$    &    $-4.604\text{e-03}$    & $-5.5\text{e-03}$    \\
SXS:BBH:0065       & 8   &  $+0.50$  &   $ 0$    &    $-4.792\text{e-03}$    & $-6.8\text{e-03}$    \\
 \end{tabular}
\end{ruledtabular}
\label{tab:shifts}
\end{table}

The second shift vector, simply denoted $(\Delta j^{0},\Delta E_{b}^{0})$, is determined as follows.
First of all we look at the dimensionless``orbital frequency'' $\hat{\Omega}\equiv M\Omega$,
that is defined (both in the EOB model and NR) as  $\hOmg = \de E_b/\de j$. 
This quantity is, by construction, independent of the constant $\Delta E_{b}^{0}$,
but depends on $\Delta j^{0}$.  
We then compute the difference $\Delta \hOmg^{\rm EOBNR}(j) \equiv \hOmg^{\rm EOB}(j)-\hOmg^{\rm NR}(j) $
and plot it versus $j$. The shift $\Delta j^{0}$ is chosen so 
that $\Delta \hOmg(j)$ is ``flattened''
as much as possible on the largest possible $j$-interval. In practice, we choose $\Delta j^{0}$ 
such that the oscillating  (because of residual, tiny, eccentricity effects in the NR waveform)
$\Delta \hOmg(j) $ averages to zero. Figure~\ref{fig:Omg} shows the oscillating  $\Delta \hOmg(j)$ for two, fiducial,
configurations chosen randomly in the data sample: $(q,\chi_{1},\chi_{2})=(2,0.6,0)$ or $(q,\chi_{1},\chi_{2})=(3,-0.50,0)$.
In both cases, the raw $\Delta \hOmg$ (blue line) shows a trend, that can be flattened by 
just choosing properly the constant $\Delta j_{0}$ (red curve). Once this is done, we inspect
the difference $\Delta E_b^{\rm EOBNR}=E_b^{\rm EOB}-E_b^{\rm NR}$ and we find it remains
practically constant on a large $j$ interval. We then choose $\Delta E_{b}^{0}$ so that the 
$\Delta E_b^{\rm EOBNR}$ is found to oscillate around zero. This algorithm for determining
the shift vector $(\Delta j_{0},\Delta E_{b}^{0})$ can be applied to all SXS spin-aligned configurations
at our disposal. Note that the shifts that are so determined are, in most cases, tiny, 
but relevant on this scale.
In some cases (see e.g. case $(q,\chi_1,\chi_2)=(8,-0.5,0)$ in Fig.~\ref{fig:sxs_ej})
the combination of the two shifts leads to an end point for the shifted $E_b(j)$ NR
curve that is close to, but visibly displaced with respect to the $(j_{f},M_{f})$
provided in the {\tt metadata.txt}. Note however that in this case the total angular
momentum changes sign during the inspiral which may be connected to subtleties in
the computation of $E_b(j)$. We leave to future work a deeper investigation of this
and similar cases.

For completeness, we list in Table~\ref{tab:shifts} the 
shift vectors $(\Delta j^0,\Delta E_b^0)$ that are used to improve the compatibility 
between $(\Delta j^0,\Delta E_b^0)$ the NR $E_b(j)$ and the EOB one. The numbers
listed in the table always indicate the crucial ``second-shift'' vector mentioned above 
(i.e., are applied to curves already shifted by $(\Delta j^f,\Delta E_b^f)$ so as the 
final point of $E_b(j)$ approximately coincides with $(j_f,M_f)$), except for those 
datasets marked by an asterisk, where $(\Delta j^0,\Delta E_b^0)$ is applied directly
to the raw data.

\bibliography{refs20151124}

\end{document}